\def\revtex{1}

\ifx\revtex\undefined

\documentclass[entropy,article,submit,oneauthor,pdftex]{Definitions/mdpi}

\firstpage{1}
\pubvolume{1}
\issuenum{1}
\articlenumber{0}
\pubyear{2021}
\copyrightyear{2021}
\datereceived{}
\dateaccepted{}
\datepublished{}
\hreflink{https://doi.org/} 

\Title{Quantum Randomness is Chimeric}

\TitleCitation{Quantum randomness is chimeric}

\Author{Karl Svozil $^{1}$\orcidA{}}

\AuthorNames{Karl Svozil}

\AuthorCitation{Svozil, K.}

\address[1]{%
$^{1}$ \quad Institute for Theoretical Physics, TU Wien, Wiedner Hauptstrasse 8-10/136, 1040 Vienna,  Austria; svozil@tuwien.ac.at; \url{http://tph.tuwien.ac.at/~svozil}}

\corres{Correspondence: svozil@tuwien.ac.at}

\abstract{If quantum mechanics is taken for granted the randomness derived from it may be vacuous or even delusional, yet sufficient for many practical purposes. ``Random'' quantum events are intimately related to the emergence of both space-time as well as the identification of physical properties through which so-called objects are aggregated. We also present a brief review of the metaphysics of indeterminism.}



\begin{document}

\else
\documentclass[%
  twocolumn,
 showpacs,
 showkeys,
 preprintnumbers,
  nofootinbib,
 amsmath,amssymb,
 aps,
 pra,
  longbibliography,
 ]{revtex4-1}

\usepackage[dvipsnames]{xcolor}

\usepackage{mathptmx}

\usepackage{amssymb,amsthm,amsmath,bm}

\usepackage{tikz}
\usetikzlibrary{calc,decorations.pathreplacing,decorations.markings,positioning,shapes,snakes}

\usepackage[breaklinks=true,colorlinks=true,anchorcolor=blue,citecolor=blue,filecolor=blue,menucolor=blue,pagecolor=blue,urlcolor=blue,linkcolor=blue]{hyperref}
\usepackage{graphicx}
\usepackage{url}

%
%
%

\begin{document}

\title{Quantum randomness is chimeric}

\author{Karl Svozil}
\email{svozil@tuwien.ac.at}
\homepage{http://tph.tuwien.ac.at/~svozil}

\affiliation{Institute for Theoretical Physics,
TU Wien,
Wiedner Hauptstrasse 8-10/136,
1040 Vienna,  Austria}

\date{\today}

\begin{abstract}
If quantum mechanics is taken for granted the randomness derived from it may be vacuous or even delusional, yet sufficient for many practical purposes. ``Random'' quantum events are intimately related to the emergence of both space-time as well as the identification of physical properties through which so-called objects are aggregated. We also present a brief review of the metaphysics of indeterminism.
\end{abstract}


\maketitle

\fi

\setcounter{MaxMatrixCols}{20}

\section{Quantum oracles for randomness}

Almost 40~years since the glamorous inception~\cite{feynman,deutsch,deutsch:92,nielsen-book10,mermin-07}
of quantum computing, and
despite numerous grandiouse claims and prospects of quantum computational advantages~\cite{svozil-2016-quantum-hokus-pokus,Calude-C&E-2020,Um-2013},
only the random generation of bit sequences by beam
splitters~\cite{svozil-qct,zeilinger:qct,stefanov-2000,svozil-2009-howto,Furst:10,PhysRevA.82.022102,Abbott:2010uq,10.1038/nature09008,Abbott_2019}
has reached a certain commercial~\cite{Quantis} maturity.
Yet, these quantum random number generators present oracles~\cite{benn-84,svozil-qct} for ``randomness'',
which
(i) inductively are imagined and extrapolated to be a finitistic version of an
essentially  transfinite concept~\cite{martin-lof}. ``Certifications'' by NIST and DIEHARD and other sophisticated
test suites are of little consolation; and other natural resources for randomness
exhibit similar performances~\cite{PhysRevA.82.022102,Abbott_2019};
and
(ii) deductively are certifiable merely relative to the principles, assumptions, and axioms---such as, for instance, complementarity or
``contextuality''~\cite{svozil-2009-howto,10.1038/nature09008,PhysRevA.89.032109,2015-AnalyticKS}---they are based upon.
It is therefore of utmost importance to carefully delineate and be aware of these latter presumptions if we want to certify and trust such devices.

In what follows, we shall discuss randomness ``extracted'' from measurements of coherent superpositions of classically mutually exclusive
states, then proceed to multipartite and mixed states.
No quantum field theoretic many-particle effects such as stimulated or spontaneous emission or decay will be mentioned.
In the later parts of the paper, we shall attempt a brief history of physical events that have been deemed ``random'' and,
in particular, their relationship to the metaphysical ideas implied.

I encourage the reader to consider some of the content speculative and challenging---not as disrespectful to proposals and operationalisations of quantum randomness,
including some earlier ones I myself contributed to~\cite{svozil-qct,svozil-2009-howto,2012-incomput-proofsCJ,Abbott:2010uq}---but as reflections on some
aspects that might be noteworthy and even troubling.
A recent ``canonical'' presentation of quantum randomness in a broad perspective can be found in Reference~\cite{Bera2017}.
One might also add that certain interpretations of Everett's relative state formulation~\cite{everett}
suggest similar conclusions, albeit for very different reasons:
that randomness is an intrinsic illusion~\cite{Vaidman2014}.

\subsection{Quantum randomness through the measurement problem}

Quantum mechanics allows the coherent superposition (or, by another denomination, linear combination) of states which correspond to mutually exclusive outcomes.
The question arises: what kind of physical meaning can be given to these ``self-contradictory'' states?
Furthermore, is it not amazing that, for such states,
there exist two types of very closely related measurements that
give vastly different results: one random and one not?

Let me explain this question in some more detail for an ideal configuration, thereby neglecting observational (or measurement) errors;
in particular, no stochastic or random errors are taken into account.
Suppose one prepares a pure quantum state\footnote{By similar arguments as the ones exposed here the randomness of mixed states are epistemic rather than ontic,
and therefore, for all practical purposes, chimeric as well.}, say, by pre-selecting certain outcomes of beam splitter experiments.
If one ``measures'' this pre-selected states again and again by serial composition of either identical beam splitters,
or ``contextual intertwined'' beam splitters, one of whose output ports ``shares'' (and corresponds to) the pre-selected state~\cite{svozil-1999-haunted-qc,svozil:040102,Griffiths2017,Griffiths2019},
then a detector registering (or post-selecting) the ``resulting'' states (``after such serial processing'') will {\em always} click {\em with certainty}.
That is to say, such an experiment reveals a strictly {\em deterministic}, absolutely predictable behavior of this pre-selected quantum state.

Even the slightest physical ``tilt'' or ``rotation'' of one of the serially composed beam splitters changes the situation entirely and dramatically:
according to the standard quantum narrative, the experiment suddenly and discontinuously ``performs indeterministically'',
such that individual events---or at least post-processed sequences of such individual events---turn out to be irreducibly random~\cite{zeil-05_nature_ofQuantum}
(relative to maybe ``mild'' side assumptions, such as independence, any bias can be eliminated by (Borel) normalization~\cite{von-neumann1,Taub:1963:JNCa,AbbottCalude10}).
Such a physical manipulation of the beam splitter---literally ``tilting'' or ``rotating'' it---translates into a unitary transformation;
that is, a generalized ``rotation'', of the state (or context)
or (by the dyadic products) the respective observable proposition(s) in Hilbert space.
A ``slight detuning'' associated with a
``small'' change of the post-selected context with respect to the pre-selected context
will not ``throw the outcomes into crazy randomness''.
Indeed, the quantum probability is a smooth function of detuning, so a ``slight detuning'' will
only introduce a ``small'' amount of indeterminism in the raw data extracted.
Nevertheles, relative to certain mild side assumptions such as independence of events,
any such ``tiny signal'' of indeterminism in the raw data can be ``amplified into crazy randomness'' by (Borel) normalization, such as von Neumann's~\cite{von-neumann1}
partitioning of the raw data sequence into subsequences of length two, and then mapping
$00 \mapsto \emptyset$,
$11 \mapsto \emptyset$,
$01 \mapsto 0$, and
$10 \mapsto 1$.
This sudden, discontinuous change from determinism into complete indeterminism by some ``smooth, continuous'' manipulation
boggles a mind prepared to ``evangelically''~\cite{CLAUSER1992,clauser-talkvie}
accept the quantum canon.

For the sake of a concrete example, take
$\vert \psi \rangle =  \psi_0 \vert 0 \rangle + \psi_1 \vert 1 \rangle = \begin{pmatrix} \psi_0,\psi_1\end{pmatrix}^\intercal$ with
$\vert \psi_0  \vert^2 + \vert \psi_1  \vert^2 =1$
and ($\intercal$ stands for transposition),
$\vert 0 \rangle = \begin{pmatrix} 1,0\end{pmatrix}^\intercal$ and
$\vert 1 \rangle = \begin{pmatrix} 0,1\end{pmatrix}^\intercal$.
Suppose
we prepare or pre-select the quantized system to be in the state
$\vert \psi \rangle = \frac{1}{\sqrt{2}}\left( \vert 0 \rangle + \vert 1 \rangle \right)= \frac{1}{\sqrt{2}}\begin{pmatrix}1,1\end{pmatrix}^\intercal$,
and
we prefer to measure an observable $\vert \psi \rangle \langle  \psi \vert $
(that appears ``rotated'' or transformed relative to the observables $\vert 0 \rangle \langle  0 \vert$ and
$\vert 1 \rangle \langle 1 \vert$).
In such a case the system presents itself
as being perfectly determined and value definite; with the respective outcome always occurring.
No randomness or value definiteness\footnote{Value definiteness
shall be understood as ``possessing'' a well-defined property, encodable by some mathematical entity.
In terms of (ideal) measurements value definite properties yield the respective outcomes with certainty.}
can be ascribed to such a configuration.
With respect to $\vert \psi \rangle \langle  \psi \vert $  and its perpendicular orthogonal projection operator ${\bf 1}_2 - \vert \psi \rangle \langle  \psi \vert $
there is no uncertainty, and no possibility to obtain randomness.

Randomness comes about if ``detuned experiments'' are performed, such as,
for instance, the ones ``measuring observables'' corresponding to
the orthogonal projection operators
$\vert 0 \rangle \langle  0 \vert $ and $\vert 1 \rangle \langle  1 \vert= {\bf 1}_2 - \vert 0 \rangle  \langle 0  \vert $.
This concrete example features maximally or mutually unbiased~\cite{Schwinger.60} bases;
but any ``tiny'' rotation $0 \neq \varphi \ll 1$, with $\psi_0 = \cos \varphi$ and  $\psi_1 = \sin \varphi$
suffices to yield irreducible randomness through (Borel) normalization, as mentioned earlier.

An immediate question arises: why should such ``tilted'' or ``detuned'' experiments yield any results at all, and if so, in what way do outcomes of such ``wrong experiments''
come about; and to what extent do they
reflect any intrinsic property of the pre-selected state $\vert \psi \rangle $?
It is rather mindboggling that one should get any answer at all from such queries or ``detuned'' measurements.
But this may be  as confounding as it may be deceptive: because one might get the impression that there is a physical property ``out there'',
``sticking'' and being associated with the state.
I believe that mistakenly interpreting an experimental outcome---such as a detector click---as some inherent property,
constitutes a major epistemological issue that underlies many ill-posed claims and confusions
about such quantum states.
Indeed, these misconceptions may epitomize erroneous claims upon which quantum number generators by ``quantum coin tosses'' are based.

The quantum measurement problem is relevant for any judgment or certification or opinion on quantum randomness:
``extracting'' or ``reducing'' such states as $\vert \psi \rangle $ by ``measuring'' them in the ``wrong and detuned'' basis
$\vert 0 \rangle$ and $\vert 1 \rangle$  different from  $\vert \psi \rangle $ and its orthogonal vector
lies at the heart of the quantum measurement problem.
The respective ``process'', just as taking (partial) traces,
is non-unitary because it is postulated ``many-to-one'' and irreversible.
Therefore, such ``processes'' are inconsistent with the unitary quantum evolution, which is ``one-to-one'' and reversible.
(See Section~1.8 of Ref.~\cite{mermin-07} for a nice presentation.)

This inconsistency is an old issue that has already been raised by von Neumann~\cite{v-neumann-49,v-neumann-55},
Schr\"odinger~\cite{schrodinger}, London and Bauer~\cite{london-Bauer-1939,london-Bauer-1983}, Everett~\cite{everett,Barrett-2011,everett-1956} and Wigner~\cite{wigner:mb}.
It can be developed as a ``nesting'' or ``inverse Russian doll'' type argument by ever-increasing the domain of unitarity; including the measurement
apparatus and the measured state, and hence the interface or cut ``between'' them.
This has been proposed and operationalized in quantum optical experiments reconstructing the coherent superposition of states after
``measurements''~\cite{PhysRevD.22.879,PhysRevA.25.2208,greenberger2,Nature351,Zajonc-91,PhysRevA.45.7729,PhysRevLett.73.1223,PhysRevLett.75.3783,hkwz},
as well as in discussions about the insurmountable practical difficulties in doing so~\cite{engrt-sg-I,engrt-sg-II}.

Strictly speaking by assuming irreversible many-to-one ``processes'' one has to go beyond quantum mechanics in an {\it ad hoc} fashion.
Presently there is no evidence suggesting that this is necessary or even consistent with empirical data.
Should quantum mechanics be extended against all experimental evidence,
just because it is theoretically convenient and saves primitive notions of ``measurement''?

\subsection{Objectification by emergent context translation}

In what follows, it will be argued that any kind of measurement---in particular, also associated with ``detuned experiments''---constitutes an object or reality construction,
whereby the conventionality of measurement plays an essential role.
In this process, the very notion of objects or physical properties becomes conventionalized.
Objects or the properties constituting them may be real or chimeric; in the latter, chimeric case those experiments
relate to properties the system is fantasized about but not encoded in~\cite{zeil-99}.
In a metaphorical sense, this is like map-making or the creation of an encyclopedia, in which entries are constituted as facts or fiction,
or in any other way that is supposed to be consensical or intentional.

The term {\em chimeric} will be associated with coherent superpositions or linear combinations of different (mutually orthogonal) states,
{\em relative} to those states or their associated observable propositions involved.
For instance, $\vert \psi \rangle =  \psi_0 \vert 0 \rangle + \psi_1 \vert 1 \rangle$ with nonzero $\psi_1$ and $\psi_2$ is chimeric relative to
the propositions $\vert 0 \rangle \langle 0 \vert$ and $\vert 1 \rangle \langle 1 \vert$;
but is value definite or ``real'' and not chimeric relative to $\vert \psi \rangle \langle \psi \vert$.
States are not chimeric relative to the propositions associated with those exact states,
that is, $\vert \psi \rangle$ is ``real'' and not chimeric relative to $\vert \psi \rangle \langle \psi \vert$.

The emergent process of ``creating chimeras'' will be called {\em objectification} or object emergence or (re)construction.
Objectification is related to an ancient  conundrum~\cite{Yanofsky-object}:
the {\em Ship of Theseus}, or more generally, what is in Philosophy called ``the problem of identity''~\cite{Gallois-SEP,Gallois-POI}.
In the physical measurement process, it is the question of how, through ``mediation'' of its environment and the measurement apparatus,
a physical state or system which initially is unprepared to answer a particular query---or,
stated differently, is value indefinite and
chimeric---``translates'' the respective ``detuned'' query such that it is can respond to the request.
Through this ``context translation'' it may have acquired signals and information exterior to itself,
which may render the answer stochastic relative to itself (because of an influx from the open environment)
and to the experimental means available~\cite{svozil-2003-garda,svozil-2013-omelette} (containing or severing that open environment).

One might object that this stance reiterates a well-known fact: that quantum measurement introduce stochasticity.
The point of departure from this common view is the emphasis on the ``nesting'' aspect of the situation,
as outlined already by Everett~\cite{everett} and Wigner~\cite{wigner:mb}; but unlike them, more in the spirit of statistical physics:
in a Maxwellian view~\cite{Myrvold2011237}
the stochastic behavior (and entropy increase) originates from sampling---from not looking at the micro-physical level but at some
``aggregates''---rather than taking this for granted.

This has consequences for the stochasticity of chimeras:
they are not only based on some property intrinsic to the object but on the combined context by which
the object, as well as the apparatus, is defined~\cite{bohr-1949}.
Stochasticity enters by the many degrees of freedom of such a combined system.
This kind of  emergence of an ``experimental outcome'' associated with a counter reading of a (macroscopic) measurement apparatus has already been modeled
(i)
by a coupling of the object with the apparatus and its environment~\cite{Verrucchi-18},
and
(ii)
by ``attenuating'' a quantum signal from a state to cloning a ``noisy multitude'' of this state~\cite{glauber-collected-cat,Glauber-cat-86}  (it is always possible to clone two fixed orthogonal states)
``as much as possible'' (that is, nothing at all)  within the framework of the no-cloning theorem (cf.~Section~2.1 of Ref.~\cite{mermin-07}).

For the sake of understanding on which basis claims of absolute randomness are raised
beyond evangelical confessions~\cite{zeil-05_nature_ofQuantum,clauser-talkvie}
let me reconstruct current ``best-practice arguments'' for quantum indeterminacy~\cite{pitowsky:218} and
value indefiniteness~\cite{2015-AnalyticKS,svozil-2018-whycontexts} and their counterfactual~\cite{specker-60,specker-ges} character.
There ``exist'' collections of (counterfactual~\cite{specker-60}) observables comprising intertwining contexts (formalized by orthonormal bases or maximal operators  in dimension three or higher)
with two terminal point states---one serving as pre-selection or preparation, the other one for postselection or ``measurement''---with the following inconsistent properties:
upon pre-selection or preparation of a particular state   $\vert \Psi \rangle $,
(i) one such collection of observables enforces the {\em nonoccurrence} of some post-selected state $\vert \Phi \rangle $, associated with a certain {\em negative} experimental result;
(ii) another one such collection of observables enforces the {\em occurrence}
of some post-selected state $\vert \Phi' \rangle $,
associated with a certain {\em positive} experimental result~\cite{svozil-2006-omni,2018-minimalYIYS};
(iii) both post- and pre-selected states are the same, say,
$\vert \Psi \rangle = \begin{pmatrix} 1,0,0\end{pmatrix}^\intercal$ and
$\vert \Phi \rangle= \vert \Phi' \rangle =(1/2) \begin{pmatrix} \sqrt{2},1,1\end{pmatrix}^\intercal$~\cite{2015-AnalyticKS,svozil-2018-whycontexts,svozil-2020-c}.
Figure~\ref{2021-rto-Baba-Taher} sketches such a configuration.
The classical inconsistency arises from the fact that, depending on the arrangement of the quantum observables, the same observable must either be false (snake-like decorated curve)
and true (zigzag-like decorated curve) at the same time---a complete contradiction
amounting to the absurd prediction that a detector associated with such a binary observable simultaneously registers
a click and does not do so.
Relative to the assumptions made
$\vert \Phi \rangle$ given $\vert \Psi \rangle$
cannot have a classical value definite truth assignment:
any such truth assignment would need to be undefined at least for $\vert \Phi \rangle$.
This yields the truth assignment as a partial function, a notion well known in theoretical computer science~\cite{Kleene1936}
The argument can be extended to any state not collinear with or orthogonal to the pre-selected state $\vert \Psi \rangle$~\cite{2015-AnalyticKS}.

\ifx\revtex\undefined

\begin{figure}[H]
\includegraphics[width=10.5 cm]{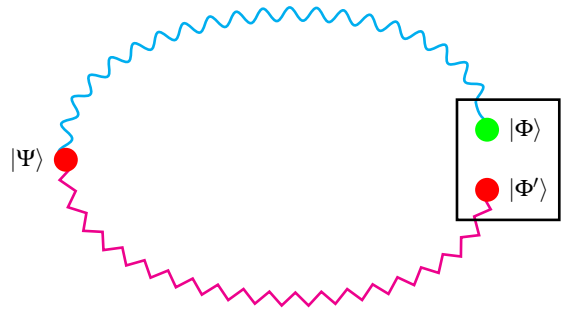}

\else

\begin{figure}
        \begin{center}
                \begin{tikzpicture}  [scale=0.8]

                        \newdimen\ms
                        \ms=0.1cm

                        \tikzstyle{every path}=[line width=1pt]

                        \tikzstyle{c1}=[draw=gray,fill=white,circle,inner sep={\ms/1}]
                        \tikzstyle{c2}=[color=blue,fill,circle,inner sep={\ms/8},minimum size=2*\ms]
                        \tikzstyle{c3}=[color=red,fill,circle,inner sep={\ms/8},minimum size=2*\ms]

                        \newdimen\R
                        \R=30mm     



                        %

                        \coordinate (a1) at (0,0);
                        \coordinate (a8) at (7,0.5);
                        \coordinate (a14) at (7,-0.5);

                        \draw [color=cyan, decoration = snake,decorate] (a1) to [bend left=90]  (a8);
                        \draw [color=magenta, decoration = zigzag,decorate] (a1) to [bend right=90]  (a14);


                        \draw (a1) coordinate[c1,draw=red,fill=red,label=left:$\vert \Psi \rangle$];
                        \draw (a8) coordinate[c1,draw=green,fill=green,label=right:$\vert \Phi \rangle$];
                        \draw (a14) coordinate[c1,draw=red,fill=red,label=right:$\vert \Phi'\rangle$];

                        \draw (6.5,-1) rectangle (8.2,1);

                \end{tikzpicture}
        \end{center}
\fi
        \caption{\label{2021-rto-Baba-Taher}
                Serial composition of two gadget hypergraphs with terminal points $\vert \Psi \rangle$ and approaching
$\vert \Phi  \rangle \leftrightarrow \vert \Phi' \rangle$.
                The snake-like decorated curve indicates a classical true-implies-false relation.
                The zigzag-like decorated curve indicates a classical true-implies-true relation.
        }
\end{figure}

Another implicit assumption that is seldom mentioned because it is assumed evident is the
omni-existence of the collection of complementary observables
(because the argument involves different contexts).
Indeed, the coexistence of counterfactual, complementary observables is (mostly implicitly) assumed without further discussion.
One common response to critical doubts about their existence is that ``they can be measured''.
That is, a particular state $\vert \psi \rangle$ can be prepared or pre-selected
and subsequently, the proposition corresponding to another ``mismatching'' state  $\vert \varphi \rangle$
(which should neither be orthogonal to nor collinear with $\vert \psi \rangle$)
can be measured or postselected.
This, of course, is omni-realism pure and simple.

Coming back to the argument sketched in Figure~\ref{2021-rto-Baba-Taher}, it is evident that, due to pre-selection or preparation, the state $\vert \Psi \rangle$
and its associated observable proposition $\vert \Psi \rangle \langle \Psi \vert$ is value definite relative to measurements $\vert \Psi \rangle \langle \Psi \vert$.
But should this be assumed for all the other observables entering the argument?
In particular, should value definiteness be expected from some state $\vert \Phi \rangle$ given  $\vert \Psi \rangle$?
Because $\vert \Phi \rangle$, and all other observables, entering as counterfactual ``intermediaries'' in the argument,
need to be in a coherent superposition of states different from the pre-selected state $\vert \Psi \rangle$ and other states,
which makes them chimeric
relative to  $\vert \Psi \rangle$.

\subsection{Information theoretic approach to quantum randomness}

A related information-theoretic argument for ``irreducible''~\cite{zeil-05_nature_ofQuantum}
quantum randomness contends that a quantum system can ``carry'' only a
finite amount of information~\cite{zeil-99,zeil-bruk-99a,zeil-bruk-99}--- namely (maximally) about the occurrence of a single proposition within a single context.
Therefore, the
{\em ``$\ldots$~reason for the irreducible randomness in
quantum measurement $\ldots$~is the simple fact that an elementary system cannot carry enough information to provide definite answers to all questions
that could be asked experimentally.''}
Stated differently, {\em ``there are less available answers than possible questions''}~\cite{Grangier2018}.
Any query attempting to forcefully retreive ``more'' information from such a quantized system is confronted with this ``underdetermination'',
resulting in ontological indeterminism.

Alternatively, one might argue that
this ``insistance on the enforced retrieval of information the quantum system is unprepared to hold'' results in a context translation.
Typical examples are ``detuned experiments'' mentioned earlier, associated with an influx of information from the environment and,
in particular, the measurement apparatus.
This effectively results in epistemic quantum indeterminism.

One could still maintain that, through nesting~\cite{everett,wigner:mb} and the effects
of the translational environment the number of degrees of freedom during the measurement cannot be bounded from above
and approaches infinity, resulting in nonseparable hyper-Hilbert spaces~\cite{vonNeumann1939}, a situation which might yield
a sort of irreducible randomness  based on the diverging complexity of the environment~\cite{Auffves2019}.
Note that even classically the hypothetical invocation of infinity ``in the limit produces'' provable random sequences, such as Chaitin's
halting probability Omega~\cite{calude-dinneen06}.

\subsection{Entanglement and emergence of space-time}

Einstein's primary intent~\cite{einstei-letter-to-schr,Meyenn-2011,Howard1985171} in writing a paper
with Podolsky and Rosen~\cite{epr} (EPR) was to present a {\em separation principle} or {\em separation hypothesis}:
given any two (space-like) separated subsystems $A$ and $B$ of a joint system $(A\;B)$, then
$B$ (my translation, see also\cite{Howard1985171})
{\em ``and everything related to its content is independent of what happens with regard to''} $A$.
Thereby,
Einstein's presumption has been that, after any interaction between $A$ and $B$ in the past
(quoted from the same letter,  my translation, see also~\cite{Howard1985171})
{\em ``the real state of $(A\;B)$ consists of the real state of $A$ and the real state of $B$,
which two states have nothing to do with one another''.}

This latter assumption, at least for Einstein, is one pillar of the EPR argument.
However, suppose that we are not inclined to follow Einstein's critique of quantum mechanics but
propose that, rather than quantum theory, space-time physics, and relativity theory
would need to adapt in case there is a collision with quantum mechanics.
Then the separation principle should be considered
incorrect and not be applied for entangled quantum states introduced by
Schr\"odinger~\cite{schrodinger,CambridgeJournals:2027212,CambridgeJournals:1737068}
around the time of the EPR paper.
In particular, there exist entangled states of two subsystems $A$ and $B$ which are indecomposable; that is,
they cannot be written as the product of the states of the two ``separated'' systems
$A$ and $B$; more formally,
$\vert \Psi(A\; B) \rangle \neq \vert \psi (A) \rangle \otimes \vert \phi (B) \rangle$,
where $\otimes$ stands for the tensor product.

This inseparability, as discussed by Schr\"odinger in the measurement context (between object and measurement apparatus)
has been re-interpreted in terms of relational properties~\cite{zeil-99} for multi-partite configurations.
It comprises two parts---a restrictive and an extensive property for classical physical systems:
(i) quantum mechanics limits the amount of information encodable in a quantized system from above, and
(ii) it allows the storage, resampling~\cite{svozil-2016-sampling} or scrambling of such limited information ``across quanta''.
Both properties can be viewed as direct consequences of the unitary transformations postulated as
formalizations of quantum state evolution, because entangled systems are merely
``a unitary transformation apart'' from separable states~\cite[Section~12.8.2]{svozil-pac}.

Let us pursue a very radical, iconoclastic deviation from the Kantian idea
that space-time is an {\it a priori} theatric frame, a sort of scaffolding,
in which physics takes place. Rather, suppose that
\begin{itemize}
\item[(i)]
in reversing Einstein's verdict mentioned earlier,
for (maximally) entangled states of a composite system $(A\;B)$,
its constituents share a common identity---that is,
they ``are tied together'' and can be considered ``being aspects of a single entity'' and,
in particular, ``not spatio-temporally separated at all'';
so much so that any individuality or separateness vanishes.

\item[(ii)]
Space-time needs to be {\em derived} from quantum effects as an (emergent) {\em epiphenomenon},
a secondary effect or byproduct that arises vis-\`a-vis  quantized systems
and does not stand separate from or independent of them.
\end{itemize}

In this view distances are a matter of disentanglement and gradual:
two events such as detector clicks are ``apart'' if their corresponding states
are (for all practical purposes) factorizable and decomposable, and thus disentangled.
Spacio-temporal separations and distances are to be understood more like the second law
of thermodynamics~\cite{Myrvold2011237}: they are not absolute but relative
to the (entanglement) means involved.
This creates a ``patchwork'' of clocks and rulers, associated with the respective entanglements.
Such emergent space-time frames need not necessarily be consistent with one another,
but rather form a mesh of spatial-temporal networks.

Most radically, what may be considered ``far apart'' in the old Kantian-Einsteinian
framework maybe not be separated at all in the new scheme.
For most practical purposes~\cite{bell:a1,bell-a},
the two notions of spatial-temporal distances may coincide.
Because entanglement and ``nonlocality'' with respect to the old ``absolute''
theatrical framework of space-time (for all practical purposes) ``happens locally'' and---again according
to the {\it Ancien R\'egime} in terms of Kantian-Einsteinian space-time frames---not ``far away''.


This radical departure from the Kant-Einsteinian framework of space-time by
emergence from entanglement has been discussed in entanglement-induced
gravity~\cite{VanRaamsdonk2010,VANRAAMSDONK2010b,Faulkner2014,Swingle:2014uza,Jacobson2016,Cao2017,Swingle2018,Musser2018}.
See also Ref.~~\cite{Knuth-Bahreyni} for another approach to emergent space-time.
This research program is a new and active area of research.

A lot of questions arise immediately.
One issue that need to be addressed is that of the finite speed of light,
as compared to instantaneous entanglement: can some finite speed of information transfer
be derived from an infinite property? One Ansatz is given in Ref.~\cite{Couch2020}.
What is (inertial) motion, and the type of kinematics resulting from entanglement?
Entanglement swapping comes to mind immediately, but this lacks any notion of inertia.
Indeed, we might be tempted to speculate that the absence of inertia,
rather than being a problematic feature, might be an advantage, suggesting
possibilities of inertialess motion~\cite{Knuth-e21100939}, and motion beyond the relativistic speed limit.
It might not appear too unreasonable to speculate, that,
if entanglement swapping takes place instantaneously, so maybe motion or signaling in space and time,
even despite the following discussion.

\subsection{Peaceful coexistence}

The argument stated by Einstein in his letter~\cite{einstei-letter-to-schr,Meyenn-2011,Howard1985171} to Schr\"odinger quoted earlier
amounts to the aforementioned separation principle: measurement of a subsystem $A$ of $(A\;B)$
cannot alter the state of the subsystem  $B$;
in particular, not if the two subsystems are spatially separated.
As noted earlier, Einstein attacked quantum mechanics for failing this principle for entangled multi-partite states.
But as our approach considers the emergence of space-time as secondary to quantization,
rather than questioning the validity of quantum mechanics
we might as well respond with an ``upside-down'' question: why not?  Why is space-time not challenged by these issues?
To answer such questions it might be prudent to compare a similar classical EPR-type configuration
with classical and more general resources.
We can imagine at least two scenarios:
\begin{itemize}
\item[(i)]
{\em Value {\em definiteness} of the individual constituents $A$ and $B$ and the fixing of their respective {\em local} shares at creation point:}
for this scenario Peres gave a most insightful analysis~\cite{peres222}.
Classical ``singlet'' states (e.g., obtained by the preservation
of angular momentum) may exhibit certain (dis-)similar behaviors as compared to the quantum case.
Classically the joint system $(A\;B)$ ``carries'' some ``common share''---e.g.,
a hidden parameter such as the opposite angular momentum pseudovectors
of the particles~\cite{peres,toner-bacon-03,svozil-2004-brainteaser} along one and the same direction.
These angular momentum pseudovectors are fixed and value definite for both parties or subsystems  $A$ and $B$
already after their interaction.
Therefore, the local information can in principle be used to produce local ``copies'' or ``clones'' of $A$ and $B$.
This is consistent with relativity theory because those shares remain fixed after their creation,
so that whatever manipulation happens on one side does not alter the respective state or share on the other side.
\item[(ii)]
{\em Value {\em indefiniteness} of the individual constituents $A$ and $B$ but the fixing of their respective {\em global} shares at creation point:}
This may for instance be achieved by assuming a global value definite share or state of $(A\;B)$;
and yet  by not allowing or ``granting'' definite states to the individual constituents $A$ and $B$.
Therefore, any attempt to copy them fails because of the absence of value definiteness.
Quantum mechanics ``guarantees'' or realizes such a scenario by demanding that
any entangled quantized pair $(A\;B)$ exhibits a relational encoding.
The states of the individual constituents $A$ and $B$ are not value definite: they lack
``definiteness'' or ``memory'' or information about individual properties
of its constituents---the value definiteness ``resides'' in the relational (not the individual),
holistic, global, ``collective'' properties among the constituents~\cite{zeil-99}.
If such individual properties are ``enforced'' upon the constituents through measurement, they react with
a context translation which, through nesting, introduces stochasticity
because of the many degrees of freedom introduced from the ``outside'' environment.
As a result one obtains outcome independence although one still obtains parameter dependence;
but the latter is only ``recoverable'' after the outcomes from both sides are compared~\cite{shimony2,shimony3};
locality prevails~\cite{Griffiths2010,Griffiths2020}.
\end{itemize}

{\it Per se} both scenarios could be extended to any type
of two-partite expectation functions,
which need not be linear as in the classical case, but can take on any functional form;
in particular also the quantum ``stronger''-than-classical,
nonlinear (trigonometric because of the projective character
of the quantum probabilities) form.
Indeed, by the same argument expectations and correlations might be even ``stronger'' than classical and quantum
ones~\cite{popescu-97,svozil-krenn,svozil-2004-brainteaser,popescu-2014}
without violating Einstein locality.

Some argue that random outcomes ``save'' quantum mechanics from violating relativistic causality.
Because if it were possible to somehow use the relational encoding of entangled inseparable states,
either by duplicating nonorthogonal states~\cite{herbert}, or by stimulated emmission~\cite{svozil-slash},
then $B$ could infer information on $A$'s settings even {\em before} knowing $A$'s outcome {\it post factum},
{\it posterior} and in retrospect (after combining the knowledge of both outcomes).
The random outcomes on $A$'s side assure that $B$ cannot know what happens at the former side.
This argument can be extended to stronger-than-quantum correlations.

However, this kind of ``peaceful coexistence''~\cite{shimony2,shimony3} may also be seen as a characterization
of the second scenario~(ii) discussed earlier.
In particular, if one is considering the ``common share'' accessible to $A$ and $B$:
it is, say, a pure entangled state of $(A\;B)$; more formally, it is an indecomposable vector.
As it is not decomposable, there is no meaning associated with individual properties of $A$ and $B$.
In this form, quantum entanglement defines spatio-temporal proximity, yet cannot produce
any means of communication between the entangled parties:
the ``more entangled'' the parties get the ``less individual'' properties they carry.
Their common share, such as indecomposable vectors, cannot give rise to any form of classical communication between the
entangled parties as it is useless.

I, therefore, suggest that rather than speaking about a ``peaceful coexistence'' between relativity and quantum theory
we should speak of this no-signaling constraint as an unavoidable feature of emergent space-time from entanglement.
The value-definiteness of the common indecomposable vector share of $(A\;B)$; that is, in a
value indefiniteness of the individual states of $A$ and $B$ results in stochasticity if individuality is forced
upon those subsystems; very much in the same way as stochasticity emerges (by context translation) from
coherent superpositions or linear combinations of states, when measured ``along with the detuned, twisted contexts''; as
sketched earlier.

\section{Historic perception of randomness}


In what follows, randomness will be discussed in the historic context.
This is important because of the lessons one could learn for the contemporary debate and perception of lawlessness and randomness.
According to an influential narrative,
the European Enlightenment
developed as a courageous, thorough, and highly successful\footnote{The criterion of success is taken relative to and in terms of full-spectrum dominance compared to alternative worldviews grounded in esoteric thought.}
exorcism of  transcendence;
in particular, the rejection of law-defying {\em miracles}~\cite{Swinburne-Miracle};
moreover, the empirical sciences ``established natural laws'' of regular, reliable tempo-spacial coincidences
which appear to be trustworthy and therefore of great utility.

The denial of any direct breach or ``rupture''
of the laws of nature~\cite[Sect.~III,~10]{frank,franke}
has pushed the boundaries of conceivable transcendental real-time interventions,
and, in particular, divine providence, to the fringe
of ``gaps''~\cite[Sect.~III,~12]{frank,franke} in the laws of nature---
indeterminate situations where applicable laws,
and thus the {\em Principle of Sufficient Reason}~\cite{sep-sufficient-reason} have not (yet?) been identified.

As effective as the formal~\cite{wigner} and natural sciences are in terms of utility,
they turn out to be as means and context relative\footnote{Means relativity of an entity such as an idea
or a physical theory is
the dependence (eg., validity, existence) of this entity on the means, conventions, or assumptions employed.
Context relativity relates to whatever are the circumstances that form the setting for an event
in terms of which it can be fully understood.
Perhaps means and context relativity are equivalent notions, yet the emphasis lies on different aspects of a situation.}
as any construct of thought:
those imaginations of the human mind
cannot deliver any ``Archimedean point''
or ``ontological anchor'' upon which an ``objective reality''
(whatever that is) can be based.
Indeed, it is my idealistic~\cite{berkeley,stace,stace1,Goldschmidt2017-idealism}  observation, or,
rather stance or conviction,
that all our physical narratives~\cite{descartes-meditation,Nietzsche-WahrheitLuege,derrida-Royle},
doubles~\cite{Arthaud,Arthaud-en}, images~\cite{hertz-94,hertz-94e} and---more optimistically---representations~\cite{plato-republic}
of what we experience as ``Nature''
are metaphysical---or at least amalgamated with metaphysical components---and
ultimately can be denounced as being suspended in our free thought.
Therefore, historically we experience a succession of incongruent, incommensurable~\cite{Feyerabend-62,fey-philpapers1,kuhn,Oberheim2005,sep-incommensurability,Pigliucci2018}
scientific research programs~\cite{lakatosch,lakatos_1978};
a lineup which should make us humble when it comes to the mind-boggling effectiveness~\cite{wigner} of some of our formalisms in predicting, programming,
manipulating and instrumentalizing physical systems\footnote{The desperation, if not
nihilism, that results from the deconstruction of long-held beliefs and narratives
has been very vividly described
by Schopenhauer~\cite{schopenhauer-dwawuv-VI},
as well as through Nietzsche's
{\it \"Ubermensch}~\cite{Nietzsche-ZarathustraI,Nietzsche-EcceHomo}
and Camus' {\it {S}isyphe}~\cite{camus-mos}.}.

An obvious counter-response to such idealistic positions is to contend that physics is firmly grounded in
empirical data are drawn from observation of experimental outcomes.
Support of theoretical physical models in the form of corroboration or falsification~\cite{popper,popper-en}
by empirical evidence~\cite{sep-francis-bacon}  is indispensable.
As an extreme demand, physical theory should strive to include only operational entities which are physically realizable
in terms of achievable actions and measurements~\cite{bridgman27,bridgman,bridgman36,bridgman50,bridgman52}.

However, the history of science presents ample evidence that it has never been possible to resort to empirical evidence
for the advancement or discrimination of theoretical models alone~\cite{kuhn,lakatosch,lakatos_1978}.
Indeed, as stated by Einstein~\cite{einstei-letter-to-schr} (reprinted as Letter 206 in~\cite{Meyenn-2011}, my translation),
there is a metaphysical circularity because
{\em ``the real difficulty lies in the fact that physics is a kind of metaphysics;
Physics describes `reality'. But we don't know what `reality' is;
we only know it through the physical description!''}
And although both the prediction and the willful reproduction of phenomena
appears to be the cornerstone of current natural sciences,
the ``empirical evidence'' relating to ``scientific facts'' is often indirect and fragile,
deserving a nuanced and careful analysis~\cite{Hume-Treatise,Hume-Enquiry}.

I shall offer three examples for the type of problems encountered in quantum mechanics;
all three related to the occurrence of certain ``clicks'' of detectors.
Arguably the occurrence or non-occurrence of such a click is the most elementary, binary observable one could think of.
But while the (non)registration of detector clicks may be considered undisputable
(for all practical purposes~\cite{bell-a}, and nonwithstanding quantum erasures or haunted measurements~\cite{PhysRevD.22.879,PhysRevA.25.2208,greenberger2,Nature351,Zajonc-91,PhysRevA.45.7729,PhysRevLett.73.1223,PhysRevLett.75.3783,hkwz})
the  ``meaning'' of such clicks~\cite{svozil-2017-b} remain open to a great variety of  perceptions, interpretations, and understandings.

The first example is about  measurements~\cite{aspect-81} of violations of classical locality with time-varying analyzers~\cite{aspect-82b}
if the periodic switching is synchronized with photon emmissions~\cite{zeilinger-86}.
A second example is about a debate~\cite{Kimble-aposterioriQT,Bouwm-aposterioriQTReply} on quantum teleportation~\cite{Bouwmeester1997,BBCJPW}.
A third example is about the contingencies~\cite{svozil-2020-c}
arising from counterfactual arguments of Hardy-type configurations~\cite{2018-minimalYIYS,svozil-2020-hardy}.
These cases document well the different claims and aspects derived from single detector clicks, as perceived by different participating discussants.


Other aspects related to very general limits on symbolic representations need to be acknowledged.
Any formalization of physical (in)determinism by (in)computability,
and physical randomness as algorithmic incompressibility,
and general induction~\cite{go-67,blum75blum,angluin:83,ad-91,li:92}
would require transfinite means not available~\cite{gandy1}
in this Universe~\cite{svozil-93,svozil-unev,svozil-07-physical_unknowables}.
This is because the associated
formal proofs are blocked by the aforementioned
G\"odel-Turing-type incompleteness/incomputability results.

Therefore, one cannot expect that the formal and natural sciences
offer absolute corroboration of any type of semantic statements.
All they allow is the systematic exploitation of syntax and narratives
which are true relative to the chosen means and purposes.

In what follows, we shall first discuss what general options of randomness can be imagined;
and then proceed with a discussion of their concrete physical {\it modi operandi.}

\subsection{Bowler type scenario of a clockwork universe}

The assumption of a ``clockwork universe''---that is, ``stuff'' such as
matter, energy,  together with its assorted evolution laws which are uniformly valid and unique (leaving no room for alternatives)---
entails a ``bowler''-type situation\footnote{In what follows, ``god'' or ``deity'' is understood as an entity creating existence; a sort of ``programmer of the Universe.''}.
The Principle of Sufficient Reason~\cite{sep-sufficient-reason} rules; nothing occurs without a ``reason'' or ``cause''.
Once this universe is created {\it ex nihilo} and put into motion
there is no further or additional interference with it; as all necessary and sufficient conditions exist to
determine its evolution uniquely and completely from a ``previous'' state into a ``later''
one\footnote{In such a scenario free will appears to be illusory and subjectively,
as per assumption choices are merely fictitious and
delusional}.

\subsubsection{How could physics facilitate and support such a view?}

Here are some arguments that may be put forward {\em in favour of} a bowler-type clockwork universe:

\begin{itemize}

\item[(i)]
The description of a unique physical state as a {\em function} of some operational physical quantity such as time---
indeed, the very notion of {\em total function} (as opposed to partiality~\cite{Kleene1936}),
{\em Laplace's demon}, {\em causal~\cite{Norton-2003-cafs} determinism}
and the Principle of Sufficient Reason are scientific tropes and schemes
signifying clockwork universes.
They were widely held in pre-statistical physics and quantum areas until around {\it fin de si\`ecle}.

In ordinary differential equations of
classical continuum mechanics and classical electrodynamics
the semantic notion of ``determinism''
is formalized by the {\em uniqueness} of the solutions, which
are guaranteed by a Lipschitz continuity
condition~\cite[Chapter~17]{svozil-pac}.

\item[(ii)]
The quantum state evolution is postulated to be unique and deterministic\footnote{Formally it is represented by a unitary transformation, that is,
a generalized rotation mapping one orthonormal basis into another one.
Such a state evolution is one-to-one and thus reversible and unique.
However, if the preparation context differs from the measurement context,
the quantum state does not identify outcomes uniquely,
thereby allowing one particular kind of quantum indeterminacy.}.
However, in general---in the case of coherent superposition or mixed states---the
quantum state is not operationally accessible.
Therefore this sort of
quantum determinacy cannot be given any direct empirical meaning.

\item[(iii)]
{\em Deterministic chaos} is characterized by a unique initial value---
a ``seed''
supposed to be taken from the mathematical continuum and thus
incomputable and even random\footnote{Randomness of an infinite string
is taken to be algorithmically incompressible~\cite{martin-lof}.}
with probability one---whose information or digits are ``revealed''
by some suitable deterministic temporal evolution.
To be suitable a temporal evolution needs to be very sensitive to changes of
initial seeds such that very small
fluctuations may produce very large effects.
This is like Maxwell's gap scenario discussed later.

Like quantum evolution, deterministic chaos might be considered both an argument
for and against classical determinism: because
the assumption of the continuum renders almost all seeds formally random~\cite{martin-lof},
thereby passing all statistical tests of randomness; in particular ``elementary'' test such as
Borel normality, certifying that all sequences of arbitrary length occur
with the expected frequency\footnote{Unfortunately
Borel normality is no guarantee of randomness because very regular sequences,
for instance,
the Champernowne constant~\cite{Sloane_oeis.org/A033307} $C_{10}$ in base $10$
is just the sequence obtained by concatenating
successive numbers (encoded in base $10$),
turn out to be normal.}, but also much stronger ones.

In this respect classical machinery designed to use
extreme sensitivities of the temporal evolution to the initial seed,
 such as the Athenian~\cite{dow_aristotlekleroteria_1939}
$\kappa \lambda \eta \rho \omega \tau \eta \rho \iota o \nu$
({\it kleroterion}),
for all practical purposes is not inferior to a quantum oracle
for randomness, such as {\it QUANTIS}~\cite{Quantis},
based on the ``evangelical'' belief of irreducible quantum randomness~\cite{zeil-05_nature_ofQuantum}.

\item[(iv)]
In  system science or virtual physics, this modus could be referred to as a very restricted {\em virtual reality,}
{\em computational gaming environment} or {\em simulation}~\cite{toffoli:79,fredkin,svozil-nat-acad,Bostrom-sim} ({\it aka} simulacrum),
whereby it is  assumed that there is
no interference from ``the outside'' ({\it aka} beyond): the respective universe is hermetic.
No participation is possible; only passive (without interference) observation.
\end{itemize}

\subsubsection{How could physics contradict such a view?}

Here are some arguments that may be put forward {\em against} a bowler-type clockwork universe:

\begin{itemize}

\item[(i)]
Classical gaps are characterized by {\em instabilities}  at {\em singular points}, such that very small
fluctuations may produce very large effects.
To quote Maxwell~\cite[pp.~211,212]{Campbell-1882},
{\em ``for example, the rock loosed by frost and balanced on a singular point of the mountain-side, the little spark which
kindles the great forest~$\ldots$ At these
points, influences whose physical magnitude is too small to be taken account of by a finite being, may produce
results of the greatest importance''.}

\item[(ii)]
In some physical situations the
Lipschitz continuity is violated, yielding no unique solutions.
The Norton dome~\cite{Norton-dome-2008,vanStrien2014} is a
contemporary example of such a situation.

\item[(iii)]
Spontaneous symmetry breaking,
a physical (re)source of nonuniqueness,
is a spontaneous process
by which a physical system in a symmetric state ends up in an asymmetric state.
This is facilitated by some appropriate ``Mexican hat'' potential,
not dissimilar to Norton's dome or
Maxwell's~\cite[pp.~211,212]{Campbell-1882}
{\em ``rock loosed by frost and balanced on a singular point''} mentioned earlier.

In particle physics the Higgs mechanism, the spontaneous symmetry breaking of gauge symmetries,
plays an important role in the origin of particle masses in
the standard model of particle physics.
All of these ruptures or breaches of uniqueness depend on the assumptions and models involved.

\item[(iv)]
Quantum indeterminacy, in particular, complementarity, contextuality ({\it aka} value indefiniteness),
and aspects (such as the exact decay time) of the occurrence of certain single events are postulated to signify indeterminism.
\end{itemize}

Because of both formal and empirical reasons these scenarios might no be interrelated and not separate:
for instance, one might suspect that Maxwell's
instabilities  at  singular points could be formalized by ``Mexican hat'' type potentials discussed in spontaneous symmetry breaking,
or by ordinary differential equations yielding Norton dome-type configurations.
One might even speculate that all violations of Lipschitz continuity amount to some kind of symmetry breaking.

Empirically one might argue that, for all practical purposes~\cite{bell-a}, Maxwell's scenario and Norton dome-type configurations
(related to violations of Lipschitz continuity) or spontaneous symmetry breaking, never ``actually'' happen.
Because for all practical purposes a rock loosed by frost is never (with probability zero)
symmetrically balanced at a singular point; rather the position of its center of gravity will
fluctuate around the tip, thereby spoiling symmetry.
Also one may argue that, due to (vacuum) fluctuations, singular points make no operational sense; they
are (over)idealized concepts invented by the human mind for mere convenience.
In particular, microscopic quantum zero-point fluctuations, and  thermal fluctuations~\cite{Smoluchovski-1912}
ultimately spoil symmetries.
Therefore, all such exploitations of such singularities might confuse epistemic convenience with an ontology that has no physical, operational grounds.

\subsection{Scenario of a stochastic, disorganized universe}

The ``converse'' of a Laplacean determinism governed by a unique state evolution
``tied to'' previous states, as mentioned in the previous section, is one in which any given state is
independent\footnote{Two events $A$ and $B$
are statistically independent if their joint probability $P(A\cap B)$ can be written
as the product of their single probabilities $P(A)$ and $P(B)$; that is,
$P(A\cap B)= P(A)P(B)$.
It turns out that this results in a journey down a rabbit hole, as the concept of probability
is a nontrivial one~\cite{Uffink2011-UFFSPS}.}
of the respective previous (and future) states.
In such a most extreme scenario among many conceivable
degrees of stochasticity
the universe is ``completely'' stochastic and disorganized on the most fundamental level.
For the embedded observer's intrinsic perspective, due to irreducible contingency and chance,
it appears as if such a world is constantly created anew by throwing some sort of
dice\footnote{This may be considered an extreme form of {\it creatio continua.}
However, {\em extrinsically}---that is, from an external, extrinsic, perspective---this may be considered {\em creatio ex nihilo} as
no active, real-time participation is assumed.
Indeed, one may speculate that
if the temporal ordering of events (and causality) turns out to be epistemic---
an intrinsically emerging concept/observable of  (self-)cognition/observation---
then any differentiation
based on temporal creation---such as {\it creatio continua {\em versus} ex nihilo}---turns out to be a ``red herring.''
Alas, without granting ``time'' some ontology, also differentiation between a ``bowling'' or ``curling'' god collapse.\label{2019-cob-lcc-cen}}.

Whether and how some sort of structural continuity of existence can emerge and be maintained under such circumstances is a fascinating question.
As in such a scenario space and time, as much as notions of causality and the laws, are emergent concepts, continuity might emerge with them.

Indeed, one might speculate that ``the laws'' are some sort of
expressions of chaos\footnote{This is not dissimilar to the
impossible choice not to communicate~\cite{Watzlawick-1967}.}, the formation of matter and genes are expressions of these laws,
the individuals carrying those genes are expressions thereof~\cite{Hamilton-1963}, and that the ideas about the world are expressions of these individuals.
In that transitive way, the Universe contemplates itself through our ideas---ideas such as religion, mathematics, ethics, and so on.

Contemporary physics supports such a view in postulating that many elementary events
 --  such as the spontaneous or stimulated emission of photons---occur acausally, irreducibly pure and simple~\cite{born-26-1,zeil-05_nature_ofQuantum}.
Indeed, both classical statistical physics at finite resolution\footnote{A Laplacian demon with unbounded resources might be able to determine
future states from present ones with arbitrary precision.}, and quantum mechanics, support such a view.

The Viennese {\it fin de si\`ecle} physicist Exner~\cite{Hiebert2000,Stoeltzner-1999}, motivated by statistical physics
and the radiation law~\cite{schweidler-1905},
suggested that~\cite[p.~7,18]{Exner-1908}
{\em ``$\ldots$~laws do not exist in nature,
those are only formulated by man, he makes use of it
as a linguistic and computational aid
and only wants to say
that the processes in nature run as if matter, like a sentient being, would obey these laws.
$\ldots$
So we must understand all so-called exact laws
only as average laws, which are not valid with absolute certainty,
but with the higher probability
the more individual processes they result from.
All physical laws
go back to molecular processes of random nature
and from them follows the result according to the laws
of probability calculus~$\ldots$~.''}

Even in totally ``random'' datasets, some sort of structure must necessarily emerge
by the law of large numbers:
for instance, if two dice are thrown sufficiently often, the number seven appears to be the most likely sum of their two faces.
Modern arguments for the emergence of laws from chaos employ,
among other methods~\cite{armstrong_1983,vanFraassen1989-VANLAS,calude-meyerstein,lawlses_rosen2010,calude2013theeinai,chaos_multiverse2017,Mueller-2017,Cabello-2018-BornRule},
Ramsey theory, for structure formation and structural continuity through spurious correlations~\cite{svozil-2018-was}.
It is irrelevant whether these events occur ``absolutely randomly''---indeed,
as has been pointed out earlier, on an individual level and with finitistic means, ``absolute randomness''
appears to be a vacuous concept.

\subsection{The intermediate curler case}

Intuitively the curler case~\cite{Clark-2017-GodAsCurler}
is one in which the natural laws---whatever their form and origin---predominate,
but there are situations in which such laws do not exist, or if laws exist they are violated.
The first ``weak'' case of indeterminism can be realized by gaps\footnote{As mentioned earlier~\cite[Sect.~III,~10]{frank,franke}
``stronger'' forms of curling involve a ``rupture'' of the laws of nature, as they
are in direct violations of those laws
as mentioned in Voltaire's Philosophical Dictionary~\cite[Chapter~330]{voltaire-dict}.
Although nobody can {\it a priori}exclude such latter cases we shall henceforth stick with
Hume's attitude towards miracles~\cite[Section~X]{Hume-Enquiry}   and neglect them.}.

Theologically this could be perceived as a mild form of {\it creatio continua}\footnote{{\it Cf.} my earlier remarks on {\it creatio continua}
in footnote~\ref{2019-cob-lcc-cen}.}: god has created laws that are not violated,
but god also left  ``some room'' to communicate  {\it via} gaps.

A ``god of the gaps'' has been rephrased in many ways.
This concept is also quite popular since, on the one hand, the obvious regularities of experience and life express correlations or laws which appear evident:
the daily cycle of the sun, the yearly cycle of the seasons, life, death; apples and other stuff falling down and not up, and so on.
So denial of regularities appears futile.
On the other hand, humans experience fate and uncontrollable circumstances quite often.
In a similar reaction, the primitive mind (re)interpreted such ``evidence'' as god's signal.

As more and more ``fateful'' behaviors became ``understood'' and even controllable\footnote{Think of conditions treatments and also
volcanic eruptions,
floods or weather phenomena such as lightning and thunder.} it is not unreasonable to speculate that,
maybe, eventually, there will be no such gaps left---in which case
one recovers the bowler, {\em ex nihilo,} scenario.
Alternatively some ``pure'' gaps in the causal fabric of our universe
might ``turn out''---that is, relative to the assumptions and means employed---
to be irreducible and final: those gaps cannot be eliminated and might remain forever.
In secular terms, this could be suspected to signify irreducible indeterminism or randomness~\cite{zeil-05_nature_ofQuantum}.
But there exist other, possibly transcendental, interpretations involving {\em intentionality} across gaps.

That these latter scenarios are not purely speculative can be demonstrated by an interactive gaming scenario:
If one is considering an interactive virtual reality environment~\cite{simula,permutationcity} one usually assumes that the virtual reality is
``sustained'' or ``supported'' by a computational process ``running'' on some kind of computer whose physical characteristics
are not directly related\footnote{To be feasible and nonmonotonic it can be assumed without loss of generality that both the
universe in which the simulation is implemented and the simulated universe are capable of universal computation in
the sense of Chuch-Turing.} to the simulacrum.
To be interactive the two universes need to be intertwined and connected
by some sort of (bidirectional) gap through which
information flows in both ``directions''\footnote{This could result in a sort of {\em dialogue} between those realms---a ``backflow'' from the simulacrum to the universe in which the
simulation takes place---such that the former simulacrum
performs ``empirical studies'' on the latter, thereby fully and actively participating in it.
In this very speculative scenario, ``transcendence becomes immanence.''
Think of evolving artificial intelligence in a computer simulation becoming aware of its situation and asking online players questions
about its situation and the general setup.
However, as symmetric as an exchange through the interface may appear, it is asymmetric in one aspect: whereas the simulacrum cannot exist without the world
in which the simulation takes place
the latter can exist without the former.}.
For an intrinsic~\cite{svozil-94} observer embedded~\cite{toffoli:79}
in the virtual environment and bound by its operational means {\em the capacity to send an arbitrary signal through the interface
 --  from the simulating universe ({\it aka} ``the beyond'')  to the simulacrum---can only be realized by a gap.}
Because without a gap, the signal must remain immanent;
that is, it reduces to either lawful or chaotic behavior.

Gaps potentially allow some ``transcendental'' exchange of signals but do not necessarily imply such a
conversation or dialogue.
Therefore, gaps are a necessary but not a sufficient condition for transcendence---just because gaps have been located does not imply the existence of  ``active'' transcendental entities.

From a theological perspective, gaps can realize individual (human) soul/mind-body dualism~\cite{eccles:papal},
and also divine providence~\cite[Sect.~III,~9-16]{frank,franke}.

How does physics support gaps? Can physics rule them out?
The following is an update and extension of Frank's discussion on physical gaps.
\begin{itemize}

\item[(i)]
As has been mentioned earlier, in the classical domain of ordinary differential equations some breach of the
Lipschitz continuity condition~\cite[Chapter~17]{svozil-pac}
could cause nonunique solutions.
Often such types of gaps are identified with instabilities
at their singular points~\cite[pp.~211,212]{Campbell-1882}, \cite[Sect.~III,~13]{frank,franke}.

\item[(ii)]
As has also been discussed earlier, quantum complementarity, and, as an extension thereof, quantum contextuality ({\it aka} value indefiniteness) can
be interpreted as the impossibility to co-represent~\cite{peres,kochen1,2015-AnalyticKS}
certain (even finite) sets of---necessarily {\em counterfactual} because they are complementary
 --  quantum observables,
relative to the asssumptions\footnote{One assumption entering those proofs are the (context) independence
of outcomes of measurements for ``intertwine'' observables occurring in more than one context.
For reasons of being able to intertwine contexts formalized by orthonormal bases
this can only happen in vector spaces of dimension higher than two.}.
This is problematic as the corresponding experimental protocols
(``prepare a pure state and measure a different one'') seem to suggest that they
``reveal'' some pre-existing property---indicated by the (non)occurrence of a detector click.
This could be misleading, as the respective click might
either be subject to debate and interpretation\footnote{A debate~\cite{Kimble-aposterioriQT,Bouwm-aposterioriQTReply}
on the alleged ``{\it a posteriori} teleportation'' is an example for such a nonunique
semantic perception of syntactically undisputed detector clicks.}
or merely signify the capacity of the measurement apparatus
to ``translate an improper question;'' introducing stochastic noise~\cite{svozil-2003-garda}.
This appears to be related to notorious inconsistencies
in quantum physics proper~\cite{v-neumann-49,v-neumann-55,everett,wigner:mb,everett-collw} due to the
assumption of irreversible quantum measurements.

\item[(iii)]
Aspects of certain individual, single events in quantized systems such
as the time of emission or absorption of single quanta of light,
are postulated to be indeterministic.
\end{itemize}

\section{The (un)known (un)knowns}

The relativity of the considerations on the respective assumptions and means invested or taken for granted results in
an echo-chamber of sorts: whatever one puts in one gets out.
As mentioned earlier there is no ``firm (meta)physical ground,'' no  undisputable
``Archimedean ontological anchor'' upon which such speculations can be based.
And the tendency of the mind to rationalize, project~\cite{Jaynes1990,Freud-1912,Freud-1912-e} and empathically
embrace opinions that are favorable to one's ego-investments
increases delusions about particular beliefs and corroborations thereof even further.

At this point, the Reader might get frustrated: a negative message (akin to a negative theology)
has been delivered\footnote{One positive side effect might be the abandoning of what the Vienna Circle (in a Humean tradition) called
``meaningless pseudo-statements''~\cite{Hahn1930,Carnap1931,Carnap-1931-engl}
targeting a particular hocus-pocus, abracadabra delusional (thought) rituals delivered by sophistic philosophers and an orthodox clergy.
However, one has to be very careful not to ``throw the baby out with the bathwater.''
Shortly after these bold rejections of metaphysical entities,
it turned out that their program based on empirical evidence and formal logic proposed could not be carried out as completely as
desired~\cite{godel1,turing-36,smullyan-92,Smullyan1993-SMURTF,book:486992,chaitin3}.}.
Alas, unfortunately, this is all that can be safely stated.

Therefore, we should accept the sober fact that there is certainty only in our uncertainty.
This has been expressed by many insightful individuals of many philosophical traditions and religions and at various times.
Aurelius Augustinus, for instance, writes~\cite[Book~XI, chapter~25.32]{Augustinus-Confessiones},
{\em ``Do I perhaps not know how to express what I do know?
Woe is me: I do not even know what it is I do not know!''}

\section{Summary}

Quantum randomness appears epistemic:
identical pre- and post-selected states and observables yield definite outcomes
because the vector or projection operator shares are identical.
If there is a mismatch between preparation and measurement, then the measurement apparatus, as part of the environment,
may ``contribute'' to the respective outcomes by context translation.
Therefore, randomness extracted from coherent superpositions or linear combinations of the quantum state might be based on
the complexity of the environment rather than on the intrinsic, ontologic ``oracle'' nature of the state.
``Objectification''---the emergence of a property which the original state is not encoded in---is
associated with this influx of information from the environment.

This readily extends into entanglement: relationaly encoded quantum shares (that can be pure entangled states
represented by inseparable vectors) will not be able to render
individual value definiteness of its constituents that is necessary for communication between those constituents.
This relates to the concept of emergent space-time from
separation through nonentanglement, and inseparability by entanglement.

In the second part of the paper, a wealth of historic resources on random physical outcomes has been reviewed.
The emphasis has been on the ``evangelical'' side of the perception of value indefiniteness,
as it has emerged historically.

\ifx\revtex\undefined

\funding{This research was funded in whole, or in part, by the Austrian Science Fund (FWF), Project No. I 4579-N. For the purpose of open access, the author has applied a CC BY public copyright licence to any Author Accepted Manuscript version arising from this submission.}

\acknowledgments{I kindly acknowledge discussions with and suggestions by Cristian Calude, Kelly James Clark, Silvia Jonas, Jeffrey Koperski, Irem Kurtsal, Emil Salim, Mohammad Hadi Shekarriz, and Noson S. Yanofsky.}

\conflictsofinterest{The author declares no conflict of interest.
The funders had no role in the design of the study; in the collection, analyses, or interpretation of data; in the writing of the manuscript, or in the decision to publish the~results.}

\else

\begin{acknowledgments}

This research was funded in whole, or in part, by the Austrian Science Fund (FWF), Project No. I 4579-N. For the purpose of open access, the author has applied a CC BY public copyright licence to any Author Accepted Manuscript version arising from this submission.


I kindly acknowledge discussions with and suggestions by
Cristian Calude, Kelly James Clark, Philippe Grangier, Silvia Jonas, Jeffrey Koperski, Irem Kurtsal, Emil Salim, Mohammad Hadi Shekarriz, and Noson S. Yanofsky.


All misconceptions and errors are mine.

The author declares no conflict of interest.
\end{acknowledgments}

\fi

\ifx\revtex\undefined

\end{paracol}
\reftitle{References}


\externalbibliography{yes}
\bibliography{svozil}

\begin{thebibliography}{221}%
\makeatletter
\providecommand \@ifxundefined [1]{%
 \@ifx{#1\undefined}
}%
\providecommand \@ifnum [1]{%
 \ifnum #1\expandafter \@firstoftwo
 \else \expandafter \@secondoftwo
 \fi
}%
\providecommand \@ifx [1]{%
 \ifx #1\expandafter \@firstoftwo
 \else \expandafter \@secondoftwo
 \fi
}%
\providecommand \natexlab [1]{#1}%
\providecommand \enquote  [1]{``#1''}%
\providecommand \bibnamefont  [1]{#1}%
\providecommand \bibfnamefont [1]{#1}%
\providecommand \citenamefont [1]{#1}%
\providecommand \href@noop [0]{\@secondoftwo}%
\providecommand \href [0]{\begingroup \@sanitize@url \@href}%
\providecommand \@href[1]{\@@startlink{#1}\@@href}%
\providecommand \@@href[1]{\endgroup#1\@@endlink}%
\providecommand \@sanitize@url [0]{\catcode `\\12\catcode `\$12\catcode
  `\&12\catcode `\#12\catcode `\^12\catcode `\_12\catcode `\%12\relax}%
\providecommand \@@startlink[1]{}%
\providecommand \@@endlink[0]{}%
\providecommand \url  [0]{\begingroup\@sanitize@url \@url }%
\providecommand \@url [1]{\endgroup\@href {#1}{\urlprefix }}%
\providecommand \urlprefix  [0]{URL }%
\providecommand \Eprint [0]{\href }%
\providecommand \doibase [0]{http://dx.doi.org/}%
\providecommand \selectlanguage [0]{\@gobble}%
\providecommand \bibinfo  [0]{\@secondoftwo}%
\providecommand \bibfield  [0]{\@secondoftwo}%
\providecommand \translation [1]{[#1]}%
\providecommand \BibitemOpen [0]{}%
\providecommand \bibitemStop [0]{}%
\providecommand \bibitemNoStop [0]{.\EOS\space}%
\providecommand \EOS [0]{\spacefactor3000\relax}%
\providecommand \BibitemShut  [1]{\csname bibitem#1\endcsname}%
\let\auto@bib@innerbib\@empty
\bibitem [{\citenamefont {Feynman}(1982)}]{feynman}%
  \BibitemOpen
  \bibfield  {author} {\bibinfo {author} {\bibfnamefont {Richard~Phillips}\
  \bibnamefont {Feynman}},\ }\bibfield  {title} {\enquote {\bibinfo {title}
  {Simulating physics with computers},}\ }\href {\doibase 10.1007/BF02650179}
  {\bibfield  {journal} {\bibinfo  {journal} {International Journal of
  Theoretical Physics}\ }\textbf {\bibinfo {volume} {21}},\ \bibinfo {pages}
  {467--488} (\bibinfo {year} {1982})},\ \bibinfo {note} {physics of
  computation, Part II (Dedham, Mass., 1981)}\BibitemShut {NoStop}%
\bibitem [{\citenamefont {Deutsch}(1985)}]{deutsch}%
  \BibitemOpen
  \bibfield  {author} {\bibinfo {author} {\bibfnamefont {David}\ \bibnamefont
  {Deutsch}},\ }\bibfield  {title} {\enquote {\bibinfo {title} {Quantum theory,
  the {C}hurch-{T}uring principle and the universal quantum computer},}\ }\href
  {\doibase 10.1098/rspa.1985.0070} {\bibfield  {journal} {\bibinfo  {journal}
  {Proceedings of the Royal Society of London. Series A, Mathematical and
  Physical Sciences (1934-1990)}\ }\textbf {\bibinfo {volume} {400}},\ \bibinfo
  {pages} {97--117} (\bibinfo {year} {1985})}\BibitemShut {NoStop}%
\bibitem [{\citenamefont {Deutsch}\ and\ \citenamefont
  {Jozsa}(1992)}]{deutsch:92}%
  \BibitemOpen
  \bibfield  {author} {\bibinfo {author} {\bibfnamefont {David}\ \bibnamefont
  {Deutsch}}\ and\ \bibinfo {author} {\bibfnamefont {Richard}\ \bibnamefont
  {Jozsa}},\ }\bibfield  {title} {\enquote {\bibinfo {title} {Rapid solution of
  problems by quantum computation},}\ }\href {\doibase 10.1098/rspa.1992.0167}
  {\bibfield  {journal} {\bibinfo  {journal} {Proceedings of the Royal Society:
  Mathematical and Physical Sciences (1990-1995)}\ }\textbf {\bibinfo {volume}
  {439}},\ \bibinfo {pages} {553--558} (\bibinfo {year} {1992})}\BibitemShut
  {NoStop}%
\bibitem [{\citenamefont {Nielsen}\ and\ \citenamefont
  {Chuang}(2010)}]{nielsen-book10}%
  \BibitemOpen
  \bibfield  {author} {\bibinfo {author} {\bibfnamefont {Michael~A.}\
  \bibnamefont {Nielsen}}\ and\ \bibinfo {author} {\bibfnamefont {I.~L.}\
  \bibnamefont {Chuang}},\ }\href {\doibase 10.1017/CBO9780511976667} {\emph
  {\bibinfo {title} {Quantum Computation and Quantum Information}}}\ (\bibinfo
  {publisher} {Cambridge University Press},\ \bibinfo {address} {Cambridge},\
  \bibinfo {year} {2010})\ \bibinfo {note} {10th Anniversary
  Edition}\BibitemShut {NoStop}%
\bibitem [{\citenamefont {Mermin}(2007)}]{mermin-07}%
  \BibitemOpen
  \bibfield  {author} {\bibinfo {author} {\bibfnamefont {David~N.}\
  \bibnamefont {Mermin}},\ }\href {\doibase 10.1017/CBO9780511813870} {\emph
  {\bibinfo {title} {Quantum Computer Science}}}\ (\bibinfo  {publisher}
  {Cambridge University Press},\ \bibinfo {address} {Cambridge},\ \bibinfo
  {year} {2007})\BibitemShut {NoStop}%
\bibitem [{\citenamefont {Svozil}(2016)}]{svozil-2016-quantum-hokus-pokus}%
  \BibitemOpen
  \bibfield  {author} {\bibinfo {author} {\bibfnamefont {Karl}\ \bibnamefont
  {Svozil}},\ }\bibfield  {title} {\enquote {\bibinfo {title} {Quantum
  hocus-pocus},}\ }\href {\doibase 10.3354/esep00171} {\bibfield  {journal}
  {\bibinfo  {journal} {Ethics in Science and Environmental Politics (ESEP)}\
  }\textbf {\bibinfo {volume} {16}},\ \bibinfo {pages} {25--30} (\bibinfo
  {year} {2016})},\ \Eprint {http://arxiv.org/abs/arXiv:1605.08569}
  {arXiv:1605.08569} \BibitemShut {NoStop}%
\bibitem [{\citenamefont {Calude}\ and\ \citenamefont
  {Calude}(2020)}]{Calude-C&E-2020}%
  \BibitemOpen
  \bibfield  {author} {\bibinfo {author} {\bibfnamefont {Cristian~S.}\
  \bibnamefont {Calude}}\ and\ \bibinfo {author} {\bibfnamefont {Elena}\
  \bibnamefont {Calude}},\ }\bibfield  {title} {\enquote {\bibinfo {title} {The
  road to quantum computational supremacy},}\ }in\ \href {\doibase
  10.1007/978-3-030-36568-4\_22} {\emph {\bibinfo {booktitle} {{JBCC} 2017:
  {F}rom Analysis to Visualization}}},\ \bibinfo {series} {Springer Proceedings
  in Mathematics \& Statistics}, Vol.\ \bibinfo {volume} {313}\ (\bibinfo
  {publisher} {Springer, Cham},\ \bibinfo {year} {2020})\ pp.\ \bibinfo {pages}
  {349--367},\ \bibinfo {note} {{J}onathan {M}. {B}orwein Commemorative
  Conference. A celebration of the life and legacy of {J}onathan {M}.
  {B}orwein, {C}allaghan, {A}ustralia, {S}eptember 2017},\ \Eprint
  {http://arxiv.org/abs/arXiv:1712.01356} {arXiv:1712.01356} \BibitemShut
  {NoStop}%
\bibitem [{\citenamefont {Um}\ \emph {et~al.}(2013)\citenamefont {Um},
  \citenamefont {Zhang}, \citenamefont {Zhang}, \citenamefont {Wang},
  \citenamefont {Yangchao}, \citenamefont {Deng}, \citenamefont {Duan},\ and\
  \citenamefont {Kim}}]{Um-2013}%
  \BibitemOpen
  \bibfield  {author} {\bibinfo {author} {\bibfnamefont {Mark}\ \bibnamefont
  {Um}}, \bibinfo {author} {\bibfnamefont {Xiang}\ \bibnamefont {Zhang}},
  \bibinfo {author} {\bibfnamefont {Junhua}\ \bibnamefont {Zhang}}, \bibinfo
  {author} {\bibfnamefont {Ye}~\bibnamefont {Wang}}, \bibinfo {author}
  {\bibfnamefont {Shen}\ \bibnamefont {Yangchao}}, \bibinfo {author}
  {\bibfnamefont {D.~L}\ \bibnamefont {Deng}}, \bibinfo {author} {\bibfnamefont
  {Lu-Ming}\ \bibnamefont {Duan}}, \ and\ \bibinfo {author} {\bibfnamefont
  {Kihwan}\ \bibnamefont {Kim}},\ }\bibfield  {title} {\enquote {\bibinfo
  {title} {Experimental certification of random numbers via quantum
  contextuality},}\ }\href {\doibase 10.1038/srep01627} {\bibfield  {journal}
  {\bibinfo  {journal} {Scientific Reports}\ }\textbf {\bibinfo {volume} {3}},\
  \bibinfo {pages} {1--7} (\bibinfo {year} {2013})}\BibitemShut {NoStop}%
\bibitem [{\citenamefont {Svozil}(1990)}]{svozil-qct}%
  \BibitemOpen
  \bibfield  {author} {\bibinfo {author} {\bibfnamefont {Karl}\ \bibnamefont
  {Svozil}},\ }\bibfield  {title} {\enquote {\bibinfo {title} {The quantum coin
  toss---testing microphysical undecidability},}\ }\href {\doibase
  10.1016/0375-9601(90)90408-G} {\bibfield  {journal} {\bibinfo  {journal}
  {Physics Letters A}\ }\textbf {\bibinfo {volume} {143}},\ \bibinfo {pages}
  {433--437} (\bibinfo {year} {1990})}\BibitemShut {NoStop}%
\bibitem [{\citenamefont {Jennewein}\ \emph {et~al.}(2000)\citenamefont
  {Jennewein}, \citenamefont {Achleitner}, \citenamefont {Weihs}, \citenamefont
  {Weinfurter},\ and\ \citenamefont {Zeilinger}}]{zeilinger:qct}%
  \BibitemOpen
  \bibfield  {author} {\bibinfo {author} {\bibfnamefont {Thomas}\ \bibnamefont
  {Jennewein}}, \bibinfo {author} {\bibfnamefont {Ulrich}\ \bibnamefont
  {Achleitner}}, \bibinfo {author} {\bibfnamefont {Gregor}\ \bibnamefont
  {Weihs}}, \bibinfo {author} {\bibfnamefont {Harald}\ \bibnamefont
  {Weinfurter}}, \ and\ \bibinfo {author} {\bibfnamefont {Anton}\ \bibnamefont
  {Zeilinger}},\ }\bibfield  {title} {\enquote {\bibinfo {title} {A fast and
  compact quantum random number generator},}\ }\href {\doibase
  10.1063/1.1150518} {\bibfield  {journal} {\bibinfo  {journal} {Review of
  Scientific Instruments}\ }\textbf {\bibinfo {volume} {71}},\ \bibinfo {pages}
  {1675--1680} (\bibinfo {year} {2000})},\ \Eprint
  {http://arxiv.org/abs/arXiv:quant-ph/9912118} {arXiv:quant-ph/9912118}
  \BibitemShut {NoStop}%
\bibitem [{\citenamefont {Stefanov}\ \emph {et~al.}(2000)\citenamefont
  {Stefanov}, \citenamefont {Gisin}, \citenamefont {Guinnard}, \citenamefont
  {Guinnard},\ and\ \citenamefont {Zbinden}}]{stefanov-2000}%
  \BibitemOpen
  \bibfield  {author} {\bibinfo {author} {\bibfnamefont {Andr{\'{e}}}\
  \bibnamefont {Stefanov}}, \bibinfo {author} {\bibfnamefont {Nicolas}\
  \bibnamefont {Gisin}}, \bibinfo {author} {\bibfnamefont {Olivier}\
  \bibnamefont {Guinnard}}, \bibinfo {author} {\bibfnamefont {Laurent}\
  \bibnamefont {Guinnard}}, \ and\ \bibinfo {author} {\bibfnamefont {Hugo}\
  \bibnamefont {Zbinden}},\ }\bibfield  {title} {\enquote {\bibinfo {title}
  {Optical quantum random number generator},}\ }\href {\doibase
  10.1080/095003400147908} {\bibfield  {journal} {\bibinfo  {journal} {Journal
  of Modern Optics}\ }\textbf {\bibinfo {volume} {47}},\ \bibinfo {pages}
  {595--598} (\bibinfo {year} {2000})}\BibitemShut {NoStop}%
\bibitem [{\citenamefont {Svozil}(2009{\natexlab{a}})}]{svozil-2009-howto}%
  \BibitemOpen
  \bibfield  {author} {\bibinfo {author} {\bibfnamefont {Karl}\ \bibnamefont
  {Svozil}},\ }\bibfield  {title} {\enquote {\bibinfo {title} {Three criteria
  for quantum random-number generators based on beam splitters},}\ }\href
  {\doibase 10.1103/PhysRevA.79.054306} {\bibfield  {journal} {\bibinfo
  {journal} {Physical Review A}\ }\textbf {\bibinfo {volume} {79}},\ \bibinfo
  {eid} {054306} (\bibinfo {year} {2009}{\natexlab{a}})},\ \Eprint
  {http://arxiv.org/abs/arXiv:quant-ph/0903.2744} {arXiv:quant-ph/0903.2744}
  \BibitemShut {NoStop}%
\bibitem [{\citenamefont {F\"{u}rst}\ \emph {et~al.}(2010)\citenamefont
  {F\"{u}rst}, \citenamefont {Weier}, \citenamefont {Nauerth}, \citenamefont
  {Marangon}, \citenamefont {Kurtsiefer},\ and\ \citenamefont
  {Weinfurter}}]{Furst:10}%
  \BibitemOpen
  \bibfield  {author} {\bibinfo {author} {\bibfnamefont {Martin}\ \bibnamefont
  {F\"{u}rst}}, \bibinfo {author} {\bibfnamefont {Henning}\ \bibnamefont
  {Weier}}, \bibinfo {author} {\bibfnamefont {Sebastian}\ \bibnamefont
  {Nauerth}}, \bibinfo {author} {\bibfnamefont {Davide~G.}\ \bibnamefont
  {Marangon}}, \bibinfo {author} {\bibfnamefont {Christian}\ \bibnamefont
  {Kurtsiefer}}, \ and\ \bibinfo {author} {\bibfnamefont {Harald}\ \bibnamefont
  {Weinfurter}},\ }\bibfield  {title} {\enquote {\bibinfo {title} {High speed
  optical quantum random number generation},}\ }\href {\doibase
  10.1364/OE.18.013029} {\bibfield  {journal} {\bibinfo  {journal} {Optics
  Express}\ }\textbf {\bibinfo {volume} {18}},\ \bibinfo {pages} {13029--13037}
  (\bibinfo {year} {2010})}\BibitemShut {NoStop}%
\bibitem [{\citenamefont {Calude}\ \emph {et~al.}(2010)\citenamefont {Calude},
  \citenamefont {Dinneen}, \citenamefont {Dumitrescu},\ and\ \citenamefont
  {Svozil}}]{PhysRevA.82.022102}%
  \BibitemOpen
  \bibfield  {author} {\bibinfo {author} {\bibfnamefont {Cristian~S.}\
  \bibnamefont {Calude}}, \bibinfo {author} {\bibfnamefont {Michael~J.}\
  \bibnamefont {Dinneen}}, \bibinfo {author} {\bibfnamefont {Monica}\
  \bibnamefont {Dumitrescu}}, \ and\ \bibinfo {author} {\bibfnamefont {Karl}\
  \bibnamefont {Svozil}},\ }\bibfield  {title} {\enquote {\bibinfo {title}
  {Experimental evidence of quantum randomness incomputability},}\ }\href
  {\doibase 10.1103/PhysRevA.82.022102} {\bibfield  {journal} {\bibinfo
  {journal} {Physical Review A}\ }\textbf {\bibinfo {volume} {82}},\ \bibinfo
  {pages} {022102} (\bibinfo {year} {2010})}\BibitemShut {NoStop}%
\bibitem [{\citenamefont {Abbott}\ \emph
  {et~al.}(2014{\natexlab{a}})\citenamefont {Abbott}, \citenamefont {Calude},\
  and\ \citenamefont {Svozil}}]{Abbott:2010uq}%
  \BibitemOpen
  \bibfield  {author} {\bibinfo {author} {\bibfnamefont {Alastair~A.}\
  \bibnamefont {Abbott}}, \bibinfo {author} {\bibfnamefont {Cristian~S.}\
  \bibnamefont {Calude}}, \ and\ \bibinfo {author} {\bibfnamefont {Karl}\
  \bibnamefont {Svozil}},\ }\bibfield  {title} {\enquote {\bibinfo {title} {A
  quantum random number generator certified by value indefiniteness},}\ }\href
  {\doibase 10.1017/S0960129512000692} {\bibfield  {journal} {\bibinfo
  {journal} {Mathematical Structures in Computer Science}\ }\textbf {\bibinfo
  {volume} {24}},\ \bibinfo {pages} {e240303} (\bibinfo {year}
  {2014}{\natexlab{a}})},\ \Eprint {http://arxiv.org/abs/arXiv:1012.1960}
  {arXiv:1012.1960} \BibitemShut {NoStop}%
\bibitem [{\citenamefont {Pironio}\ \emph {et~al.}(2010)\citenamefont
  {Pironio}, \citenamefont {Ac{\'i}n}, \citenamefont {Massar}, \citenamefont
  {{Boyer de la Giroday}}, \citenamefont {Matsukevich}, \citenamefont {Maunz},
  \citenamefont {Olmschenk}, \citenamefont {Hayes}, \citenamefont {Luo},
  \citenamefont {Manning},\ and\ \citenamefont {Monroe}}]{10.1038/nature09008}%
  \BibitemOpen
  \bibfield  {author} {\bibinfo {author} {\bibfnamefont {S.}~\bibnamefont
  {Pironio}}, \bibinfo {author} {\bibfnamefont {A.}~\bibnamefont {Ac{\'i}n}},
  \bibinfo {author} {\bibfnamefont {S.}~\bibnamefont {Massar}}, \bibinfo
  {author} {\bibfnamefont {A.}~\bibnamefont {{Boyer de la Giroday}}}, \bibinfo
  {author} {\bibfnamefont {D.~N.}\ \bibnamefont {Matsukevich}}, \bibinfo
  {author} {\bibfnamefont {P.}~\bibnamefont {Maunz}}, \bibinfo {author}
  {\bibfnamefont {S.}~\bibnamefont {Olmschenk}}, \bibinfo {author}
  {\bibfnamefont {D.}~\bibnamefont {Hayes}}, \bibinfo {author} {\bibfnamefont
  {L.}~\bibnamefont {Luo}}, \bibinfo {author} {\bibfnamefont {T.~A.}\
  \bibnamefont {Manning}}, \ and\ \bibinfo {author} {\bibfnamefont
  {C.}~\bibnamefont {Monroe}},\ }\bibfield  {title} {\enquote {\bibinfo {title}
  {Random numbers certified by {B}ell's theorem},}\ }\href {\doibase
  10.1038/nature09008} {\bibfield  {journal} {\bibinfo  {journal} {Nature}\
  }\textbf {\bibinfo {volume} {464}},\ \bibinfo {pages} {1021--1024} (\bibinfo
  {year} {2010})}\BibitemShut {NoStop}%
\bibitem [{\citenamefont {Abbott}\ \emph {et~al.}(2019)\citenamefont {Abbott},
  \citenamefont {Calude}, \citenamefont {Dinneen},\ and\ \citenamefont
  {Huang}}]{Abbott_2019}%
  \BibitemOpen
  \bibfield  {author} {\bibinfo {author} {\bibfnamefont {Alastair~A}\
  \bibnamefont {Abbott}}, \bibinfo {author} {\bibfnamefont {Cristian~S}\
  \bibnamefont {Calude}}, \bibinfo {author} {\bibfnamefont {Michael~J}\
  \bibnamefont {Dinneen}}, \ and\ \bibinfo {author} {\bibfnamefont {Nan}\
  \bibnamefont {Huang}},\ }\bibfield  {title} {\enquote {\bibinfo {title}
  {Experimentally probing the algorithmic randomness and incomputability of
  quantum randomness},}\ }\href {\doibase 10.1088/1402-4896/aaf36a} {\bibfield
  {journal} {\bibinfo  {journal} {Physica Scripta}\ }\textbf {\bibinfo {volume}
  {94}},\ \bibinfo {pages} {045103} (\bibinfo {year} {2019})},\ \Eprint
  {http://arxiv.org/abs/arXiv:1806.08762} {arXiv:1806.08762} \BibitemShut
  {NoStop}%
\bibitem [{\citenamefont {{ID Quantique SA}}(2001-2010)}]{Quantis}%
  \BibitemOpen
  \bibfield  {author} {\bibinfo {author} {\bibnamefont {{ID Quantique SA}}},\
  }\href
  {https://www.idquantique.com/random-number-generation/products/quantis-random-number-generator/}
  {\emph {\bibinfo {title} {{QUANTIS}. Quantum number generator}}}\ (\bibinfo
  {publisher} {idQuantique},\ \bibinfo {address} {Geneva, Switzerland},\
  \bibinfo {year} {2001-2010})\ \bibinfo {note} {accessed on Sep 8,
  2019}\BibitemShut {NoStop}%
\bibitem [{\citenamefont {Bennett}\ and\ \citenamefont
  {Brassard}(1984)}]{benn-84}%
  \BibitemOpen
  \bibfield  {author} {\bibinfo {author} {\bibfnamefont {Charles~H.}\
  \bibnamefont {Bennett}}\ and\ \bibinfo {author} {\bibfnamefont
  {G.}~\bibnamefont {Brassard}},\ }\bibfield  {title} {\enquote {\bibinfo
  {title} {Quantum cryptography: Public key distribution and coin tossing},}\
  }in\ \href {https://arxiv.org/abs/2003.06557} {\emph {\bibinfo {booktitle}
  {Proceedings of the IEEE International Conference on Computers, Systems, and
  Signal Processing, Bangalore, India}}}\ (\bibinfo  {publisher} {IEEE Computer
  Society Press},\ \bibinfo {year} {1984})\ pp.\ \bibinfo {pages} {175--179},\
  \Eprint {http://arxiv.org/abs/arXiv:2003.06557} {arXiv:2003.06557}
  \BibitemShut {NoStop}%
\bibitem [{\citenamefont {Martin-L{\"{o}}f}(1966)}]{martin-lof}%
  \BibitemOpen
  \bibfield  {author} {\bibinfo {author} {\bibfnamefont {Per}\ \bibnamefont
  {Martin-L{\"{o}}f}},\ }\bibfield  {title} {\enquote {\bibinfo {title} {The
  definition of random sequences},}\ }\href {\doibase
  10.1016/S0019-9958(66)80018-9} {\bibfield  {journal} {\bibinfo  {journal}
  {Information and Control}\ }\textbf {\bibinfo {volume} {9}},\ \bibinfo
  {pages} {602--619} (\bibinfo {year} {1966})}\BibitemShut {NoStop}%
\bibitem [{\citenamefont {Abbott}\ \emph
  {et~al.}(2014{\natexlab{b}})\citenamefont {Abbott}, \citenamefont {Calude},\
  and\ \citenamefont {Svozil}}]{PhysRevA.89.032109}%
  \BibitemOpen
  \bibfield  {author} {\bibinfo {author} {\bibfnamefont {Alastair~A.}\
  \bibnamefont {Abbott}}, \bibinfo {author} {\bibfnamefont {Cristian~S.}\
  \bibnamefont {Calude}}, \ and\ \bibinfo {author} {\bibfnamefont {Karl}\
  \bibnamefont {Svozil}},\ }\bibfield  {title} {\enquote {\bibinfo {title}
  {Value-indefinite observables are almost everywhere},}\ }\href {\doibase
  10.1103/PhysRevA.89.032109} {\bibfield  {journal} {\bibinfo  {journal}
  {Physical Review A}\ }\textbf {\bibinfo {volume} {89}},\ \bibinfo {pages}
  {032109} (\bibinfo {year} {2014}{\natexlab{b}})},\ \Eprint
  {http://arxiv.org/abs/arXiv:1309.7188} {arXiv:1309.7188} \BibitemShut
  {NoStop}%
\bibitem [{\citenamefont {Abbott}\ \emph {et~al.}(2015)\citenamefont {Abbott},
  \citenamefont {Calude},\ and\ \citenamefont {Svozil}}]{2015-AnalyticKS}%
  \BibitemOpen
  \bibfield  {author} {\bibinfo {author} {\bibfnamefont {Alastair~A.}\
  \bibnamefont {Abbott}}, \bibinfo {author} {\bibfnamefont {Cristian~S.}\
  \bibnamefont {Calude}}, \ and\ \bibinfo {author} {\bibfnamefont {Karl}\
  \bibnamefont {Svozil}},\ }\bibfield  {title} {\enquote {\bibinfo {title} {A
  variant of the {K}ochen-{S}pecker theorem localising value indefiniteness},}\
  }\href {\doibase 10.1063/1.4931658} {\bibfield  {journal} {\bibinfo
  {journal} {Journal of Mathematical Physics}\ }\textbf {\bibinfo {volume}
  {56}},\ \bibinfo {eid} {102201} (\bibinfo {year} {2015})},\ \Eprint
  {http://arxiv.org/abs/arXiv:1503.01985} {arXiv:1503.01985} \BibitemShut
  {NoStop}%
\bibitem [{\citenamefont {Abbott}\ \emph {et~al.}(2012)\citenamefont {Abbott},
  \citenamefont {Calude}, \citenamefont {Conder},\ and\ \citenamefont
  {Svozil}}]{2012-incomput-proofsCJ}%
  \BibitemOpen
  \bibfield  {author} {\bibinfo {author} {\bibfnamefont {Alastair~A.}\
  \bibnamefont {Abbott}}, \bibinfo {author} {\bibfnamefont {Cristian~S.}\
  \bibnamefont {Calude}}, \bibinfo {author} {\bibfnamefont {Jonathan}\
  \bibnamefont {Conder}}, \ and\ \bibinfo {author} {\bibfnamefont {Karl}\
  \bibnamefont {Svozil}},\ }\bibfield  {title} {\enquote {\bibinfo {title}
  {Strong {K}ochen-{S}pecker theorem and incomputability of quantum
  randomness},}\ }\href {\doibase 10.1103/PhysRevA.86.062109} {\bibfield
  {journal} {\bibinfo  {journal} {Physical Review A}\ }\textbf {\bibinfo
  {volume} {86}},\ \bibinfo {pages} {062109} (\bibinfo {year} {2012})},\
  \Eprint {http://arxiv.org/abs/arXiv:1207.2029} {arXiv:1207.2029} \BibitemShut
  {NoStop}%
\bibitem [{\citenamefont {Bera}\ \emph {et~al.}(2017)\citenamefont {Bera},
  \citenamefont {Ac\'in}, \citenamefont {Ku{\'s}}, \citenamefont {Mitchell},\
  and\ \citenamefont {Lewenstein}}]{Bera2017}%
  \BibitemOpen
  \bibfield  {author} {\bibinfo {author} {\bibfnamefont {Manabendra~Nath}\
  \bibnamefont {Bera}}, \bibinfo {author} {\bibfnamefont {Antonio}\
  \bibnamefont {Ac\'in}}, \bibinfo {author} {\bibfnamefont {Marek}\
  \bibnamefont {Ku{\'s}}}, \bibinfo {author} {\bibfnamefont {Morgan~W}\
  \bibnamefont {Mitchell}}, \ and\ \bibinfo {author} {\bibfnamefont {Maciej}\
  \bibnamefont {Lewenstein}},\ }\bibfield  {title} {\enquote {\bibinfo {title}
  {Randomness in quantum mechanics: philosophy, physics and technology},}\
  }\href {\doibase 10.1088/1361-6633/aa8731} {\bibfield  {journal} {\bibinfo
  {journal} {Reports on Progress in Physics}\ }\textbf {\bibinfo {volume}
  {80}},\ \bibinfo {pages} {124001} (\bibinfo {year} {2017})},\ \Eprint
  {http://arxiv.org/abs/arXiv:1611.02176} {arXiv:1611.02176} \BibitemShut
  {NoStop}%
\bibitem [{\citenamefont {{Everett III}}(1957)}]{everett}%
  \BibitemOpen
  \bibfield  {author} {\bibinfo {author} {\bibfnamefont {Hugh}\ \bibnamefont
  {{Everett III}}},\ }\bibfield  {title} {\enquote {\bibinfo {title}
  {`{R}elative {S}tate' formulation of quantum mechanics},}\ }\href {\doibase
  10.1103/RevModPhys.29.454} {\bibfield  {journal} {\bibinfo  {journal}
  {Reviews of Modern Physics}\ }\textbf {\bibinfo {volume} {29}},\ \bibinfo
  {pages} {454--462} (\bibinfo {year} {1957})}\BibitemShut {NoStop}%
\bibitem [{\citenamefont {Vaidman}(2014)}]{Vaidman2014}%
  \BibitemOpen
  \bibfield  {author} {\bibinfo {author} {\bibfnamefont {Lev}\ \bibnamefont
  {Vaidman}},\ }\bibfield  {title} {\enquote {\bibinfo {title} {Quantum theory
  and determinism},}\ }\href {\doibase 10.1007/s40509-014-0008-4} {\bibfield
  {journal} {\bibinfo  {journal} {Quantum Studies: Mathematics and
  Foundations}\ }\textbf {\bibinfo {volume} {1}},\ \bibinfo {pages} {5--38}
  (\bibinfo {year} {2014})},\ \Eprint {http://arxiv.org/abs/arXiv:1405.4222}
  {arXiv:1405.4222} \BibitemShut {NoStop}%
\bibitem [{\citenamefont {Svozil}(1999)}]{svozil-1999-haunted-qc}%
  \BibitemOpen
  \bibfield  {author} {\bibinfo {author} {\bibfnamefont {Karl}\ \bibnamefont
  {Svozil}},\ }\href {https://arxiv.org/abs/quant-ph/9907015} {\enquote
  {\bibinfo {title} {``{H}aunted'' quantum contextuality},}\ } (\bibinfo {year}
  {1999}),\ \Eprint {http://arxiv.org/abs/arXiv:quant-ph/9907015}
  {arXiv:quant-ph/9907015} \BibitemShut {NoStop}%
\bibitem [{\citenamefont {Svozil}(2009{\natexlab{b}})}]{svozil:040102}%
  \BibitemOpen
  \bibfield  {author} {\bibinfo {author} {\bibfnamefont {Karl}\ \bibnamefont
  {Svozil}},\ }\bibfield  {title} {\enquote {\bibinfo {title} {Proposed direct
  test of a certain type of noncontextuality in quantum mechanics},}\ }\href
  {\doibase 10.1103/PhysRevA.80.040102} {\bibfield  {journal} {\bibinfo
  {journal} {Physical Review A}\ }\textbf {\bibinfo {volume} {80}},\ \bibinfo
  {eid} {040102} (\bibinfo {year} {2009}{\natexlab{b}})}\BibitemShut {NoStop}%
\bibitem [{\citenamefont {Griffiths}(2017)}]{Griffiths2017}%
  \BibitemOpen
  \bibfield  {author} {\bibinfo {author} {\bibfnamefont {Robert~B.}\
  \bibnamefont {Griffiths}},\ }\bibfield  {title} {\enquote {\bibinfo {title}
  {What quantum measurements measure},}\ }\href {\doibase
  10.1103/physreva.96.032110} {\bibfield  {journal} {\bibinfo  {journal}
  {Physical Review A}\ }\textbf {\bibinfo {volume} {96}} (\bibinfo {year}
  {2017}),\ 10.1103/physreva.96.032110},\ \Eprint
  {http://arxiv.org/abs/arXiv:1704.08725} {arXiv:1704.08725} \BibitemShut
  {NoStop}%
\bibitem [{\citenamefont {Griffiths}(2019)}]{Griffiths2019}%
  \BibitemOpen
  \bibfield  {author} {\bibinfo {author} {\bibfnamefont {Robert~B.}\
  \bibnamefont {Griffiths}},\ }\bibfield  {title} {\enquote {\bibinfo {title}
  {Quantum measurements and contextuality},}\ }\href {\doibase
  10.1098/rsta.2019.0033} {\bibfield  {journal} {\bibinfo  {journal}
  {Philosophical Transactions of the Royal Society A: Mathematical, Physical
  and Engineering Sciences}\ }\textbf {\bibinfo {volume} {377}},\ \bibinfo
  {pages} {20190033} (\bibinfo {year} {2019})},\ \Eprint
  {http://arxiv.org/abs/arXiv:1902.05633} {arXiv:1902.05633} \BibitemShut
  {NoStop}%
\bibitem [{\citenamefont {Zeilinger}(2005)}]{zeil-05_nature_ofQuantum}%
  \BibitemOpen
  \bibfield  {author} {\bibinfo {author} {\bibfnamefont {Anton}\ \bibnamefont
  {Zeilinger}},\ }\bibfield  {title} {\enquote {\bibinfo {title} {The message
  of the quantum},}\ }\href {\doibase 10.1038/438743a} {\bibfield  {journal}
  {\bibinfo  {journal} {Nature}\ }\textbf {\bibinfo {volume} {438}},\ \bibinfo
  {pages} {743} (\bibinfo {year} {2005})}\BibitemShut {NoStop}%
\bibitem [{\citenamefont {von Neumann}(1951)}]{von-neumann1}%
  \BibitemOpen
  \bibfield  {author} {\bibinfo {author} {\bibfnamefont {John}\ \bibnamefont
  {von Neumann}},\ }\bibfield  {title} {\enquote {\bibinfo {title} {Various
  techniques used in connection with random digits},}\ }in\ \href
  {http://fsaad.mit.edu/assets/VonNeumann-ams12p36-38.pdf} {\emph {\bibinfo
  {booktitle} {Monte Carlo Method}}},\ \bibinfo {series} {National Bureau of
  Standards Applied Mathematics Series}, Vol.~\bibinfo {volume} {12},\ \bibinfo
  {editor} {edited by\ \bibinfo {editor} {\bibfnamefont {A.~S.}\ \bibnamefont
  {Householder}}, \bibinfo {editor} {\bibfnamefont {G.~E.}\ \bibnamefont
  {Forsythe}}, \ and\ \bibinfo {editor} {\bibfnamefont {H.~H.}\ \bibnamefont
  {Germond}}}\ (\bibinfo  {publisher} {US Government Printing Office},\
  \bibinfo {address} {Washington, DC},\ \bibinfo {year} {1951})\ Chap.~\bibinfo
  {chapter} {13}, pp.\ \bibinfo {pages} {36--38},\ \bibinfo {note} {reprinted
  in {\sl John {von Neumann}, Collected Works, Vol. V}, A. H. Taub, editor,
  Pergamon Press, New York, 1963, p. 768--770}\BibitemShut {NoStop}%
\bibitem [{\citenamefont {{von Neumann}}(1963)}]{Taub:1963:JNCa}%
  \BibitemOpen
  \bibfield  {author} {\bibinfo {author} {\bibfnamefont {John}\ \bibnamefont
  {{von Neumann}}},\ }\href@noop {} {\emph {\bibinfo {title} {{John {von
  Neumann}}: Collected Works. {Volume V}: {Design} of Computers, Theory of
  Automata and Numerical Analysis}}}\ (\bibinfo  {publisher} {Pergamon},\
  \bibinfo {address} {New York},\ \bibinfo {year} {1963})\ pp.\ \bibinfo
  {pages} {ix + 784}\BibitemShut {NoStop}%
\bibitem [{\citenamefont {Abbott}\ and\ \citenamefont
  {Calude}(2012)}]{AbbottCalude10}%
  \BibitemOpen
  \bibfield  {author} {\bibinfo {author} {\bibfnamefont {Alastair~A.}\
  \bibnamefont {Abbott}}\ and\ \bibinfo {author} {\bibfnamefont {Cristian~S.}\
  \bibnamefont {Calude}},\ }\bibfield  {title} {\enquote {\bibinfo {title} {Von
  {N}eumann normalisation of a quantum random number generator},}\ }\href
  {\doibase 10.3233/COM-2012-001} {\bibfield  {journal} {\bibinfo  {journal}
  {Computability}\ }\textbf {\bibinfo {volume} {1}},\ \bibinfo {pages} {59--83}
  (\bibinfo {year} {2012})},\ \Eprint
  {http://arxiv.org/abs/https://arxiv.org/abs/1101.4711}
  {https://arxiv.org/abs/1101.4711} \BibitemShut {NoStop}%
\bibitem [{\citenamefont {Clauser}(1992)}]{CLAUSER1992}%
  \BibitemOpen
  \bibfield  {author} {\bibinfo {author} {\bibfnamefont {John~F.}\ \bibnamefont
  {Clauser}},\ }\bibfield  {title} {\enquote {\bibinfo {title} {Early history
  of {B}ell's theorem theory and experiment},}\ }in\ \href {\doibase
  10.1142/9789814436687_0020} {\emph {\bibinfo {booktitle} {Foundations of
  Quantum Mechanics}}}\ (\bibinfo  {publisher} {World Scientific},\ \bibinfo
  {year} {1992})\ pp.\ \bibinfo {pages} {168--174},\ \bibinfo {note} {santa Fe
  Workshop, Santa Fe, New Mexico 27-31 May 1991, book
  DOI:10.1142/1671}\BibitemShut {NoStop}%
\bibitem [{\citenamefont {Clauser}(2002)}]{clauser-talkvie}%
  \BibitemOpen
  \bibfield  {author} {\bibinfo {author} {\bibfnamefont {John~F.}\ \bibnamefont
  {Clauser}},\ }\bibfield  {title} {\enquote {\bibinfo {title} {Early history
  of {B}ell's theorem},}\ }in\ \href {\doibase 10.1007/978-3-662-05032-3\_6}
  {\emph {\bibinfo {booktitle} {Quantum (Un)speakables: {F}rom {B}ell to
  Quantum Information}}},\ \bibinfo {editor} {edited by\ \bibinfo {editor}
  {\bibfnamefont {Reinhold}\ \bibnamefont {Bertlmann}}\ and\ \bibinfo {editor}
  {\bibfnamefont {Anton}\ \bibnamefont {Zeilinger}}}\ (\bibinfo  {publisher}
  {Springer},\ \bibinfo {address} {Berlin},\ \bibinfo {year} {2002})\ pp.\
  \bibinfo {pages} {61--96}\BibitemShut {NoStop}%
\bibitem [{\citenamefont {Schwinger}(1960)}]{Schwinger.60}%
  \BibitemOpen
  \bibfield  {author} {\bibinfo {author} {\bibfnamefont {Julian}\ \bibnamefont
  {Schwinger}},\ }\bibfield  {title} {\enquote {\bibinfo {title} {Unitary
  operators bases},}\ }\href {\doibase 10.1073/pnas.46.4.570} {\bibfield
  {journal} {\bibinfo  {journal} {Proceedings of the National Academy of
  Sciences (PNAS)}\ }\textbf {\bibinfo {volume} {46}},\ \bibinfo {pages}
  {570--579} (\bibinfo {year} {1960})}\BibitemShut {NoStop}%
\bibitem [{\citenamefont {{von Neumann}}(1932, 1996)}]{v-neumann-49}%
  \BibitemOpen
  \bibfield  {author} {\bibinfo {author} {\bibfnamefont {John}\ \bibnamefont
  {{von Neumann}}},\ }\href {\doibase 10.1007/978-3-642-61409-5} {\emph
  {\bibinfo {title} {{M}athematische {G}rundlagen der {Q}uantenmechanik}}},\
  \bibinfo {edition} {2nd}\ ed.\ (\bibinfo  {publisher} {Springer},\ \bibinfo
  {address} {Berlin, Heidelberg},\ \bibinfo {year} {1932, 1996})\ \bibinfo
  {note} {{E}nglish translation in~\cite{v-neumann-55}}\BibitemShut {NoStop}%
\bibitem [{\citenamefont {{von Neumann}}(1955)}]{v-neumann-55}%
  \BibitemOpen
  \bibfield  {author} {\bibinfo {author} {\bibfnamefont {John}\ \bibnamefont
  {{von Neumann}}},\ }\href {http://press.princeton.edu/titles/2113.html}
  {\emph {\bibinfo {title} {Mathematical Foundations of Quantum Mechanics}}}\
  (\bibinfo  {publisher} {Princeton University Press},\ \bibinfo {address}
  {Princeton, NJ},\ \bibinfo {year} {1955})\ \bibinfo {note} {{G}erman original
  in~\cite{v-neumann-49}}\BibitemShut {NoStop}%
\bibitem [{\citenamefont
  {Schr{\"{o}}dinger}(1935{\natexlab{a}})}]{schrodinger}%
  \BibitemOpen
  \bibfield  {author} {\bibinfo {author} {\bibfnamefont {Erwin}\ \bibnamefont
  {Schr{\"{o}}dinger}},\ }\bibfield  {title} {\enquote {\bibinfo {title} {{D}ie
  gegenw\"artige {S}ituation in der {Q}uantenmechanik},}\ }\href {\doibase
  10.1007/BF01491891, 10.1007/BF01491914, 10.1007/BF01491987} {\bibfield
  {journal} {\bibinfo  {journal} {Naturwissenschaften}\ }\textbf {\bibinfo
  {volume} {23}},\ \bibinfo {pages} {807--812, 823--828, 844--849} (\bibinfo
  {year} {1935}{\natexlab{a}})}\BibitemShut {NoStop}%
\bibitem [{\citenamefont {London}\ and\ \citenamefont
  {Bauer}(1939)}]{london-Bauer-1939}%
  \BibitemOpen
  \bibfield  {author} {\bibinfo {author} {\bibfnamefont {Fritz}\ \bibnamefont
  {London}}\ and\ \bibinfo {author} {\bibfnamefont {Edmond}\ \bibnamefont
  {Bauer}},\ }\href@noop {} {\emph {\bibinfo {title} {La theorie de
  l'observation en m\'ecanique quantique; {N}o.~775 of Actualit\'es
  scientifiques et industrielles: Expos\'es de physique g\'en\'erale, publi\'es
  sous la direction de {P}aul {L}angevin}}}\ (\bibinfo  {publisher} {Hermann},\
  \bibinfo {address} {Paris},\ \bibinfo {year} {1939})\ \bibinfo {note}
  {english translation in~\cite{london-Bauer-1983}}\BibitemShut {NoStop}%
\bibitem [{\citenamefont {London}\ and\ \citenamefont
  {Bauer}(1983)}]{london-Bauer-1983}%
  \BibitemOpen
  \bibfield  {author} {\bibinfo {author} {\bibfnamefont {Fritz}\ \bibnamefont
  {London}}\ and\ \bibinfo {author} {\bibfnamefont {Edmond}\ \bibnamefont
  {Bauer}},\ }\bibfield  {title} {\enquote {\bibinfo {title} {The theory of
  observation in quantum mechanics},}\ }in\ \href@noop {} {\emph {\bibinfo
  {booktitle} {Quantum Theory and Measurement}}}\ (\bibinfo  {publisher}
  {Princeton University Press},\ \bibinfo {address} {Princeton, NJ},\ \bibinfo
  {year} {1983})\ pp.\ \bibinfo {pages} {217--259},\ \bibinfo {note}
  {consolidated translation of French
  original~\cite{london-Bauer-1939}}\BibitemShut {NoStop}%
\bibitem [{\citenamefont {Barrett}(2011)}]{Barrett-2011}%
  \BibitemOpen
  \bibfield  {author} {\bibinfo {author} {\bibfnamefont {Jeffrey~A.}\
  \bibnamefont {Barrett}},\ }\bibfield  {title} {\enquote {\bibinfo {title}
  {{E}verett's pure wave mechanics and the notion of worlds},}\ }\href
  {\doibase 10.1007/s13194-011-0023-9} {\bibfield  {journal} {\bibinfo
  {journal} {European Journal for Philosophy of Science}\ }\textbf {\bibinfo
  {volume} {1}},\ \bibinfo {pages} {277--302} (\bibinfo {year}
  {2011})}\BibitemShut {NoStop}%
\bibitem [{\citenamefont {{Everett III}}(1956,2012)}]{everett-1956}%
  \BibitemOpen
  \bibfield  {author} {\bibinfo {author} {\bibfnamefont {Hugh}\ \bibnamefont
  {{Everett III}}},\ }\bibfield  {title} {\enquote {\bibinfo {title} {The
  theory of the universal wave function},}\ }in\ \href
  {http://press.princeton.edu/titles/9770.html} {\emph {\bibinfo {booktitle}
  {The {E}verett Interpretation of Quantum Mechanics: Collected Works 1955-1980
  with Commentary}}},\ \bibinfo {editor} {edited by\ \bibinfo {editor}
  {\bibfnamefont {Jeffrey~A.}\ \bibnamefont {Barrett}}\ and\ \bibinfo {editor}
  {\bibfnamefont {Peter}\ \bibnamefont {Byrne}}}\ (\bibinfo  {publisher}
  {Princeton University Press},\ \bibinfo {address} {Princeton, NJ},\ \bibinfo
  {year} {1956,2012})\ pp.\ \bibinfo {pages} {72--172}\BibitemShut {NoStop}%
\bibitem [{\citenamefont {Wigner}(1961, 1962, 1995)}]{wigner:mb}%
  \BibitemOpen
  \bibfield  {author} {\bibinfo {author} {\bibfnamefont {Eugene~P.}\
  \bibnamefont {Wigner}},\ }\bibfield  {title} {\enquote {\bibinfo {title}
  {Remarks on the mind-body question},}\ }in\ \href {\doibase
  10.1007/978-3-642-78374-6\_20} {\emph {\bibinfo {booktitle} {The Scientist
  Speculates}}},\ \bibinfo {editor} {edited by\ \bibinfo {editor}
  {\bibfnamefont {I.~J.}\ \bibnamefont {Good}}}\ (\bibinfo  {publisher}
  {Heinemann, Basic Books, and Springer-Verlag},\ \bibinfo {address} {London,
  New York, and Berlin},\ \bibinfo {year} {1961, 1962, 1995})\ pp.\ \bibinfo
  {pages} {284--302}\BibitemShut {NoStop}%
\bibitem [{\citenamefont {Peres}(1980)}]{PhysRevD.22.879}%
  \BibitemOpen
  \bibfield  {author} {\bibinfo {author} {\bibfnamefont {Asher}\ \bibnamefont
  {Peres}},\ }\bibfield  {title} {\enquote {\bibinfo {title} {Can we undo
  quantum measurements?}}\ }\href {\doibase 10.1103/PhysRevD.22.879} {\bibfield
   {journal} {\bibinfo  {journal} {Physical Review D}\ }\textbf {\bibinfo
  {volume} {22}},\ \bibinfo {pages} {879--883} (\bibinfo {year}
  {1980})}\BibitemShut {NoStop}%
\bibitem [{\citenamefont {Scully}\ and\ \citenamefont
  {Dr\"uhl}(1982)}]{PhysRevA.25.2208}%
  \BibitemOpen
  \bibfield  {author} {\bibinfo {author} {\bibfnamefont {Marlan~O.}\
  \bibnamefont {Scully}}\ and\ \bibinfo {author} {\bibfnamefont {Kai}\
  \bibnamefont {Dr\"uhl}},\ }\bibfield  {title} {\enquote {\bibinfo {title}
  {Quantum eraser: A proposed photon correlation experiment concerning
  observation and ``delayed choice'' in quantum mechanics},}\ }\href {\doibase
  10.1103/PhysRevA.25.2208} {\bibfield  {journal} {\bibinfo  {journal}
  {Physical Review A}\ }\textbf {\bibinfo {volume} {25}},\ \bibinfo {pages}
  {2208--2213} (\bibinfo {year} {1982})}\BibitemShut {NoStop}%
\bibitem [{\citenamefont {Greenberger}\ and\ \citenamefont
  {YaSin}(1989)}]{greenberger2}%
  \BibitemOpen
  \bibfield  {author} {\bibinfo {author} {\bibfnamefont {Daniel~M.}\
  \bibnamefont {Greenberger}}\ and\ \bibinfo {author} {\bibfnamefont {Alaine}\
  \bibnamefont {YaSin}},\ }\bibfield  {title} {\enquote {\bibinfo {title}
  {``{H}aunted'' measurements in quantum theory},}\ }\href {\doibase
  10.1007/BF00731905} {\bibfield  {journal} {\bibinfo  {journal} {Foundation of
  Physics}\ }\textbf {\bibinfo {volume} {19}},\ \bibinfo {pages} {679--704}
  (\bibinfo {year} {1989})}\BibitemShut {NoStop}%
\bibitem [{\citenamefont {Scully}\ \emph {et~al.}(1991)\citenamefont {Scully},
  \citenamefont {Englert},\ and\ \citenamefont {Walther}}]{Nature351}%
  \BibitemOpen
  \bibfield  {author} {\bibinfo {author} {\bibfnamefont {Marian~O.}\
  \bibnamefont {Scully}}, \bibinfo {author} {\bibfnamefont {Berthold-Georg}\
  \bibnamefont {Englert}}, \ and\ \bibinfo {author} {\bibfnamefont {Herbert}\
  \bibnamefont {Walther}},\ }\bibfield  {title} {\enquote {\bibinfo {title}
  {Quantum optical tests of complementarity},}\ }\href {\doibase
  10.1038/351111a0} {\bibfield  {journal} {\bibinfo  {journal} {Nature}\
  }\textbf {\bibinfo {volume} {351}},\ \bibinfo {pages} {111--116} (\bibinfo
  {year} {1991})}\BibitemShut {NoStop}%
\bibitem [{\citenamefont {Zajonc}\ \emph {et~al.}(1991)\citenamefont {Zajonc},
  \citenamefont {Wang}, \citenamefont {Zou},\ and\ \citenamefont
  {Mandel}}]{Zajonc-91}%
  \BibitemOpen
  \bibfield  {author} {\bibinfo {author} {\bibfnamefont {A.~G.}\ \bibnamefont
  {Zajonc}}, \bibinfo {author} {\bibfnamefont {L.~J.}\ \bibnamefont {Wang}},
  \bibinfo {author} {\bibfnamefont {X.~Y.}\ \bibnamefont {Zou}}, \ and\
  \bibinfo {author} {\bibfnamefont {L.}~\bibnamefont {Mandel}},\ }\bibfield
  {title} {\enquote {\bibinfo {title} {Quantum eraser},}\ }\href {\doibase
  10.1038/353507b0} {\bibfield  {journal} {\bibinfo  {journal} {Nature}\
  }\textbf {\bibinfo {volume} {353}},\ \bibinfo {pages} {507--508} (\bibinfo
  {year} {1991})}\BibitemShut {NoStop}%
\bibitem [{\citenamefont {Kwiat}\ \emph {et~al.}(1992)\citenamefont {Kwiat},
  \citenamefont {Steinberg},\ and\ \citenamefont {Chiao}}]{PhysRevA.45.7729}%
  \BibitemOpen
  \bibfield  {author} {\bibinfo {author} {\bibfnamefont {Paul~G.}\ \bibnamefont
  {Kwiat}}, \bibinfo {author} {\bibfnamefont {Aephraim~M.}\ \bibnamefont
  {Steinberg}}, \ and\ \bibinfo {author} {\bibfnamefont {Raymond~Y.}\
  \bibnamefont {Chiao}},\ }\bibfield  {title} {\enquote {\bibinfo {title}
  {Observation of a ``quantum eraser:'' a revival of coherence in a two-photon
  interference experiment},}\ }\href {\doibase 10.1103/PhysRevA.45.7729}
  {\bibfield  {journal} {\bibinfo  {journal} {Physical Review A}\ }\textbf
  {\bibinfo {volume} {45}},\ \bibinfo {pages} {7729--7739} (\bibinfo {year}
  {1992})}\BibitemShut {NoStop}%
\bibitem [{\citenamefont {Pfau}\ \emph {et~al.}(1994)\citenamefont {Pfau},
  \citenamefont {Sp\"alter}, \citenamefont {Kurtsiefer}, \citenamefont
  {Ekstrom},\ and\ \citenamefont {Mlynek}}]{PhysRevLett.73.1223}%
  \BibitemOpen
  \bibfield  {author} {\bibinfo {author} {\bibfnamefont {T.}~\bibnamefont
  {Pfau}}, \bibinfo {author} {\bibfnamefont {S.}~\bibnamefont {Sp\"alter}},
  \bibinfo {author} {\bibfnamefont {Ch.}\ \bibnamefont {Kurtsiefer}}, \bibinfo
  {author} {\bibfnamefont {C.~R.}\ \bibnamefont {Ekstrom}}, \ and\ \bibinfo
  {author} {\bibfnamefont {J.}~\bibnamefont {Mlynek}},\ }\bibfield  {title}
  {\enquote {\bibinfo {title} {Loss of spatial coherence by a single
  spontaneous emission},}\ }\href {\doibase 10.1103/PhysRevLett.73.1223}
  {\bibfield  {journal} {\bibinfo  {journal} {Physical Review Letters}\
  }\textbf {\bibinfo {volume} {73}},\ \bibinfo {pages} {1223--1226} (\bibinfo
  {year} {1994})}\BibitemShut {NoStop}%
\bibitem [{\citenamefont {Chapman}\ \emph {et~al.}(1995)\citenamefont
  {Chapman}, \citenamefont {Hammond}, \citenamefont {Lenef}, \citenamefont
  {Schmiedmayer}, \citenamefont {Rubenstein}, \citenamefont {Smith},\ and\
  \citenamefont {Pritchard}}]{PhysRevLett.75.3783}%
  \BibitemOpen
  \bibfield  {author} {\bibinfo {author} {\bibfnamefont {Michael~S.}\
  \bibnamefont {Chapman}}, \bibinfo {author} {\bibfnamefont {Troy~D.}\
  \bibnamefont {Hammond}}, \bibinfo {author} {\bibfnamefont {Alan}\
  \bibnamefont {Lenef}}, \bibinfo {author} {\bibfnamefont {J\"org}\
  \bibnamefont {Schmiedmayer}}, \bibinfo {author} {\bibfnamefont {Richard~A.}\
  \bibnamefont {Rubenstein}}, \bibinfo {author} {\bibfnamefont {Edward}\
  \bibnamefont {Smith}}, \ and\ \bibinfo {author} {\bibfnamefont {David~E.}\
  \bibnamefont {Pritchard}},\ }\bibfield  {title} {\enquote {\bibinfo {title}
  {Photon scattering from atoms in an atom interferometer: Coherence lost and
  regained},}\ }\href {\doibase 10.1103/PhysRevLett.75.3783} {\bibfield
  {journal} {\bibinfo  {journal} {Physical Review Letters}\ }\textbf {\bibinfo
  {volume} {75}},\ \bibinfo {pages} {3783--3787} (\bibinfo {year}
  {1995})}\BibitemShut {NoStop}%
\bibitem [{\citenamefont {Herzog}\ \emph {et~al.}(1995)\citenamefont {Herzog},
  \citenamefont {Kwiat}, \citenamefont {Weinfurter},\ and\ \citenamefont
  {Zeilinger}}]{hkwz}%
  \BibitemOpen
  \bibfield  {author} {\bibinfo {author} {\bibfnamefont {Thomas~J.}\
  \bibnamefont {Herzog}}, \bibinfo {author} {\bibfnamefont {Paul~G.}\
  \bibnamefont {Kwiat}}, \bibinfo {author} {\bibfnamefont {Harald}\
  \bibnamefont {Weinfurter}}, \ and\ \bibinfo {author} {\bibfnamefont {Anton}\
  \bibnamefont {Zeilinger}},\ }\bibfield  {title} {\enquote {\bibinfo {title}
  {Complementarity and the quantum eraser},}\ }\href {\doibase
  10.1103/PhysRevLett.75.3034} {\bibfield  {journal} {\bibinfo  {journal}
  {Physical Review Letters}\ }\textbf {\bibinfo {volume} {75}},\ \bibinfo
  {pages} {3034--3037} (\bibinfo {year} {1995})}\BibitemShut {NoStop}%
\bibitem [{\citenamefont {Englert}\ \emph {et~al.}(1988)\citenamefont
  {Englert}, \citenamefont {Schwinger},\ and\ \citenamefont
  {Scully}}]{engrt-sg-I}%
  \BibitemOpen
  \bibfield  {author} {\bibinfo {author} {\bibfnamefont {Berthold-Georg}\
  \bibnamefont {Englert}}, \bibinfo {author} {\bibfnamefont {Julian}\
  \bibnamefont {Schwinger}}, \ and\ \bibinfo {author} {\bibfnamefont
  {Marlan~O.}\ \bibnamefont {Scully}},\ }\bibfield  {title} {\enquote {\bibinfo
  {title} {Is spin coherence like {H}umpty-{D}umpty? {I}. {S}implified
  treatment},}\ }\href {\doibase 10.1007/BF01909939} {\bibfield  {journal}
  {\bibinfo  {journal} {Foundations of Physics}\ }\textbf {\bibinfo {volume}
  {18}},\ \bibinfo {pages} {1045--1056} (\bibinfo {year} {1988})}\BibitemShut
  {NoStop}%
\bibitem [{\citenamefont {Schwinger}\ \emph {et~al.}(1988)\citenamefont
  {Schwinger}, \citenamefont {Scully},\ and\ \citenamefont
  {Englert}}]{engrt-sg-II}%
  \BibitemOpen
  \bibfield  {author} {\bibinfo {author} {\bibfnamefont {Julian}\ \bibnamefont
  {Schwinger}}, \bibinfo {author} {\bibfnamefont {Marlan~O.}\ \bibnamefont
  {Scully}}, \ and\ \bibinfo {author} {\bibfnamefont {Berthold-Georg}\
  \bibnamefont {Englert}},\ }\bibfield  {title} {\enquote {\bibinfo {title} {Is
  spin coherence like {H}umpty-{D}umpty? {II}. {G}eneral theory},}\ }\href
  {\doibase 10.1007/BF01384847} {\bibfield  {journal} {\bibinfo  {journal}
  {Zeitschrift f{\"u}r Physik D: Atoms, Molecules and Clusters}\ }\textbf
  {\bibinfo {volume} {10}},\ \bibinfo {pages} {135--144} (\bibinfo {year}
  {1988})}\BibitemShut {NoStop}%
\bibitem [{\citenamefont {Zeilinger}(1999)}]{zeil-99}%
  \BibitemOpen
  \bibfield  {author} {\bibinfo {author} {\bibfnamefont {Anton}\ \bibnamefont
  {Zeilinger}},\ }\bibfield  {title} {\enquote {\bibinfo {title} {A
  foundational principle for quantum mechanics},}\ }\href {\doibase
  10.1023/A:1018820410908} {\bibfield  {journal} {\bibinfo  {journal}
  {Foundations of Physics}\ }\textbf {\bibinfo {volume} {29}},\ \bibinfo
  {pages} {631--643} (\bibinfo {year} {1999})}\BibitemShut {NoStop}%
\bibitem [{\citenamefont {Yanofsky}(2019)}]{Yanofsky-object}%
  \BibitemOpen
  \bibfield  {author} {\bibinfo {author} {\bibfnamefont {Noson~S.}\
  \bibnamefont {Yanofsky}},\ }\href
  {http://www.sci.brooklyn.cuny.edu/~noson/Mind%20and%20Physics.pdf} {\enquote
  {\bibinfo {title} {The mind and the limitations of physics},}\ } (\bibinfo
  {year} {2019}),\ \bibinfo {note} {preprint, , accessed on January 14,
  2021}\BibitemShut {NoStop}%
\bibitem [{\citenamefont {Gallois}(2016)}]{Gallois-SEP}%
  \BibitemOpen
  \bibfield  {author} {\bibinfo {author} {\bibfnamefont {Andr\'e}\ \bibnamefont
  {Gallois}},\ }\bibfield  {title} {\enquote {\bibinfo {title} {Identity over
  time},}\ }in\ \href
  {https://plato.stanford.edu/archives/win2016/entries/identity-time/} {\emph
  {\bibinfo {booktitle} {The {S}tanford Encyclopedia of Philosophy}}},\
  \bibinfo {editor} {edited by\ \bibinfo {editor} {\bibfnamefont {Edward~N.}\
  \bibnamefont {Zalta}}}\ (\bibinfo  {publisher} {Metaphysics Research Lab,
  Stanford University},\ \bibinfo {year} {2016})\ \bibinfo {edition} {winter
  2016}\ ed.\BibitemShut {Stop}%
\bibitem [{\citenamefont {Gallois}(1998,2003,2011)}]{Gallois-POI}%
  \BibitemOpen
  \bibfield  {author} {\bibinfo {author} {\bibfnamefont {Andr\'e}\ \bibnamefont
  {Gallois}},\ }\href {\doibase 10.1093/acprof:oso/9780199261833.001.0001}
  {\emph {\bibinfo {title} {Occasions of Identity: {A} Study in the Metaphysics
  of Persistence, Change, and Sameness}}}\ (\bibinfo  {publisher} {Bantam
  Books},\ \bibinfo {address} {New York},\ \bibinfo {year}
  {1998,2003,2011})\BibitemShut {NoStop}%
\bibitem [{\citenamefont {Svozil}(2004)}]{svozil-2003-garda}%
  \BibitemOpen
  \bibfield  {author} {\bibinfo {author} {\bibfnamefont {Karl}\ \bibnamefont
  {Svozil}},\ }\bibfield  {title} {\enquote {\bibinfo {title} {Quantum
  information via state partitions and the context translation principle},}\
  }\href {\doibase 10.1080/09500340410001664179} {\bibfield  {journal}
  {\bibinfo  {journal} {Journal of Modern Optics}\ }\textbf {\bibinfo {volume}
  {51}},\ \bibinfo {pages} {811--819} (\bibinfo {year} {2004})},\ \Eprint
  {http://arxiv.org/abs/arXiv:quant-ph/0308110} {arXiv:quant-ph/0308110}
  \BibitemShut {NoStop}%
\bibitem [{\citenamefont {Svozil}(2014)}]{svozil-2013-omelette}%
  \BibitemOpen
  \bibfield  {author} {\bibinfo {author} {\bibfnamefont {Karl}\ \bibnamefont
  {Svozil}},\ }\bibfield  {title} {\enquote {\bibinfo {title} {Unscrambling the
  quantum omelette},}\ }\href {\doibase 10.1007/s10773-013-1995-3} {\bibfield
  {journal} {\bibinfo  {journal} {International Journal of Theoretical
  Physics}\ }\textbf {\bibinfo {volume} {53}},\ \bibinfo {pages} {3648--3657}
  (\bibinfo {year} {2014})},\ \Eprint {http://arxiv.org/abs/arXiv:1206.6024}
  {arXiv:1206.6024} \BibitemShut {NoStop}%
\bibitem [{\citenamefont {Myrvold}(2011)}]{Myrvold2011237}%
  \BibitemOpen
  \bibfield  {author} {\bibinfo {author} {\bibfnamefont {Wayne~C.}\
  \bibnamefont {Myrvold}},\ }\bibfield  {title} {\enquote {\bibinfo {title}
  {Statistical mechanics and thermodynamics: A {M}axwellian view},}\ }\href
  {\doibase 10.1016/j.shpsb.2011.07.001} {\bibfield  {journal} {\bibinfo
  {journal} {Studies in History and Philosophy of Science Part B: Studies in
  History and Philosophy of Modern Physics}\ }\textbf {\bibinfo {volume}
  {42}},\ \bibinfo {pages} {237--243} (\bibinfo {year} {2011})}\BibitemShut
  {NoStop}%
\bibitem [{\citenamefont {Bohr}(1949)}]{bohr-1949}%
  \BibitemOpen
  \bibfield  {author} {\bibinfo {author} {\bibfnamefont {Niels}\ \bibnamefont
  {Bohr}},\ }\bibfield  {title} {\enquote {\bibinfo {title} {Discussion with
  {E}instein on epistemological problems in atomic physics},}\ }in\ \href
  {\doibase 10.1016/S1876-0503(08)70379-7} {\emph {\bibinfo {booktitle}
  {{A}lbert {E}instein: Philosopher-Scientist}}},\ \bibinfo {editor} {edited
  by\ \bibinfo {editor} {\bibfnamefont {P.~A.}\ \bibnamefont {Schilpp}}}\
  (\bibinfo  {publisher} {The Library of Living Philosophers},\ \bibinfo
  {address} {Evanston, Ill.},\ \bibinfo {year} {1949})\ pp.\ \bibinfo {pages}
  {200--241}\BibitemShut {NoStop}%
\bibitem [{\citenamefont {Foti}\ \emph {et~al.}(2019)\citenamefont {Foti},
  \citenamefont {Heinosaari}, \citenamefont {Maniscalco},\ and\ \citenamefont
  {Verrucchi}}]{Verrucchi-18}%
  \BibitemOpen
  \bibfield  {author} {\bibinfo {author} {\bibfnamefont {Caterina}\
  \bibnamefont {Foti}}, \bibinfo {author} {\bibfnamefont {Teiko}\ \bibnamefont
  {Heinosaari}}, \bibinfo {author} {\bibfnamefont {Sabrina}\ \bibnamefont
  {Maniscalco}}, \ and\ \bibinfo {author} {\bibfnamefont {Paola}\ \bibnamefont
  {Verrucchi}},\ }\bibfield  {title} {\enquote {\bibinfo {title} {Whenever a
  quantum environment emerges as a classical system, it behaves like a
  measuring apparatus},}\ }\href {\doibase 10.22331/q-2019-08-26-179}
  {\bibfield  {journal} {\bibinfo  {journal} {Quantum}\ }\textbf {\bibinfo
  {volume} {3}},\ \bibinfo {pages} {179} (\bibinfo {year} {2019})},\ \Eprint
  {http://arxiv.org/abs/arXiv:1810.10261} {arXiv:1810.10261} \BibitemShut
  {NoStop}%
\bibitem [{\citenamefont {Glauber}(2007)}]{glauber-collected-cat}%
  \BibitemOpen
  \bibfield  {author} {\bibinfo {author} {\bibfnamefont {Roy~J.}\ \bibnamefont
  {Glauber}},\ }\enquote {\bibinfo {title} {Amplifiers, attenuators and
  {S}chr\"odingers cat},}\ in\ \href {\doibase 10.1002/9783527610075.ch14}
  {\emph {\bibinfo {booktitle} {Quantum Theory of Optical Coherence}}}\
  (\bibinfo  {publisher} {Wiley-VCH Verlag GmbH \& Co. KGaA},\ \bibinfo {year}
  {2007})\ pp.\ \bibinfo {pages} {537--576}\BibitemShut {NoStop}%
\bibitem [{\citenamefont {Glauber}(1986)}]{Glauber-cat-86}%
  \BibitemOpen
  \bibfield  {author} {\bibinfo {author} {\bibfnamefont {Roy~J.}\ \bibnamefont
  {Glauber}},\ }\bibfield  {title} {\enquote {\bibinfo {title} {Amplifiers,
  attenuators, and schr\"odinger's cat},}\ }\href {\doibase
  10.1111/j.1749-6632.1986.tb12437.x} {\bibfield  {journal} {\bibinfo
  {journal} {Annals of the New York Academy of Sciences}\ }\textbf {\bibinfo
  {volume} {480}},\ \bibinfo {pages} {336--372} (\bibinfo {year}
  {1986})}\BibitemShut {NoStop}%
\bibitem [{\citenamefont {Pitowsky}(1998)}]{pitowsky:218}%
  \BibitemOpen
  \bibfield  {author} {\bibinfo {author} {\bibfnamefont {Itamar}\ \bibnamefont
  {Pitowsky}},\ }\bibfield  {title} {\enquote {\bibinfo {title} {Infinite and
  finite {G}leason's theorems and the logic of indeterminacy},}\ }\href
  {\doibase 10.1063/1.532334} {\bibfield  {journal} {\bibinfo  {journal}
  {Journal of Mathematical Physics}\ }\textbf {\bibinfo {volume} {39}},\
  \bibinfo {pages} {218--228} (\bibinfo {year} {1998})}\BibitemShut {NoStop}%
\bibitem [{\citenamefont
  {Svozil}(2018{\natexlab{a}})}]{svozil-2018-whycontexts}%
  \BibitemOpen
  \bibfield  {author} {\bibinfo {author} {\bibfnamefont {Karl}\ \bibnamefont
  {Svozil}},\ }\bibfield  {title} {\enquote {\bibinfo {title} {New forms of
  quantum value indefiniteness suggest that incompatible views on contexts are
  epistemic},}\ }\href {\doibase 10.3390/e20060406} {\bibfield  {journal}
  {\bibinfo  {journal} {Entropy}\ }\textbf {\bibinfo {volume} {20}},\ \bibinfo
  {pages} {406(22)} (\bibinfo {year} {2018}{\natexlab{a}})},\ \Eprint
  {http://arxiv.org/abs/arXiv:1804.10030} {arXiv:1804.10030} \BibitemShut
  {NoStop}%
\bibitem [{\citenamefont {Specker}(1960)}]{specker-60}%
  \BibitemOpen
  \bibfield  {author} {\bibinfo {author} {\bibfnamefont {Ernst}\ \bibnamefont
  {Specker}},\ }\bibfield  {title} {\enquote {\bibinfo {title} {{D}ie {L}ogik
  nicht gleichzeitig entscheidbarer {A}ussagen},}\ }\href {\doibase
  10.1111/j.1746-8361.1960.tb00422.x} {\bibfield  {journal} {\bibinfo
  {journal} {Dialectica}\ }\textbf {\bibinfo {volume} {14}},\ \bibinfo {pages}
  {239--246} (\bibinfo {year} {1960})},\ \bibinfo {note} {english traslation at
  {https://arxiv.org/abs/1103.4537}},\ \Eprint
  {http://arxiv.org/abs/arXiv:1103.4537} {arXiv:1103.4537} \BibitemShut
  {NoStop}%
\bibitem [{\citenamefont {Specker}(1990)}]{specker-ges}%
  \BibitemOpen
  \bibfield  {author} {\bibinfo {author} {\bibfnamefont {Ernst}\ \bibnamefont
  {Specker}},\ }\href {\doibase 10.1007/978-3-0348-9259-9} {\emph {\bibinfo
  {title} {Selecta}}}\ (\bibinfo  {publisher} {Birkh{\"{a}}user Verlag},\
  \bibinfo {address} {Basel},\ \bibinfo {year} {1990})\BibitemShut {NoStop}%
\bibitem [{\citenamefont {Svozil}(2009{\natexlab{c}})}]{svozil-2006-omni}%
  \BibitemOpen
  \bibfield  {author} {\bibinfo {author} {\bibfnamefont {Karl}\ \bibnamefont
  {Svozil}},\ }\bibfield  {title} {\enquote {\bibinfo {title} {Quantum
  scholasticism: On quantum contexts, counterfactuals, and the absurdities of
  quantum omniscience},}\ }\href {\doibase 10.1016/j.ins.2008.06.012}
  {\bibfield  {journal} {\bibinfo  {journal} {Information Sciences}\ }\textbf
  {\bibinfo {volume} {179}},\ \bibinfo {pages} {535--541} (\bibinfo {year}
  {2009}{\natexlab{c}})}\BibitemShut {NoStop}%
\bibitem [{\citenamefont {Cabello}\ \emph {et~al.}(2018)\citenamefont
  {Cabello}, \citenamefont {Portillo}, \citenamefont {Sol\'{i}s},\ and\
  \citenamefont {Svozil}}]{2018-minimalYIYS}%
  \BibitemOpen
  \bibfield  {author} {\bibinfo {author} {\bibfnamefont {Ad\'an}\ \bibnamefont
  {Cabello}}, \bibinfo {author} {\bibfnamefont {Jos\'{e}~R.}\ \bibnamefont
  {Portillo}}, \bibinfo {author} {\bibfnamefont {Alberto}\ \bibnamefont
  {Sol\'{i}s}}, \ and\ \bibinfo {author} {\bibfnamefont {Karl}\ \bibnamefont
  {Svozil}},\ }\bibfield  {title} {\enquote {\bibinfo {title} {Minimal
  true-implies-false and true-implies-true sets of propositions in
  noncontextual hidden-variable theories},}\ }\href {\doibase
  10.1103/PhysRevA.98.012106} {\bibfield  {journal} {\bibinfo  {journal}
  {Physical Review A}\ }\textbf {\bibinfo {volume} {98}},\ \bibinfo {pages}
  {012106} (\bibinfo {year} {2018})},\ \Eprint
  {http://arxiv.org/abs/arXiv:1805.00796} {arXiv:1805.00796} \BibitemShut
  {NoStop}%
\bibitem [{\citenamefont {Svozil}(2020{\natexlab{a}})}]{svozil-2020-c}%
  \BibitemOpen
  \bibfield  {author} {\bibinfo {author} {\bibfnamefont {Karl}\ \bibnamefont
  {Svozil}},\ }\bibfield  {title} {\enquote {\bibinfo {title} {Classical
  predictions for intertwined quantum observables are contingent and thus
  inconclusive},}\ }\href {\doibase 10.3390/quantum2020018} {\bibfield
  {journal} {\bibinfo  {journal} {Quantum Reports}\ }\textbf {\bibinfo {volume}
  {2}},\ \bibinfo {pages} {278--292} (\bibinfo {year} {2020}{\natexlab{a}})},\
  \Eprint {http://arxiv.org/abs/arXiv:1808.00813} {arXiv:1808.00813}
  \BibitemShut {NoStop}%
\bibitem [{\citenamefont {Kleene}(1936)}]{Kleene1936}%
  \BibitemOpen
  \bibfield  {author} {\bibinfo {author} {\bibfnamefont {Stephen~Cole}\
  \bibnamefont {Kleene}},\ }\bibfield  {title} {\enquote {\bibinfo {title}
  {General recursive functions of natural numbers},}\ }\href {\doibase
  10.1007/BF01565439} {\bibfield  {journal} {\bibinfo  {journal} {Mathematische
  Annalen}\ }\textbf {\bibinfo {volume} {112}},\ \bibinfo {pages} {727--742}
  (\bibinfo {year} {1936})}\BibitemShut {NoStop}%
\bibitem [{\citenamefont {Brukner}\ and\ \citenamefont
  {Zeilinger}(1999{\natexlab{a}})}]{zeil-bruk-99a}%
  \BibitemOpen
  \bibfield  {author} {\bibinfo {author} {\bibfnamefont {{\v{C}}aslav}\
  \bibnamefont {Brukner}}\ and\ \bibinfo {author} {\bibfnamefont {Anton}\
  \bibnamefont {Zeilinger}},\ }\bibfield  {title} {\enquote {\bibinfo {title}
  {Operationally invariant information in quantum measurements},}\ }\href
  {\doibase 10.1103/PhysRevLett.83.3354} {\bibfield  {journal} {\bibinfo
  {journal} {Physical Review Letters}\ }\textbf {\bibinfo {volume} {83}},\
  \bibinfo {pages} {3354--3357} (\bibinfo {year} {1999}{\natexlab{a}})},\
  \Eprint {http://arxiv.org/abs/quant-ph/0005084} {quant-ph/0005084}
  \BibitemShut {NoStop}%
\bibitem [{\citenamefont {Brukner}\ and\ \citenamefont
  {Zeilinger}(1999{\natexlab{b}})}]{zeil-bruk-99}%
  \BibitemOpen
  \bibfield  {author} {\bibinfo {author} {\bibfnamefont {{\v{C}}aslav}\
  \bibnamefont {Brukner}}\ and\ \bibinfo {author} {\bibfnamefont {Anton}\
  \bibnamefont {Zeilinger}},\ }\bibfield  {title} {\enquote {\bibinfo {title}
  {Malus' law and quantum information},}\ }\href
  {https://www.univie.ac.at/qfp/publications3/pdffiles/1999-08.pdf} {\bibfield
  {journal} {\bibinfo  {journal} {Acta Physica Slovaca}\ }\textbf {\bibinfo
  {volume} {49}},\ \bibinfo {pages} {647--652} (\bibinfo {year}
  {1999}{\natexlab{b}})}\BibitemShut {NoStop}%
\bibitem [{\citenamefont {Grangier}\ and\ \citenamefont
  {Auff{\`{e}}ves}(2018)}]{Grangier2018}%
  \BibitemOpen
  \bibfield  {author} {\bibinfo {author} {\bibfnamefont {P.}~\bibnamefont
  {Grangier}}\ and\ \bibinfo {author} {\bibfnamefont {A.}~\bibnamefont
  {Auff{\`{e}}ves}},\ }\bibfield  {title} {\enquote {\bibinfo {title} {What is
  quantum in quantum randomness?}}\ }\href {\doibase 10.1098/rsta.2017.0322}
  {\bibfield  {journal} {\bibinfo  {journal} {Philosophical Transactions of the
  Royal Society A: Mathematical, Physical and Engineering Sciences}\ }\textbf
  {\bibinfo {volume} {376}},\ \bibinfo {pages} {20170322} (\bibinfo {year}
  {2018})},\ \Eprint {http://arxiv.org/abs/arXiv:1804.04807} {arXiv:1804.04807}
  \BibitemShut {NoStop}%
\bibitem [{\citenamefont {{von Neumann}}(1939)}]{vonNeumann1939}%
  \BibitemOpen
  \bibfield  {author} {\bibinfo {author} {\bibfnamefont {John}\ \bibnamefont
  {{von Neumann}}},\ }\bibfield  {title} {\enquote {\bibinfo {title} {On
  infinite direct products},}\ }\href
  {http://www.numdam.org/item/CM_1939__6__1_0/} {\bibfield  {journal} {\bibinfo
   {journal} {Compositio Mathematica}\ }\textbf {\bibinfo {volume} {6}},\
  \bibinfo {pages} {1--77} (\bibinfo {year} {1939})},\ \bibinfo {note}
  {reprinted in {\sl John {von Neumann}, Collected Works, Vol. III}, A. H.
  Taub, editor, Pergamon Press, New York, 1961, nr. 6, p. 323--399}\BibitemShut
  {NoStop}%
\bibitem [{\citenamefont {Auff{\`{e}}ves}\ and\ \citenamefont
  {Grangier}(2019)}]{Auffves2019}%
  \BibitemOpen
  \bibfield  {author} {\bibinfo {author} {\bibfnamefont {Alexia}\ \bibnamefont
  {Auff{\`{e}}ves}}\ and\ \bibinfo {author} {\bibfnamefont {Philippe}\
  \bibnamefont {Grangier}},\ }\bibfield  {title} {\enquote {\bibinfo {title} {A
  generic model for quantum measurements},}\ }\href {\doibase
  10.3390/e21090904} {\bibfield  {journal} {\bibinfo  {journal} {Entropy}\
  }\textbf {\bibinfo {volume} {21}},\ \bibinfo {pages} {904} (\bibinfo {year}
  {2019})},\ \Eprint {http://arxiv.org/abs/arXiv:1907.11261} {arXiv:1907.11261}
  \BibitemShut {NoStop}%
\bibitem [{\citenamefont {Calude}\ and\ \citenamefont
  {Dinneen}(2007)}]{calude-dinneen06}%
  \BibitemOpen
  \bibfield  {author} {\bibinfo {author} {\bibfnamefont {Cristian~S.}\
  \bibnamefont {Calude}}\ and\ \bibinfo {author} {\bibfnamefont {Michael~J.}\
  \bibnamefont {Dinneen}},\ }\bibfield  {title} {\enquote {\bibinfo {title}
  {Exact approximations of omega numbers},}\ }\href {\doibase
  10.1142/S0218127407018130} {\bibfield  {journal} {\bibinfo  {journal}
  {International Journal of Bifurcation and Chaos}\ }\textbf {\bibinfo {volume}
  {17}},\ \bibinfo {pages} {1937--1954} (\bibinfo {year} {2007})},\ \bibinfo
  {note} {{CDMTCS} report series 293}\BibitemShut {NoStop}%
\bibitem [{\citenamefont {Einstein}(1935)}]{einstei-letter-to-schr}%
  \BibitemOpen
  \bibfield  {author} {\bibinfo {author} {\bibfnamefont {Albert}\ \bibnamefont
  {Einstein}},\ }\href {http://alberteinstein.info/vufind1/Record/EAR000024019}
  {\enquote {\bibinfo {title} {Letter to {S}chr\"odinger},}\ } (\bibinfo {year}
  {1935}),\ \bibinfo {note} {old Lyme, dated 19.6.35, Einstein Archives 22-047
  (searchable by document nr. 22-47), Reprinted as
  letter~206~\cite[537-539]{Meyenn-2011}}\BibitemShut {NoStop}%
\bibitem [{\citenamefont {{von Meyenn}}(2011)}]{Meyenn-2011}%
  \BibitemOpen
  \bibfield  {author} {\bibinfo {author} {\bibfnamefont {Karl}\ \bibnamefont
  {{von Meyenn}}},\ }\href {\doibase 10.1007/978-3-642-04335-2} {\emph
  {\bibinfo {title} {{E}ine {E}ntdeckung von ganz au{\ss}erordentlicher
  {T}ragweite. {S}chr{\"{o}}dingers {B}riefwechsel zur {W}ellenmechanik und zum
  {K}atzenparadoxon}}}\ (\bibinfo  {publisher} {Springer},\ \bibinfo {address}
  {Heidelberg, Dordrecht, London, New York},\ \bibinfo {year}
  {2011})\BibitemShut {NoStop}%
\bibitem [{\citenamefont {Howard}(1985)}]{Howard1985171}%
  \BibitemOpen
  \bibfield  {author} {\bibinfo {author} {\bibfnamefont {Don}\ \bibnamefont
  {Howard}},\ }\bibfield  {title} {\enquote {\bibinfo {title} {{E}instein on
  locality and separability},}\ }\href {\doibase 10.1016/0039-3681(85)90001-9}
  {\bibfield  {journal} {\bibinfo  {journal} {Studies in History and Philosophy
  of Science Part A}\ }\textbf {\bibinfo {volume} {16}},\ \bibinfo {pages}
  {171--201} (\bibinfo {year} {1985})}\BibitemShut {NoStop}%
\bibitem [{\citenamefont {Einstein}\ \emph {et~al.}(1935)\citenamefont
  {Einstein}, \citenamefont {Podolsky},\ and\ \citenamefont {Rosen}}]{epr}%
  \BibitemOpen
  \bibfield  {author} {\bibinfo {author} {\bibfnamefont {Albert}\ \bibnamefont
  {Einstein}}, \bibinfo {author} {\bibfnamefont {Boris}\ \bibnamefont
  {Podolsky}}, \ and\ \bibinfo {author} {\bibfnamefont {Nathan}\ \bibnamefont
  {Rosen}},\ }\bibfield  {title} {\enquote {\bibinfo {title} {Can
  quantum-mechanical description of physical reality be considered complete?}}\
  }\href {\doibase 10.1103/PhysRev.47.777} {\bibfield  {journal} {\bibinfo
  {journal} {Physical Review}\ }\textbf {\bibinfo {volume} {47}},\ \bibinfo
  {pages} {777--780} (\bibinfo {year} {1935})}\BibitemShut {NoStop}%
\bibitem [{\citenamefont
  {Schr{\"{o}}dinger}(1936)}]{CambridgeJournals:2027212}%
  \BibitemOpen
  \bibfield  {author} {\bibinfo {author} {\bibfnamefont {Erwin}\ \bibnamefont
  {Schr{\"{o}}dinger}},\ }\bibfield  {title} {\enquote {\bibinfo {title}
  {Probability relations between separated systems},}\ }\href {\doibase
  10.1017/S0305004100019137} {\bibfield  {journal} {\bibinfo  {journal}
  {Mathematical Proceedings of the Cambridge Philosophical Society}\ }\textbf
  {\bibinfo {volume} {32}},\ \bibinfo {pages} {446--452} (\bibinfo {year}
  {1936})}\BibitemShut {NoStop}%
\bibitem [{\citenamefont
  {Schr{\"{o}}dinger}(1935{\natexlab{b}})}]{CambridgeJournals:1737068}%
  \BibitemOpen
  \bibfield  {author} {\bibinfo {author} {\bibfnamefont {Erwin}\ \bibnamefont
  {Schr{\"{o}}dinger}},\ }\bibfield  {title} {\enquote {\bibinfo {title}
  {Discussion of probability relations between separated systems},}\ }\href
  {\doibase 10.1017/S0305004100013554} {\bibfield  {journal} {\bibinfo
  {journal} {Mathematical Proceedings of the Cambridge Philosophical Society}\
  }\textbf {\bibinfo {volume} {31}},\ \bibinfo {pages} {555--563} (\bibinfo
  {year} {1935}{\natexlab{b}})}\BibitemShut {NoStop}%
\bibitem [{\citenamefont {Svozil}(2018{\natexlab{b}})}]{svozil-2016-sampling}%
  \BibitemOpen
  \bibfield  {author} {\bibinfo {author} {\bibfnamefont {Karl}\ \bibnamefont
  {Svozil}},\ }\bibfield  {title} {\enquote {\bibinfo {title} {A note on the
  statistical sampling aspect of delayed choice entanglement swapping},}\ }in\
  \href {\doibase 10.1142/9789813276895\_0001} {\emph {\bibinfo {booktitle}
  {Probing the Meaning of Quantum Mechanics}}}\ (\bibinfo  {publisher} {World
  Scientific},\ \bibinfo {address} {Singapore},\ \bibinfo {year} {2018})\ pp.\
  \bibinfo {pages} {1--9},\ \Eprint {http://arxiv.org/abs/arXiv:1608.04984}
  {arXiv:1608.04984} \BibitemShut {NoStop}%
\bibitem [{\citenamefont {Svozil}(2018{\natexlab{c}})}]{svozil-pac}%
  \BibitemOpen
  \bibfield  {author} {\bibinfo {author} {\bibfnamefont {Karl}\ \bibnamefont
  {Svozil}},\ }\href {\doibase 10.1007/978-3-319-70815-7} {\emph {\bibinfo
  {title} {Physical (A)Causality}}},\ \bibinfo {series} {Fundamental Theories
  of Physics}, Vol.\ \bibinfo {volume} {192}\ (\bibinfo  {publisher} {Springer
  International Publishing},\ \bibinfo {address} {Cham, Heidelberg, New York,
  Dordrecht, London},\ \bibinfo {year} {2018})\BibitemShut {NoStop}%
\bibitem [{\citenamefont {Bell}(1992)}]{bell:a1}%
  \BibitemOpen
  \bibfield  {author} {\bibinfo {author} {\bibfnamefont {John~S.}\ \bibnamefont
  {Bell}},\ }\bibfield  {title} {\enquote {\bibinfo {title} {Against
  `measurement'},}\ }\href@noop {} {\bibfield  {journal} {\bibinfo  {journal}
  {Physikalische Bl{\"{a}}tter}\ }\textbf {\bibinfo {volume} {48}},\ \bibinfo
  {pages} {267} (\bibinfo {year} {1992})}\BibitemShut {NoStop}%
\bibitem [{\citenamefont {Bell}(1990)}]{bell-a}%
  \BibitemOpen
  \bibfield  {author} {\bibinfo {author} {\bibfnamefont {John~Stuard}\
  \bibnamefont {Bell}},\ }\bibfield  {title} {\enquote {\bibinfo {title}
  {Against `measurement'},}\ }\href {\doibase 10.1088/2058-7058/3/8/26}
  {\bibfield  {journal} {\bibinfo  {journal} {Physics World}\ }\textbf
  {\bibinfo {volume} {3}},\ \bibinfo {pages} {33--41} (\bibinfo {year}
  {1990})}\BibitemShut {NoStop}%
\bibitem [{\citenamefont
  {Van~Raamsdonk}(2010{\natexlab{a}})}]{VanRaamsdonk2010}%
  \BibitemOpen
  \bibfield  {author} {\bibinfo {author} {\bibfnamefont {Mark}\ \bibnamefont
  {Van~Raamsdonk}},\ }\bibfield  {title} {\enquote {\bibinfo {title} {Building
  up spacetime with quantum entanglement},}\ }\href {\doibase
  10.1007/s10714-010-1034-0} {\bibfield  {journal} {\bibinfo  {journal}
  {General Relativity and Gravitation}\ }\textbf {\bibinfo {volume} {42}},\
  \bibinfo {pages} {2323--2329} (\bibinfo {year} {2010}{\natexlab{a}})},\
  \Eprint {http://arxiv.org/abs/arXiv:1005.3035} {arXiv:1005.3035} \BibitemShut
  {NoStop}%
\bibitem [{\citenamefont
  {Van~Raamsdonk}(2010{\natexlab{b}})}]{VANRAAMSDONK2010b}%
  \BibitemOpen
  \bibfield  {author} {\bibinfo {author} {\bibfnamefont {Mark}\ \bibnamefont
  {Van~Raamsdonk}},\ }\bibfield  {title} {\enquote {\bibinfo {title} {Building
  up spacetime with quantum entanglement},}\ }\href {\doibase
  10.1142/s0218271810018529} {\bibfield  {journal} {\bibinfo  {journal}
  {International Journal of Modern Physics D}\ }\textbf {\bibinfo {volume}
  {19}},\ \bibinfo {pages} {2429--2435} (\bibinfo {year}
  {2010}{\natexlab{b}})},\ \Eprint {http://arxiv.org/abs/arXiv:1005.3035}
  {arXiv:1005.3035} \BibitemShut {NoStop}%
\bibitem [{\citenamefont {Faulkner}\ \emph {et~al.}(2014)\citenamefont
  {Faulkner}, \citenamefont {Guica}, \citenamefont {Hartman}, \citenamefont
  {Myers},\ and\ \citenamefont {Raamsdonk}}]{Faulkner2014}%
  \BibitemOpen
  \bibfield  {author} {\bibinfo {author} {\bibfnamefont {Thomas}\ \bibnamefont
  {Faulkner}}, \bibinfo {author} {\bibfnamefont {Monica}\ \bibnamefont
  {Guica}}, \bibinfo {author} {\bibfnamefont {Thomas}\ \bibnamefont {Hartman}},
  \bibinfo {author} {\bibfnamefont {Robert~C.}\ \bibnamefont {Myers}}, \ and\
  \bibinfo {author} {\bibfnamefont {Mark~Van}\ \bibnamefont {Raamsdonk}},\
  }\bibfield  {title} {\enquote {\bibinfo {title} {Gravitation from
  entanglement in holographic {CFTs}},}\ }\href {\doibase
  10.1007/jhep03(2014)051} {\bibfield  {journal} {\bibinfo  {journal} {Journal
  of High Energy Physics}\ }\textbf {\bibinfo {volume} {2014}} (\bibinfo {year}
  {2014}),\ 10.1007/jhep03(2014)051},\ \Eprint
  {http://arxiv.org/abs/arXiv:1312.7856} {arXiv:1312.7856} \BibitemShut
  {NoStop}%
\bibitem [{\citenamefont {Swingle}\ and\ \citenamefont
  {Van~Raamsdonk}(2014)}]{Swingle:2014uza}%
  \BibitemOpen
  \bibfield  {author} {\bibinfo {author} {\bibfnamefont {Brian}\ \bibnamefont
  {Swingle}}\ and\ \bibinfo {author} {\bibfnamefont {Mark}\ \bibnamefont
  {Van~Raamsdonk}},\ }\href@noop {} {\enquote {\bibinfo {title} {Universality
  of gravity from entanglement},}\ } (\bibinfo {year} {2014})\BibitemShut
  {NoStop}%
\bibitem [{\citenamefont {Jacobson}(2016)}]{Jacobson2016}%
  \BibitemOpen
  \bibfield  {author} {\bibinfo {author} {\bibfnamefont {Ted}\ \bibnamefont
  {Jacobson}},\ }\bibfield  {title} {\enquote {\bibinfo {title} {Entanglement
  equilibrium and the {E}instein equation},}\ }\href {\doibase
  10.1103/physrevlett.116.201101} {\bibfield  {journal} {\bibinfo  {journal}
  {Physical Review Letters}\ }\textbf {\bibinfo {volume} {116}},\ \bibinfo
  {pages} {201101} (\bibinfo {year} {2016})},\ \Eprint
  {http://arxiv.org/abs/arXiv:1505.04753} {arXiv:1505.04753} \BibitemShut
  {NoStop}%
\bibitem [{\citenamefont {Cao}\ \emph {et~al.}(2017)\citenamefont {Cao},
  \citenamefont {Carroll},\ and\ \citenamefont {Michalakis}}]{Cao2017}%
  \BibitemOpen
  \bibfield  {author} {\bibinfo {author} {\bibfnamefont {ChunJun}\ \bibnamefont
  {Cao}}, \bibinfo {author} {\bibfnamefont {Sean~M.}\ \bibnamefont {Carroll}},
  \ and\ \bibinfo {author} {\bibfnamefont {Spyridon}\ \bibnamefont
  {Michalakis}},\ }\bibfield  {title} {\enquote {\bibinfo {title} {Space from
  hilbert space: Recovering geometry from bulk entanglement},}\ }\href
  {\doibase 10.1103/physrevd.95.024031} {\bibfield  {journal} {\bibinfo
  {journal} {Physical Review D}\ }\textbf {\bibinfo {volume} {95}},\ \bibinfo
  {pages} {024031} (\bibinfo {year} {2017})},\ \Eprint
  {http://arxiv.org/abs/arXiv:1606.08444} {arXiv:1606.08444} \BibitemShut
  {NoStop}%
\bibitem [{\citenamefont {Swingle}(2018)}]{Swingle2018}%
  \BibitemOpen
  \bibfield  {author} {\bibinfo {author} {\bibfnamefont {Brian}\ \bibnamefont
  {Swingle}},\ }\bibfield  {title} {\enquote {\bibinfo {title} {Spacetime from
  entanglement},}\ }\href {\doibase 10.1146/annurev-conmatphys-033117-054219}
  {\bibfield  {journal} {\bibinfo  {journal} {Annual Review of Condensed Matter
  Physics}\ }\textbf {\bibinfo {volume} {9}},\ \bibinfo {pages} {345--358}
  (\bibinfo {year} {2018})}\BibitemShut {NoStop}%
\bibitem [{\citenamefont {Musser}(2018)}]{Musser2018}%
  \BibitemOpen
  \bibfield  {author} {\bibinfo {author} {\bibfnamefont {George}\ \bibnamefont
  {Musser}},\ }\bibfield  {title} {\enquote {\bibinfo {title} {What is
  spacetime?}}\ }\href {\doibase 10.1038/d41586-018-05095-z} {\bibfield
  {journal} {\bibinfo  {journal} {Nature}\ }\textbf {\bibinfo {volume} {557}},\
  \bibinfo {pages} {S3--S6} (\bibinfo {year} {2018})}\BibitemShut {NoStop}%
\bibitem [{\citenamefont {Knuth}\ and\ \citenamefont
  {Bahreyni}(2012)}]{Knuth-Bahreyni}%
  \BibitemOpen
  \bibfield  {author} {\bibinfo {author} {\bibfnamefont {Kevin~H.}\
  \bibnamefont {Knuth}}\ and\ \bibinfo {author} {\bibfnamefont
  {N.}~\bibnamefont {Bahreyni}},\ }\bibfield  {title} {\enquote {\bibinfo
  {title} {The physics of events: A potential foundation for emergent
  space-time},}\ }\href {http://arxiv.org/abs/1209.0881} {\bibfield  {journal}
  {\bibinfo  {journal} {ArXiv e-prints}\ } (\bibinfo {year} {2012})},\ \Eprint
  {http://arxiv.org/abs/arXiv:1209.0881} {arXiv:arXiv:1209.0881 [math-ph]}
  \BibitemShut {NoStop}%
\bibitem [{\citenamefont {Couch}\ \emph {et~al.}(2020)\citenamefont {Couch},
  \citenamefont {Eccles}, \citenamefont {Nguyen}, \citenamefont {Swingle},\
  and\ \citenamefont {Xu}}]{Couch2020}%
  \BibitemOpen
  \bibfield  {author} {\bibinfo {author} {\bibfnamefont {Josiah}\ \bibnamefont
  {Couch}}, \bibinfo {author} {\bibfnamefont {Stefan}\ \bibnamefont {Eccles}},
  \bibinfo {author} {\bibfnamefont {Phuc}\ \bibnamefont {Nguyen}}, \bibinfo
  {author} {\bibfnamefont {Brian}\ \bibnamefont {Swingle}}, \ and\ \bibinfo
  {author} {\bibfnamefont {Shenglong}\ \bibnamefont {Xu}},\ }\bibfield  {title}
  {\enquote {\bibinfo {title} {Speed of quantum information spreading in
  chaotic systems},}\ }\href {\doibase 10.1103/physrevb.102.045114} {\bibfield
  {journal} {\bibinfo  {journal} {Physical Review B}\ }\textbf {\bibinfo
  {volume} {102}},\ \bibinfo {pages} {045114} (\bibinfo {year} {2020})},\
  \Eprint {http://arxiv.org/abs/arXiv:1908.06993} {arXiv:1908.06993}
  \BibitemShut {NoStop}%
\bibitem [{\citenamefont {Knuth}\ \emph {et~al.}(2019)\citenamefont {Knuth},
  \citenamefont {Powell},\ and\ \citenamefont {Reali}}]{Knuth-e21100939}%
  \BibitemOpen
  \bibfield  {author} {\bibinfo {author} {\bibfnamefont {Kevin~H.}\
  \bibnamefont {Knuth}}, \bibinfo {author} {\bibfnamefont {Robert~M.}\
  \bibnamefont {Powell}}, \ and\ \bibinfo {author} {\bibfnamefont {Peter~A.}\
  \bibnamefont {Reali}},\ }\bibfield  {title} {\enquote {\bibinfo {title}
  {Estimating flight characteristics of anomalous unidentified aerial
  vehicles},}\ }\href {\doibase 10.3390/e21100939} {\bibfield  {journal}
  {\bibinfo  {journal} {Entropy}\ }\textbf {\bibinfo {volume} {21}} (\bibinfo
  {year} {2019}),\ 10.3390/e21100939}\BibitemShut {NoStop}%
\bibitem [{\citenamefont {Peres}(1978)}]{peres222}%
  \BibitemOpen
  \bibfield  {author} {\bibinfo {author} {\bibfnamefont {Asher}\ \bibnamefont
  {Peres}},\ }\bibfield  {title} {\enquote {\bibinfo {title} {Unperformed
  experiments have no results},}\ }\href {\doibase 10.1119/1.11393} {\bibfield
  {journal} {\bibinfo  {journal} {American Journal of Physics}\ }\textbf
  {\bibinfo {volume} {46}},\ \bibinfo {pages} {745--747} (\bibinfo {year}
  {1978})}\BibitemShut {NoStop}%
\bibitem [{\citenamefont {Peres}(1993)}]{peres}%
  \BibitemOpen
  \bibfield  {author} {\bibinfo {author} {\bibfnamefont {Asher}\ \bibnamefont
  {Peres}},\ }\href@noop {} {\emph {\bibinfo {title} {Quantum Theory: Concepts
  and Methods}}}\ (\bibinfo  {publisher} {Kluwer Academic Publishers},\
  \bibinfo {address} {Dordrecht},\ \bibinfo {year} {1993})\BibitemShut
  {NoStop}%
\bibitem [{\citenamefont {Toner}\ and\ \citenamefont
  {Bacon}(2003)}]{toner-bacon-03}%
  \BibitemOpen
  \bibfield  {author} {\bibinfo {author} {\bibfnamefont {B.~F.}\ \bibnamefont
  {Toner}}\ and\ \bibinfo {author} {\bibfnamefont {D.}~\bibnamefont {Bacon}},\
  }\bibfield  {title} {\enquote {\bibinfo {title} {Communication cost of
  simulating {B}ell correlations},}\ }\href {\doibase
  10.1103/PhysRevLett.91.187904} {\bibfield  {journal} {\bibinfo  {journal}
  {Physical Review Letters}\ }\textbf {\bibinfo {volume} {91}},\ \bibinfo
  {pages} {187904} (\bibinfo {year} {2003})}\BibitemShut {NoStop}%
\bibitem [{\citenamefont {Svozil}(2005)}]{svozil-2004-brainteaser}%
  \BibitemOpen
  \bibfield  {author} {\bibinfo {author} {\bibfnamefont {Karl}\ \bibnamefont
  {Svozil}},\ }\bibfield  {title} {\enquote {\bibinfo {title} {Communication
  cost of breaking the {B}ell barrier},}\ }\href {\doibase
  10.1103/PhysRevA.72.050302} {\bibfield  {journal} {\bibinfo  {journal}
  {Physical Review A}\ }\textbf {\bibinfo {volume} {72}},\ \bibinfo {pages}
  {050302(R)} (\bibinfo {year} {2005})},\ \Eprint
  {http://arxiv.org/abs/arXiv:physics/0510050} {arXiv:physics/0510050}
  \BibitemShut {NoStop}%
\bibitem [{\citenamefont {Shimony}(1984)}]{shimony2}%
  \BibitemOpen
  \bibfield  {author} {\bibinfo {author} {\bibfnamefont {Abner}\ \bibnamefont
  {Shimony}},\ }\bibfield  {title} {\enquote {\bibinfo {title} {Controllable
  and uncontrollable non-locality},}\ }in\ \href@noop {} {\emph {\bibinfo
  {booktitle} {Proceedings of the International Symposium... Proceedings of the
  International Symposium Foundations of Quantum Mechanics in the Light of New
  Technology}}},\ \bibinfo {editor} {edited by\ \bibinfo {editor}
  {\bibfnamefont {Susumu}\ \bibnamefont {Kamefuchi}}\ and\ \bibinfo {editor}
  {\bibfnamefont {Nihon~Butsuri}\ \bibnamefont {Gakkai}}}\ (\bibinfo
  {publisher} {Physical Society of Japan},\ \bibinfo {address} {Tokyo},\
  \bibinfo {year} {1984})\ pp.\ \bibinfo {pages} {225--230},\ \bibinfo {note}
  {see also J. Jarrett, {\sl Bell's Theorem, Quantum Mechanics and Local
  Realism}, Ph. D. thesis, Univ. of Chicago, 1983; {\sl Nous}, {\bf 18}, 569
  (1984)}\BibitemShut {NoStop}%
\bibitem [{\citenamefont {Shimony}(1986)}]{shimony3}%
  \BibitemOpen
  \bibfield  {author} {\bibinfo {author} {\bibfnamefont {Abner}\ \bibnamefont
  {Shimony}},\ }\bibfield  {title} {\enquote {\bibinfo {title} {Events and
  processes in the quantum world},}\ }in\ \href@noop {} {\emph {\bibinfo
  {booktitle} {Quantum Concepts in Space and Time}}},\ \bibinfo {editor}
  {edited by\ \bibinfo {editor} {\bibfnamefont {R.}~\bibnamefont {Penrose}}\
  and\ \bibinfo {editor} {\bibfnamefont {C.~I.}\ \bibnamefont {Isham}}}\
  (\bibinfo  {publisher} {Clarendon Press},\ \bibinfo {address} {Oxford},\
  \bibinfo {year} {1986})\ pp.\ \bibinfo {pages} {182--203}\BibitemShut
  {NoStop}%
\bibitem [{\citenamefont {Griffiths}(2010)}]{Griffiths2010}%
  \BibitemOpen
  \bibfield  {author} {\bibinfo {author} {\bibfnamefont {Robert~B.}\
  \bibnamefont {Griffiths}},\ }\bibfield  {title} {\enquote {\bibinfo {title}
  {Quantum locality},}\ }\href {\doibase 10.1007/s10701-010-9512-5} {\bibfield
  {journal} {\bibinfo  {journal} {Foundations of Physics}\ }\textbf {\bibinfo
  {volume} {41}},\ \bibinfo {pages} {705--733} (\bibinfo {year}
  {2010})}\BibitemShut {NoStop}%
\bibitem [{\citenamefont {Griffiths}(2020)}]{Griffiths2020}%
  \BibitemOpen
  \bibfield  {author} {\bibinfo {author} {\bibfnamefont {Robert~B.}\
  \bibnamefont {Griffiths}},\ }\bibfield  {title} {\enquote {\bibinfo {title}
  {Nonlocality claims are inconsistent with {H}ilbert-space quantum
  mechanics},}\ }\href {\doibase 10.1103/physreva.101.022117} {\bibfield
  {journal} {\bibinfo  {journal} {Physical Review A}\ }\textbf {\bibinfo
  {volume} {101}} (\bibinfo {year} {2020}),\ 10.1103/physreva.101.022117},\
  \Eprint {http://arxiv.org/abs/arXiv:1901.07050} {arXiv:1901.07050}
  \BibitemShut {NoStop}%
\bibitem [{\citenamefont {Popescu}\ and\ \citenamefont
  {Rohrlich}(1997)}]{popescu-97}%
  \BibitemOpen
  \bibfield  {author} {\bibinfo {author} {\bibfnamefont {Sandu}\ \bibnamefont
  {Popescu}}\ and\ \bibinfo {author} {\bibfnamefont {Daniel}\ \bibnamefont
  {Rohrlich}},\ }\bibfield  {title} {\enquote {\bibinfo {title} {Action and
  passion at a distance},}\ }in\ \href {\doibase 10.1007/978-94-017-2732-7\_15}
  {\emph {\bibinfo {booktitle} {Potentiality, Entanglement and
  Passion-at-a-Distance: Quantum Mechanical Studies for {A}bner {S}himony,
  {V}olume Two ({B}oston Studies in the Philosophy of Science)}}},\ \bibinfo
  {editor} {edited by\ \bibinfo {editor} {\bibfnamefont {Robert~S.}\
  \bibnamefont {Cohen}}, \bibinfo {editor} {\bibfnamefont {Michael}\
  \bibnamefont {Horne}}, \ and\ \bibinfo {editor} {\bibfnamefont {John}\
  \bibnamefont {Stachel}}}\ (\bibinfo  {publisher} {Kluwer Academic publishers,
  Springer Netherlands},\ \bibinfo {address} {Dordrecht},\ \bibinfo {year}
  {1997})\ pp.\ \bibinfo {pages} {197--206},\ \Eprint
  {http://arxiv.org/abs/arXiv:quant-ph/9605004} {arXiv:quant-ph/9605004}
  \BibitemShut {NoStop}%
\bibitem [{\citenamefont {Krenn}\ and\ \citenamefont
  {Svozil}(1998)}]{svozil-krenn}%
  \BibitemOpen
  \bibfield  {author} {\bibinfo {author} {\bibfnamefont {G{\"{u}}nther}\
  \bibnamefont {Krenn}}\ and\ \bibinfo {author} {\bibfnamefont {Karl}\
  \bibnamefont {Svozil}},\ }\bibfield  {title} {\enquote {\bibinfo {title}
  {Stronger-than-quantum correlations},}\ }\href {\doibase
  10.1023/A:1018821314465} {\bibfield  {journal} {\bibinfo  {journal}
  {Foundations of Physics}\ }\textbf {\bibinfo {volume} {28}},\ \bibinfo
  {pages} {971--984} (\bibinfo {year} {1998})}\BibitemShut {NoStop}%
\bibitem [{\citenamefont {Popescu}(2014)}]{popescu-2014}%
  \BibitemOpen
  \bibfield  {author} {\bibinfo {author} {\bibfnamefont {Sandu}\ \bibnamefont
  {Popescu}},\ }\bibfield  {title} {\enquote {\bibinfo {title} {Nonlocality
  beyond quantum mechanics},}\ }\href {\doibase 10.1038/nphys2916} {\bibfield
  {journal} {\bibinfo  {journal} {Nature Physics}\ }\textbf {\bibinfo {volume}
  {10}},\ \bibinfo {pages} {264--270} (\bibinfo {year} {2014})}\BibitemShut
  {NoStop}%
\bibitem [{\citenamefont {Herbert}(1982)}]{herbert}%
  \BibitemOpen
  \bibfield  {author} {\bibinfo {author} {\bibfnamefont {Nick}\ \bibnamefont
  {Herbert}},\ }\bibfield  {title} {\enquote {\bibinfo {title} {{FLASH}---{A}
  superluminal communicator based upon a new kind of quantum measurement},}\
  }\href {\doibase 10.1007/BF00729622} {\bibfield  {journal} {\bibinfo
  {journal} {Foundation of Physics}\ }\textbf {\bibinfo {volume} {12}},\
  \bibinfo {pages} {1171--1179} (\bibinfo {year} {1982})}\BibitemShut {NoStop}%
\bibitem [{\citenamefont {Svozil}(1989)}]{svozil-slash}%
  \BibitemOpen
  \bibfield  {author} {\bibinfo {author} {\bibfnamefont {Karl}\ \bibnamefont
  {Svozil}},\ }\href {http://arxiv.org/abs/quant-ph/0103166} {\enquote
  {\bibinfo {title} {What is wrong with {SLASH}?}}\ } (\bibinfo {year}
  {1989}),\ \bibinfo {note} {eprint arXiv:quant-ph/0103166},\ \Eprint
  {http://arxiv.org/abs/arXiv:quant-ph/0103166} {arXiv:quant-ph/0103166}
  \BibitemShut {NoStop}%
\bibitem [{\citenamefont {Swinburne}(1970)}]{Swinburne-Miracle}%
  \BibitemOpen
  \bibfield  {author} {\bibinfo {author} {\bibfnamefont {Richard}\ \bibnamefont
  {Swinburne}},\ }\href {\doibase 10.1007/978-1-349-00776-9} {\emph {\bibinfo
  {title} {The Concept of Miracle}}},\ New Studies in the Philosophy of
  Religion\ (\bibinfo  {publisher} {Palgrave Macmillan},\ \bibinfo {address}
  {London},\ \bibinfo {year} {1970})\BibitemShut {NoStop}%
\bibitem [{\citenamefont {Frank}(1932)}]{frank}%
  \BibitemOpen
  \bibfield  {author} {\bibinfo {author} {\bibfnamefont {Philipp}\ \bibnamefont
  {Frank}},\ }\href@noop {} {\emph {\bibinfo {title} {Das {K}ausalgesetz und
  seine {G}renzen}}}\ (\bibinfo  {publisher} {Springer},\ \bibinfo {address}
  {Vienna},\ \bibinfo {year} {1932})\BibitemShut {NoStop}%
\bibitem [{\citenamefont {Frank}\ and\ \citenamefont {{R. S. Cohen
  (Editor)}}(1997)}]{franke}%
  \BibitemOpen
  \bibfield  {author} {\bibinfo {author} {\bibfnamefont {Philipp}\ \bibnamefont
  {Frank}}\ and\ \bibinfo {author} {\bibnamefont {{R. S. Cohen (Editor)}}},\
  }\href {\doibase 10.1007/978-94-011-5516-8} {\emph {\bibinfo {title} {The Law
  of Causality and its Limits (Vienna Circle Collection)}}}\ (\bibinfo
  {publisher} {Springer},\ \bibinfo {address} {Vienna},\ \bibinfo {year}
  {1997})\BibitemShut {NoStop}%
\bibitem [{\citenamefont {Melamed}\ and\ \citenamefont
  {Lin}(2020)}]{sep-sufficient-reason}%
  \BibitemOpen
  \bibfield  {author} {\bibinfo {author} {\bibfnamefont {Yitzhak~Y.}\
  \bibnamefont {Melamed}}\ and\ \bibinfo {author} {\bibfnamefont {Martin}\
  \bibnamefont {Lin}},\ }\bibfield  {title} {\enquote {\bibinfo {title}
  {Principle of sufficient reason},}\ }in\ \href
  {https://plato.stanford.edu/archives/spr2020/entries/sufficient-reason/}
  {\emph {\bibinfo {booktitle} {The {}Stanford Encyclopedia of Philosophy}}},\
  \bibinfo {editor} {edited by\ \bibinfo {editor} {\bibfnamefont {Edward~N.}\
  \bibnamefont {Zalta}}}\ (\bibinfo  {publisher} {Metaphysics Research Lab,
  Stanford University},\ \bibinfo {year} {2020})\ \bibinfo {edition} {spring
  2020}\ ed.\BibitemShut {Stop}%
\bibitem [{\citenamefont {Wigner}(1960)}]{wigner}%
  \BibitemOpen
  \bibfield  {author} {\bibinfo {author} {\bibfnamefont {Eugene~P.}\
  \bibnamefont {Wigner}},\ }\bibfield  {title} {\enquote {\bibinfo {title} {The
  unreasonable effectiveness of mathematics in the natural sciences. {R}ichard
  {C}ourant {L}ecture delivered at {N}ew {Y}ork {U}niversity, {M}ay 11,
  1959},}\ }\href {\doibase 10.1002/cpa.3160130102} {\bibfield  {journal}
  {\bibinfo  {journal} {Communications on Pure and Applied Mathematics}\
  }\textbf {\bibinfo {volume} {13}},\ \bibinfo {pages} {1--14} (\bibinfo {year}
  {1960})}\BibitemShut {NoStop}%
\bibitem [{\citenamefont {Berkeley}(1710)}]{berkeley}%
  \BibitemOpen
  \bibfield  {author} {\bibinfo {author} {\bibfnamefont {George}\ \bibnamefont
  {Berkeley}},\ }\href {http://www.gutenberg.org/etext/4723} {\emph {\bibinfo
  {title} {A Treatise Concerning the Principles of Human Knowledge}}}\
  (\bibinfo  {publisher} {Aaron Rhames, for Jeremy Pepyat, Bookseller},\
  \bibinfo {address} {Skinner--Row, Dublin},\ \bibinfo {year}
  {1710})\BibitemShut {NoStop}%
\bibitem [{\citenamefont {Stace}(1934)}]{stace}%
  \BibitemOpen
  \bibfield  {author} {\bibinfo {author} {\bibfnamefont {Walter~Terence}\
  \bibnamefont {Stace}},\ }\bibfield  {title} {\enquote {\bibinfo {title} {The
  refutation of realism},}\ }\href {\doibase 10.1093/mind/XLIII.170.145}
  {\bibfield  {journal} {\bibinfo  {journal} {Mind}\ }\textbf {\bibinfo
  {volume} {43}},\ \bibinfo {pages} {145--155} (\bibinfo {year}
  {1934})}\BibitemShut {NoStop}%
\bibitem [{\citenamefont {Stace}(1949)}]{stace1}%
  \BibitemOpen
  \bibfield  {author} {\bibinfo {author} {\bibfnamefont {Walter~Terence}\
  \bibnamefont {Stace}},\ }\bibfield  {title} {\enquote {\bibinfo {title} {The
  refutation of realism},}\ }in\ \href {\doibase 10.1093/mind/xliii.170.145}
  {\emph {\bibinfo {booktitle} {Readings in Philosophical Analysis}}},\
  \bibinfo {editor} {edited by\ \bibinfo {editor} {\bibfnamefont {Herbert}\
  \bibnamefont {Feigl}}\ and\ \bibinfo {editor} {\bibfnamefont {Wilfrid}\
  \bibnamefont {Sellars}}}\ (\bibinfo  {publisher} {Appleton-Century-Crofts},\
  \bibinfo {address} {New York},\ \bibinfo {year} {1949})\ pp.\ \bibinfo
  {pages} {364--372},\ \bibinfo {note} {previously published in {\em Mind} {\bf
  53}, 349-353 (1934)}\BibitemShut {NoStop}%
\bibitem [{\citenamefont {Goldschmidt}\ and\ \citenamefont {Pearce}(2017,
  2018)}]{Goldschmidt2017-idealism}%
  \BibitemOpen
  \bibfield  {author} {\bibinfo {author} {\bibfnamefont {Tyron}\ \bibnamefont
  {Goldschmidt}}\ and\ \bibinfo {author} {\bibfnamefont {Kenneth~L.}\
  \bibnamefont {Pearce}},\ }\href {\doibase 10.1093/oso/9780198746973.001.0001}
  {\emph {\bibinfo {title} {Idealism: New Essays in Metaphysics}}}\ (\bibinfo
  {publisher} {Oxford University Press},\ \bibinfo {address} {Oxford, UK},\
  \bibinfo {year} {2017, 2018})\BibitemShut {NoStop}%
\bibitem [{\citenamefont {Descartes}(1641, 1996
  (translation)}]{descartes-meditation}%
  \BibitemOpen
  \bibfield  {author} {\bibinfo {author} {\bibfnamefont {Rene}\ \bibnamefont
  {Descartes}},\ }\href
  {http://www.cambridge.org/catalogue/catalogue.asp?isbn=9781107108660} {\emph
  {\bibinfo {title} {Meditations on First Philosophy}}}\ (\bibinfo  {publisher}
  {Cambridge University Press},\ \bibinfo {address} {Cambridge, UK},\ \bibinfo
  {year} {1641, 1996 (translation})\ \bibinfo {note} {translated by John
  Cottingham}\BibitemShut {NoStop}%
\bibitem [{\citenamefont {Nietzsche}(1908, 2009-)}]{Nietzsche-WahrheitLuege}%
  \BibitemOpen
  \bibfield  {author} {\bibinfo {author} {\bibfnamefont {Friedrich}\
  \bibnamefont {Nietzsche}},\ }\href
  {http://www.nietzschesource.org/\#eKGWB/WL} {\emph {\bibinfo {title} {{U}eber
  {W}ahrheit und {L}\"uge im au{\ss}ermoralischen {S}inne}}}\ (\bibinfo {year}
  {1908, 2009-})\ \bibinfo {note} {digital critical edition of the complete
  works and letters, based on the critical text by G. Colli and M. Montinari,
  Berlin/New York, de Gruyter 1967-, edited by Paolo D'Iorio}\BibitemShut
  {NoStop}%
\bibitem [{\citenamefont {Derrida}(2000)}]{derrida-Royle}%
  \BibitemOpen
  \bibfield  {author} {\bibinfo {author} {\bibfnamefont {Jacques}\ \bibnamefont
  {Derrida}},\ }\bibfield  {title} {\enquote {\bibinfo {title} {Et cetera},}\
  }in\ \href {\doibase 10.1007/978-1-137-06095-2_15} {\emph {\bibinfo
  {booktitle} {Deconstructions}}},\ \bibinfo {editor} {edited by\ \bibinfo
  {editor} {\bibfnamefont {Nicholas}\ \bibnamefont {Royle}}}\ (\bibinfo
  {publisher} {Palgrave Publishers Ltd (formerly Macmillan Press Ltd)},\
  \bibinfo {address} {Houndmills, Basingstoke, Hampshire, UK and New York, NY,
  USA},\ \bibinfo {year} {2000})\ Chap.~\bibinfo {chapter} {15}\BibitemShut
  {NoStop}%
\bibitem [{\citenamefont {Artaud}(1938)}]{Arthaud}%
  \BibitemOpen
  \bibfield  {author} {\bibinfo {author} {\bibfnamefont {Antonin}\ \bibnamefont
  {Artaud}},\ }\href@noop {} {\emph {\bibinfo {title} {Le th{\'{e}}{\^{a}}tre
  et son double}}}\ (\bibinfo  {publisher} {Gallimard},\ \bibinfo {address}
  {Paris},\ \bibinfo {year} {1938})\BibitemShut {NoStop}%
\bibitem [{\citenamefont {Artaud}(2010 (1938, 1964))}]{Arthaud-en}%
  \BibitemOpen
  \bibfield  {author} {\bibinfo {author} {\bibfnamefont {Antonin}\ \bibnamefont
  {Artaud}},\ }\href
  {https://almabooks.com/product/the-theatre-and-its-double/} {\emph {\bibinfo
  {title} {The Theatre and Its Double}}}\ (\bibinfo  {publisher} {Alma Classics
  Limited},\ \bibinfo {address} {Richmond Surrey, UK},\ \bibinfo {year} {2010
  (1938, 1964)})\ \bibinfo {note} {translated by Victor Corti}\BibitemShut
  {NoStop}%
\bibitem [{\citenamefont {Hertz}(1894)}]{hertz-94}%
  \BibitemOpen
  \bibfield  {author} {\bibinfo {author} {\bibfnamefont {Heinrich}\
  \bibnamefont {Hertz}},\ }\href
  {https://archive.org/details/dieprinzipiende00hertgoog} {\emph {\bibinfo
  {title} {{P}rinzipien der {M}echanik}}}\ (\bibinfo  {publisher} {Johann
  Ambrosius Barth (Arthur Meiner)},\ \bibinfo {address} {Leipzig},\ \bibinfo
  {year} {1894})\ \bibinfo {note} {mit einem Vorewort von H. von
  Helmholtz}\BibitemShut {NoStop}%
\bibitem [{\citenamefont {Hertz}(1899)}]{hertz-94e}%
  \BibitemOpen
  \bibfield  {author} {\bibinfo {author} {\bibfnamefont {Heinrich}\
  \bibnamefont {Hertz}},\ }\href
  {https://archive.org/details/principlesofmech00hertuoft} {\emph {\bibinfo
  {title} {The principles of mechanics presented in a new form}}}\ (\bibinfo
  {publisher} {MacMillan and Co., Ltd.},\ \bibinfo {address} {London and New
  York},\ \bibinfo {year} {1899})\ \bibinfo {note} {with a foreword by H. von
  Helmholtz, translated by D. E. Jones and J. T. Walley}\BibitemShut {NoStop}%
\bibitem [{\citenamefont {Plato}(2000)}]{plato-republic}%
  \BibitemOpen
  \bibfield  {author} {\bibinfo {author} {\bibnamefont {Plato}},\ }\href
  {http://www.cambridge.org/academic/subjects/politics-international-relations/texts-political-thought/plato-republic}
  {\emph {\bibinfo {title} {The Republic}}},\ Cambridge Texts in the History of
  Political Thought\ (\bibinfo  {publisher} {Cambridge University Press},\
  \bibinfo {address} {Cambridge, UK},\ \bibinfo {year} {2000})\ \bibinfo {note}
  {translated by Tom Griffith, and edited by G. R. F. Ferrari}\BibitemShut
  {NoStop}%
\bibitem [{\citenamefont {Feyerabend}(1962)}]{Feyerabend-62}%
  \BibitemOpen
  \bibfield  {author} {\bibinfo {author} {\bibfnamefont {Paul~K.}\ \bibnamefont
  {Feyerabend}},\ }\bibfield  {title} {\enquote {\bibinfo {title} {Explanation,
  reduction, and empiricism},}\ }in\ \href
  {https://hdl.handle.net/11299/184633} {\emph {\bibinfo {booktitle}
  {Scientific explanation, space, and time}}},\ \bibinfo {series} {Minnesota
  Studies in the Philosophy of Science}, Vol.\ \bibinfo {volume} {III}\
  (\bibinfo  {publisher} {University of Minnesota Press},\ \bibinfo {address}
  {Minneapolis},\ \bibinfo {year} {1962})\ pp.\ \bibinfo {pages} {28--97},\
  \bibinfo {note} {reprinted in~\cite[pp.~44-96]{fey-philpapers1}}\BibitemShut
  {NoStop}%
\bibitem [{\citenamefont {Feyerabend}(1981)}]{fey-philpapers1}%
  \BibitemOpen
  \bibfield  {author} {\bibinfo {author} {\bibfnamefont {Paul~K.}\ \bibnamefont
  {Feyerabend}},\ }\href@noop {} {\emph {\bibinfo {title} {Realism, Rationalism
  and Scientific Method. {P}hilosophical Papers, Volume 1}}}\ (\bibinfo
  {publisher} {Cambridge University Press},\ \bibinfo {address} {Cambridge},\
  \bibinfo {year} {1981})\BibitemShut {NoStop}%
\bibitem [{\citenamefont {Kuhn}(1962,1970,1996,2012)}]{kuhn}%
  \BibitemOpen
  \bibfield  {author} {\bibinfo {author} {\bibfnamefont {Thomas~S.}\
  \bibnamefont {Kuhn}},\ }\href
  {https://press.uchicago.edu/ucp/books/book/chicago/S/bo13179781.html} {\emph
  {\bibinfo {title} {The Structure of Scientific Revolutions}}},\ \bibinfo
  {edition} {forth}\ ed.\ (\bibinfo  {publisher} {University of Chicago
  Press.},\ \bibinfo {address} {Chicago, IL, USA},\ \bibinfo {year}
  {1962,1970,1996,2012})\BibitemShut {NoStop}%
\bibitem [{\citenamefont {Oberheim}(2005)}]{Oberheim2005}%
  \BibitemOpen
  \bibfield  {author} {\bibinfo {author} {\bibfnamefont {Eric}\ \bibnamefont
  {Oberheim}},\ }\bibfield  {title} {\enquote {\bibinfo {title} {On the
  historical origins of the contemporary notion of incommensurability: {P}aul
  {F}eyerabend's assault on conceptual conservativism},}\ }\href {\doibase
  10.1016/j.shpsa.2005.04.003} {\bibfield  {journal} {\bibinfo  {journal}
  {Studies in History and Philosophy of Science Part A}\ }\textbf {\bibinfo
  {volume} {36}},\ \bibinfo {pages} {363--390} (\bibinfo {year}
  {2005})}\BibitemShut {NoStop}%
\bibitem [{\citenamefont {Oberheim}\ and\ \citenamefont
  {Hoyningen-Huene}(2018)}]{sep-incommensurability}%
  \BibitemOpen
  \bibfield  {author} {\bibinfo {author} {\bibfnamefont {Eric}\ \bibnamefont
  {Oberheim}}\ and\ \bibinfo {author} {\bibfnamefont {Paul}\ \bibnamefont
  {Hoyningen-Huene}},\ }\bibfield  {title} {\enquote {\bibinfo {title} {The
  incommensurability of scientific theories},}\ }in\ \href
  {https://plato.stanford.edu/archives/fall2018/entries/incommensurability/}
  {\emph {\bibinfo {booktitle} {The {S}tanford Encyclopedia of Philosophy}}},\
  \bibinfo {editor} {edited by\ \bibinfo {editor} {\bibfnamefont {Edward~N.}\
  \bibnamefont {Zalta}}}\ (\bibinfo  {publisher} {Metaphysics Research Lab,
  Stanford University},\ \bibinfo {year} {2018})\ \bibinfo {edition} {fall
  2018}\ ed.\BibitemShut {Stop}%
\bibitem [{\citenamefont {Pigliucci}(2010, 2018)}]{Pigliucci2018}%
  \BibitemOpen
  \bibfield  {author} {\bibinfo {author} {\bibfnamefont {Massimo}\ \bibnamefont
  {Pigliucci}},\ }\href {\doibase 10.7208/chicago/9780226496047.001.0001}
  {\emph {\bibinfo {title} {Nonsense on Stilts}}},\ \bibinfo {edition} {2nd}\
  ed.\ (\bibinfo  {publisher} {University of Chicago Press},\ \bibinfo
  {address} {Chicago, IL, USA and London, UK},\ \bibinfo {year} {2010,
  2018})\BibitemShut {NoStop}%
\bibitem [{\citenamefont {Lakatos}(1978, 2012{\natexlab{a}})}]{lakatosch}%
  \BibitemOpen
  \bibfield  {author} {\bibinfo {author} {\bibfnamefont {Imre}\ \bibnamefont
  {Lakatos}},\ }\href {\doibase 10.1017/CBO9780511621123} {\emph {\bibinfo
  {title} {The Methodology of Scientific Research Programmes. {P}hilosophical
  Papers {V}olume~1}}}\ (\bibinfo  {publisher} {Cambridge University Press},\
  \bibinfo {address} {Cambridge, England, UK},\ \bibinfo {year} {1978, 2012})\
  \bibinfo {note} {edited by John Worrall and Gregory Currie}\BibitemShut
  {NoStop}%
\bibitem [{\citenamefont {Lakatos}(1978, 2012{\natexlab{b}})}]{lakatos_1978}%
  \BibitemOpen
  \bibfield  {author} {\bibinfo {author} {\bibfnamefont {Imre}\ \bibnamefont
  {Lakatos}},\ }\bibfield  {title} {\enquote {\bibinfo {title} {Falsification
  and the methodology of scientific research programmes},}\ }in\ \href
  {\doibase 10.1017/CBO9780511621123.003} {\emph {\bibinfo {booktitle} {The
  Methodology of Scientific Research Programmes. {P}hilosophical Papers
  {V}olume~1}}}\ (\bibinfo  {publisher} {Cambridge University Press},\ \bibinfo
  {address} {Cambridge, England, UK},\ \bibinfo {year} {1978,
  2012})\BibitemShut {NoStop}%
\bibitem [{\citenamefont {Schopenhauer}(1819, 1912)}]{schopenhauer-dwawuv-VI}%
  \BibitemOpen
  \bibfield  {author} {\bibinfo {author} {\bibfnamefont {Arthur}\ \bibnamefont
  {Schopenhauer}},\ }\href {https://archive.org/details/dieweltalswilleu00scho}
  {\emph {\bibinfo {title} {{D}ie {W}elt als {W}ille und {V}orstellung.
  {E}rster {B}and}}},\ \bibinfo {edition} {3rd}\ ed.\ (\bibinfo  {publisher}
  {Georg M\"uller},\ \bibinfo {address} {M\"unchen},\ \bibinfo {year} {1819,
  1912})\BibitemShut {NoStop}%
\bibitem [{\citenamefont {Nietzsche}(1874, 1872, 1878,
  2009-{\natexlab{a}})}]{Nietzsche-ZarathustraI}%
  \BibitemOpen
  \bibfield  {author} {\bibinfo {author} {\bibfnamefont {Friedrich}\
  \bibnamefont {Nietzsche}},\ }\href
  {http://www.nietzschesource.org/\#eKGWB/Za-I} {\emph {\bibinfo {title}
  {{A}lso sprach {Z}arathustra. {E}in {B}uch f\"ur {A}lle und {K}einen}}}\
  (\bibinfo {year} {1874, 1872, 1878, 2009-})\ \bibinfo {note} {digital
  critical edition of the complete works and letters, based on the critical
  text by G. Colli and M. Montinari, Berlin/New York, de Gruyter 1967-, edited
  by Paolo D'Iorio}\BibitemShut {NoStop}%
\bibitem [{\citenamefont {Nietzsche}(1874, 1872, 1878,
  2009-{\natexlab{b}})}]{Nietzsche-EcceHomo}%
  \BibitemOpen
  \bibfield  {author} {\bibinfo {author} {\bibfnamefont {Friedrich}\
  \bibnamefont {Nietzsche}},\ }\href
  {http://www.nietzschesource.org/\#eKGWB/EH} {\emph {\bibinfo {title} {{E}cce
  homo. {W}ie man wird, was man ist}}}\ (\bibinfo {year} {1874, 1872, 1878,
  2009-})\ \bibinfo {note} {digital critical edition of the complete works and
  letters, based on the critical text by G. Colli and M. Montinari, Berlin/New
  York, de Gruyter 1967-, edited by Paolo D'Iorio}\BibitemShut {NoStop}%
\bibitem [{\citenamefont {Camus}(1942)}]{camus-mos}%
  \BibitemOpen
  \bibfield  {author} {\bibinfo {author} {\bibfnamefont {Albert}\ \bibnamefont
  {Camus}},\ }\href@noop {} {\emph {\bibinfo {title} {Le Mythe de {S}isyphe}}}\
  (\bibinfo  {publisher} {Gallimard},\ \bibinfo {address} {Paris},\ \bibinfo
  {year} {1942})\BibitemShut {NoStop}%
\bibitem [{\citenamefont {Popper}(1934)}]{popper}%
  \BibitemOpen
  \bibfield  {author} {\bibinfo {author} {\bibfnamefont {Karl~Raimund}\
  \bibnamefont {Popper}},\ }\href {\doibase 10.1007/978-3-7091-4177-9} {\emph
  {\bibinfo {title} {Logik der Forschung}}}\ (\bibinfo  {publisher}
  {Springer},\ \bibinfo {address} {Vienna},\ \bibinfo {year}
  {1934})\BibitemShut {NoStop}%
\bibitem [{\citenamefont {Popper}(1959, 1992, 2002)}]{popper-en}%
  \BibitemOpen
  \bibfield  {author} {\bibinfo {author} {\bibfnamefont {Karl~Raimund}\
  \bibnamefont {Popper}},\ }\href {\doibase 10.4324/9780203994627} {\emph
  {\bibinfo {title} {The Logic of Scientific Discovery}}},\ \bibinfo {edition}
  {2nd}\ ed.\ (\bibinfo  {publisher} {Hutchinson \& Co and Routledge},\
  \bibinfo {address} {New York and London},\ \bibinfo {year} {1959, 1992,
  2002})\BibitemShut {NoStop}%
\bibitem [{\citenamefont {Klein}\ and\ \citenamefont
  {Giglioni}(2020)}]{sep-francis-bacon}%
  \BibitemOpen
  \bibfield  {author} {\bibinfo {author} {\bibfnamefont {J\"urgen}\
  \bibnamefont {Klein}}\ and\ \bibinfo {author} {\bibfnamefont {Guido}\
  \bibnamefont {Giglioni}},\ }\bibfield  {title} {\enquote {\bibinfo {title}
  {{F}rancis {B}acon},}\ }in\ \href
  {https://plato.stanford.edu/archives/fall2020/entries/francis-bacon/} {\emph
  {\bibinfo {booktitle} {The {S}tanford Encyclopedia of Philosophy}}},\
  \bibinfo {editor} {edited by\ \bibinfo {editor} {\bibfnamefont {Edward~N.}\
  \bibnamefont {Zalta}}}\ (\bibinfo  {publisher} {Metaphysics Research Lab,
  Stanford University},\ \bibinfo {year} {2020})\ \bibinfo {edition} {fall
  2020}\ ed.\BibitemShut {Stop}%
\bibitem [{\citenamefont {Bridgman}(1927)}]{bridgman27}%
  \BibitemOpen
  \bibfield  {author} {\bibinfo {author} {\bibfnamefont {Percy~W.}\
  \bibnamefont {Bridgman}},\ }\href@noop {} {\emph {\bibinfo {title} {The Logic
  of Modern Physics}}}\ (\bibinfo {address} {New York},\ \bibinfo {year}
  {1927})\BibitemShut {NoStop}%
\bibitem [{\citenamefont {Bridgman}(1934)}]{bridgman}%
  \BibitemOpen
  \bibfield  {author} {\bibinfo {author} {\bibfnamefont {Percy~W.}\
  \bibnamefont {Bridgman}},\ }\bibfield  {title} {\enquote {\bibinfo {title} {A
  physicist's second reaction to {M}engenlehre},}\ }\href@noop {} {\bibfield
  {journal} {\bibinfo  {journal} {Scripta Mathematica}\ }\textbf {\bibinfo
  {volume} {2}},\ \bibinfo {pages} {101--117, 224--234} (\bibinfo {year}
  {1934})}\BibitemShut {NoStop}%
\bibitem [{\citenamefont {Bridgman}(1936)}]{bridgman36}%
  \BibitemOpen
  \bibfield  {author} {\bibinfo {author} {\bibfnamefont {Percy~W.}\
  \bibnamefont {Bridgman}},\ }\href@noop {} {\emph {\bibinfo {title} {The
  Nature of Physical Theory}}}\ (\bibinfo {address} {Princeton, NJ},\ \bibinfo
  {year} {1936})\BibitemShut {NoStop}%
\bibitem [{\citenamefont {Bridgman}(1950)}]{bridgman50}%
  \BibitemOpen
  \bibfield  {author} {\bibinfo {author} {\bibfnamefont {Percy~W.}\
  \bibnamefont {Bridgman}},\ }\href
  {http://www.archive.org/details/reflectionsofaph031333mbp} {\emph {\bibinfo
  {title} {Reflections of a Physicist}}}\ (\bibinfo  {publisher} {Philosophical
  Library},\ \bibinfo {address} {New York},\ \bibinfo {year}
  {1950})\BibitemShut {NoStop}%
\bibitem [{\citenamefont {Bridgman}(1952)}]{bridgman52}%
  \BibitemOpen
  \bibfield  {author} {\bibinfo {author} {\bibfnamefont {Percy~W.}\
  \bibnamefont {Bridgman}},\ }\href@noop {} {\emph {\bibinfo {title} {The
  Nature of Some of Our Physical Concepts}}}\ (\bibinfo  {publisher}
  {Philosophical Library},\ \bibinfo {address} {New York},\ \bibinfo {year}
  {1952})\BibitemShut {NoStop}%
\bibitem [{\citenamefont {Hume}(1739-40,2007)}]{Hume-Treatise}%
  \BibitemOpen
  \bibfield  {author} {\bibinfo {author} {\bibfnamefont {David}\ \bibnamefont
  {Hume}},\ }\href {\doibase 10.1093/actrade/9780199596331.book.1} {\emph
  {\bibinfo {title} {{A} Treatise of Human Nature: {V}olume 1: {T}exts}}},\
  \bibinfo {series} {The Clarendon Edition of the Works of David Hume},
  Vol.~\bibinfo {volume} {1}\ (\bibinfo  {publisher} {Clarendon Press and
  Oxford University Press},\ \bibinfo {year} {1739-40,2007})\ \bibinfo {note}
  {edited by David Fate Norton and Mary J. Norton}\BibitemShut {NoStop}%
\bibitem [{\citenamefont {Hume}(1748,2007)}]{Hume-Enquiry}%
  \BibitemOpen
  \bibfield  {author} {\bibinfo {author} {\bibfnamefont {David}\ \bibnamefont
  {Hume}},\ }\href {http://www.gutenberg.org/ebooks/9662} {\emph {\bibinfo
  {title} {An enquiry concerning human understanding}}},\ {O}xford world's
  classics\ (\bibinfo  {publisher} {Oxford University Press},\ \bibinfo {year}
  {1748,2007})\ \bibinfo {note} {edited by Peter Millican}\BibitemShut
  {NoStop}%
\bibitem [{\citenamefont {Svozil}(2020{\natexlab{b}})}]{svozil-2017-b}%
  \BibitemOpen
  \bibfield  {author} {\bibinfo {author} {\bibfnamefont {Karl}\ \bibnamefont
  {Svozil}},\ }\bibfield  {title} {\enquote {\bibinfo {title} {What is so
  special about quantum clicks?}}\ }\href {\doibase 10.3390/e22060602}
  {\bibfield  {journal} {\bibinfo  {journal} {Entropy}\ }\textbf {\bibinfo
  {volume} {22}},\ \bibinfo {pages} {602} (\bibinfo {year}
  {2020}{\natexlab{b}})},\ \Eprint {http://arxiv.org/abs/arXiv:1707.08915}
  {arXiv:1707.08915} \BibitemShut {NoStop}%
\bibitem [{\citenamefont {Aspect}\ \emph {et~al.}(1981)\citenamefont {Aspect},
  \citenamefont {Grangier},\ and\ \citenamefont {Roger}}]{aspect-81}%
  \BibitemOpen
  \bibfield  {author} {\bibinfo {author} {\bibfnamefont {Alain}\ \bibnamefont
  {Aspect}}, \bibinfo {author} {\bibfnamefont {Philippe}\ \bibnamefont
  {Grangier}}, \ and\ \bibinfo {author} {\bibfnamefont {G{\'{e}}rard}\
  \bibnamefont {Roger}},\ }\bibfield  {title} {\enquote {\bibinfo {title}
  {Experimental tests of realistic local theories via {B}ell's theorem},}\
  }\href {\doibase 10.1103/PhysRevLett.47.460} {\bibfield  {journal} {\bibinfo
  {journal} {Physical Review Letters}\ }\textbf {\bibinfo {volume} {47}},\
  \bibinfo {pages} {460--463} (\bibinfo {year} {1981})}\BibitemShut {NoStop}%
\bibitem [{\citenamefont {Aspect}\ \emph {et~al.}(1982)\citenamefont {Aspect},
  \citenamefont {Dalibard},\ and\ \citenamefont {Roger}}]{aspect-82b}%
  \BibitemOpen
  \bibfield  {author} {\bibinfo {author} {\bibfnamefont {Alain}\ \bibnamefont
  {Aspect}}, \bibinfo {author} {\bibfnamefont {Jean}\ \bibnamefont {Dalibard}},
  \ and\ \bibinfo {author} {\bibfnamefont {G{\'{e}}rard}\ \bibnamefont
  {Roger}},\ }\bibfield  {title} {\enquote {\bibinfo {title} {Experimental test
  of {B}ell's inequalities using time-varying analyzers},}\ }\href {\doibase
  10.1103/PhysRevLett.49.1804} {\bibfield  {journal} {\bibinfo  {journal}
  {Physical Review Letters}\ }\textbf {\bibinfo {volume} {49}},\ \bibinfo
  {pages} {1804--1807} (\bibinfo {year} {1982})}\BibitemShut {NoStop}%
\bibitem [{\citenamefont {Zeilinger}(1986)}]{zeilinger-86}%
  \BibitemOpen
  \bibfield  {author} {\bibinfo {author} {\bibfnamefont {Anton}\ \bibnamefont
  {Zeilinger}},\ }\bibfield  {title} {\enquote {\bibinfo {title} {Testing
  {B}ell's inequalities with periodic switching},}\ }\href {\doibase
  10.1016/0375-9601(86)90520-7} {\bibfield  {journal} {\bibinfo  {journal}
  {Physics Letters A}\ }\textbf {\bibinfo {volume} {118}},\ \bibinfo {pages}
  {1--2} (\bibinfo {year} {1986})}\BibitemShut {NoStop}%
\bibitem [{\citenamefont {Braunstein}\ and\ \citenamefont
  {Kimble}(1998)}]{Kimble-aposterioriQT}%
  \BibitemOpen
  \bibfield  {author} {\bibinfo {author} {\bibfnamefont {Samuel~L.}\
  \bibnamefont {Braunstein}}\ and\ \bibinfo {author} {\bibfnamefont {H.~J.}\
  \bibnamefont {Kimble}},\ }\bibfield  {title} {\enquote {\bibinfo {title} {A
  posteriori teleportation},}\ }\href {\doibase 10.1038/29674} {\bibfield
  {journal} {\bibinfo  {journal} {Nature}\ }\textbf {\bibinfo {volume} {394}},\
  \bibinfo {pages} {840--841} (\bibinfo {year} {1998})}\BibitemShut {NoStop}%
\bibitem [{\citenamefont {Bouwmeester}\ \emph {et~al.}(1998)\citenamefont
  {Bouwmeester}, \citenamefont {Pan}, \citenamefont {Daniell}, \citenamefont
  {Weinfurter}, \citenamefont {Zukowski},\ and\ \citenamefont
  {Zeilinger}}]{Bouwm-aposterioriQTReply}%
  \BibitemOpen
  \bibfield  {author} {\bibinfo {author} {\bibfnamefont {Dik}\ \bibnamefont
  {Bouwmeester}}, \bibinfo {author} {\bibfnamefont {Jian-Wei}\ \bibnamefont
  {Pan}}, \bibinfo {author} {\bibfnamefont {Matthew}\ \bibnamefont {Daniell}},
  \bibinfo {author} {\bibfnamefont {Harald}\ \bibnamefont {Weinfurter}},
  \bibinfo {author} {\bibfnamefont {Marek}\ \bibnamefont {Zukowski}}, \ and\
  \bibinfo {author} {\bibfnamefont {Anton}\ \bibnamefont {Zeilinger}},\
  }\bibfield  {title} {\enquote {\bibinfo {title} {Reply: {A} posteriori
  teleportation},}\ }\href {\doibase 10.1038/29678} {\bibfield  {journal}
  {\bibinfo  {journal} {Nature}\ }\textbf {\bibinfo {volume} {394}},\ \bibinfo
  {pages} {841} (\bibinfo {year} {1998})}\BibitemShut {NoStop}%
\bibitem [{\citenamefont {Bouwmeester}\ \emph {et~al.}(1997)\citenamefont
  {Bouwmeester}, \citenamefont {Pan}, \citenamefont {Mattle}, \citenamefont
  {Eibl}, \citenamefont {Weinfurter},\ and\ \citenamefont
  {Zeilinger}}]{Bouwmeester1997}%
  \BibitemOpen
  \bibfield  {author} {\bibinfo {author} {\bibfnamefont {Dik}\ \bibnamefont
  {Bouwmeester}}, \bibinfo {author} {\bibfnamefont {Jian-Wei}\ \bibnamefont
  {Pan}}, \bibinfo {author} {\bibfnamefont {Klaus}\ \bibnamefont {Mattle}},
  \bibinfo {author} {\bibfnamefont {Manfred}\ \bibnamefont {Eibl}}, \bibinfo
  {author} {\bibfnamefont {Harald}\ \bibnamefont {Weinfurter}}, \ and\ \bibinfo
  {author} {\bibfnamefont {Anton}\ \bibnamefont {Zeilinger}},\ }\bibfield
  {title} {\enquote {\bibinfo {title} {Experimental quantum teleportation},}\
  }\href {\doibase 10.1038/37539} {\bibfield  {journal} {\bibinfo  {journal}
  {Nature}\ }\textbf {\bibinfo {volume} {390}},\ \bibinfo {pages} {575--579}
  (\bibinfo {year} {1997})}\BibitemShut {NoStop}%
\bibitem [{\citenamefont {Bennett}\ \emph {et~al.}(1993)\citenamefont
  {Bennett}, \citenamefont {Brassard}, \citenamefont {Cr{\'{e}}peau},
  \citenamefont {Jozsa}, \citenamefont {Peres},\ and\ \citenamefont
  {Wootters}}]{BBCJPW}%
  \BibitemOpen
  \bibfield  {author} {\bibinfo {author} {\bibfnamefont {Charles~H.}\
  \bibnamefont {Bennett}}, \bibinfo {author} {\bibfnamefont {Gilles}\
  \bibnamefont {Brassard}}, \bibinfo {author} {\bibfnamefont {Claude}\
  \bibnamefont {Cr{\'{e}}peau}}, \bibinfo {author} {\bibfnamefont {Richard}\
  \bibnamefont {Jozsa}}, \bibinfo {author} {\bibfnamefont {Asher}\ \bibnamefont
  {Peres}}, \ and\ \bibinfo {author} {\bibfnamefont {William~K.}\ \bibnamefont
  {Wootters}},\ }\bibfield  {title} {\enquote {\bibinfo {title} {Teleporting an
  unknown quantum state via dual classical and {E}instein-{P}odolsky-{R}osen
  channels},}\ }\href {\doibase 10.1103/PhysRevLett.70.1895} {\bibfield
  {journal} {\bibinfo  {journal} {Physical Review Letters}\ }\textbf {\bibinfo
  {volume} {70}},\ \bibinfo {pages} {1895--1899} (\bibinfo {year}
  {1993})}\BibitemShut {NoStop}%
\bibitem [{\citenamefont {Svozil}(2021)}]{svozil-2020-hardy}%
  \BibitemOpen
  \bibfield  {author} {\bibinfo {author} {\bibfnamefont {Karl}\ \bibnamefont
  {Svozil}},\ }\bibfield  {title} {\enquote {\bibinfo {title} {Extensions of
  {H}ardy-type true-implies-false gadgets to classically obtain
  indistinguishability},}\ }\href {\doibase 10.1103/PhysRevA.103.022204}
  {\bibfield  {journal} {\bibinfo  {journal} {Physical Review A}\ }\textbf
  {\bibinfo {volume} {103}},\ \bibinfo {pages} {022204} (\bibinfo {year}
  {2021})},\ \Eprint {http://arxiv.org/abs/arXiv:2006.11396} {arXiv:2006.11396}
  \BibitemShut {NoStop}%
\bibitem [{\citenamefont {Gold}(1967)}]{go-67}%
  \BibitemOpen
  \bibfield  {author} {\bibinfo {author} {\bibfnamefont {Mark~E.}\ \bibnamefont
  {Gold}},\ }\bibfield  {title} {\enquote {\bibinfo {title} {Language
  identification in the limit},}\ }\href {\doibase
  10.1016/S0019-9958(67)91165-5} {\bibfield  {journal} {\bibinfo  {journal}
  {Information and Control}\ }\textbf {\bibinfo {volume} {10}},\ \bibinfo
  {pages} {447--474} (\bibinfo {year} {1967})}\BibitemShut {NoStop}%
\bibitem [{\citenamefont {Blum}\ and\ \citenamefont {Blum}(1975)}]{blum75blum}%
  \BibitemOpen
  \bibfield  {author} {\bibinfo {author} {\bibfnamefont {Lenore}\ \bibnamefont
  {Blum}}\ and\ \bibinfo {author} {\bibfnamefont {Manuel}\ \bibnamefont
  {Blum}},\ }\bibfield  {title} {\enquote {\bibinfo {title} {Toward a
  mathematical theory of inductive inference},}\ }\href {\doibase
  10.1016/S0019-9958(75)90261-2} {\bibfield  {journal} {\bibinfo  {journal}
  {Information and Control}\ }\textbf {\bibinfo {volume} {28}},\ \bibinfo
  {pages} {125--155} (\bibinfo {year} {1975})}\BibitemShut {NoStop}%
\bibitem [{\citenamefont {Angluin}\ and\ \citenamefont
  {Smith}(1983)}]{angluin:83}%
  \BibitemOpen
  \bibfield  {author} {\bibinfo {author} {\bibfnamefont {Dana}\ \bibnamefont
  {Angluin}}\ and\ \bibinfo {author} {\bibfnamefont {Carl~H.}\ \bibnamefont
  {Smith}},\ }\bibfield  {title} {\enquote {\bibinfo {title} {Inductive
  inference: Theory and methods},}\ }\href {\doibase 10.1145/356914.356918}
  {\bibfield  {journal} {\bibinfo  {journal} {ACM Computing Surveys}\ }\textbf
  {\bibinfo {volume} {15}},\ \bibinfo {pages} {237--269} (\bibinfo {year}
  {1983})}\BibitemShut {NoStop}%
\bibitem [{\citenamefont {Adleman}\ and\ \citenamefont {Blum}(1991)}]{ad-91}%
  \BibitemOpen
  \bibfield  {author} {\bibinfo {author} {\bibfnamefont {Leonard~M.}\
  \bibnamefont {Adleman}}\ and\ \bibinfo {author} {\bibfnamefont {Manuel}\
  \bibnamefont {Blum}},\ }\bibfield  {title} {\enquote {\bibinfo {title}
  {Inductive inference and unsolvability},}\ }\href {\doibase 10.2307/2275058}
  {\bibfield  {journal} {\bibinfo  {journal} {The Journal of Symbolic Logic}\
  }\textbf {\bibinfo {volume} {56}},\ \bibinfo {pages} {891--900} (\bibinfo
  {year} {1991})}\BibitemShut {NoStop}%
\bibitem [{\citenamefont {Li}\ and\ \citenamefont
  {Vit{\'{a}}nyi}(1992)}]{li:92}%
  \BibitemOpen
  \bibfield  {author} {\bibinfo {author} {\bibfnamefont {M.}~\bibnamefont
  {Li}}\ and\ \bibinfo {author} {\bibfnamefont {P.~M.~B.}\ \bibnamefont
  {Vit{\'{a}}nyi}},\ }\bibfield  {title} {\enquote {\bibinfo {title} {Inductive
  reasoning and {K}olmogorov complexity},}\ }\href {\doibase
  10.1016/0022-0000(92)90026-F} {\bibfield  {journal} {\bibinfo  {journal}
  {Journal of Computer and System Science}\ }\textbf {\bibinfo {volume} {44}},\
  \bibinfo {pages} {343--384} (\bibinfo {year} {1992})}\BibitemShut {NoStop}%
\bibitem [{\citenamefont {Gandy}(1982)}]{gandy1}%
  \BibitemOpen
  \bibfield  {author} {\bibinfo {author} {\bibfnamefont {Robin~O.}\
  \bibnamefont {Gandy}},\ }\bibfield  {title} {\enquote {\bibinfo {title}
  {Limitations to mathematical knowledge},}\ }in\ \href
  {https://www.elsevier.com/books/logic-colloquium-80/van-dalen/978-0-444-86465-9}
  {\emph {\bibinfo {booktitle} {Logic colloquium '80}}},\ \bibinfo {editor}
  {edited by\ \bibinfo {editor} {\bibfnamefont {D.}~\bibnamefont {van Dalen}},
  \bibinfo {editor} {\bibfnamefont {D.}~\bibnamefont {Lascar}}, \ and\ \bibinfo
  {editor} {\bibfnamefont {J.}~\bibnamefont {Smiley}}}\ (\bibinfo  {publisher}
  {North Holland},\ \bibinfo {address} {Amsterdam},\ \bibinfo {year} {1982})\
  pp.\ \bibinfo {pages} {129--146},\ \bibinfo {note} {papers intended for the
  {E}uropean Summer Meeting of the Association for Symbolic Logic}\BibitemShut
  {NoStop}%
\bibitem [{\citenamefont {Svozil}(1993)}]{svozil-93}%
  \BibitemOpen
  \bibfield  {author} {\bibinfo {author} {\bibfnamefont {Karl}\ \bibnamefont
  {Svozil}},\ }\href {\doibase 10.1142/1524} {\emph {\bibinfo {title}
  {Randomness \& Undecidability in Physics}}}\ (\bibinfo  {publisher} {World
  Scientific},\ \bibinfo {address} {Singapore},\ \bibinfo {year}
  {1993})\BibitemShut {NoStop}%
\bibitem [{\citenamefont {Svozil}(1996)}]{svozil-unev}%
  \BibitemOpen
  \bibfield  {author} {\bibinfo {author} {\bibfnamefont {Karl}\ \bibnamefont
  {Svozil}},\ }\bibfield  {title} {\enquote {\bibinfo {title} {Undecidability
  everywhere?}}\ }in\ \href@noop {} {\emph {\bibinfo {booktitle} {Boundaries
  and Barriers. On the Limits to Scientific Knowledge}}},\ \bibinfo {editor}
  {edited by\ \bibinfo {editor} {\bibfnamefont {J.~L.}\ \bibnamefont {Casti}}\
  and\ \bibinfo {editor} {\bibfnamefont {A.}~\bibnamefont {Karlquist}}}\
  (\bibinfo  {publisher} {Addison-Wesley},\ \bibinfo {address} {Reading, MA},\
  \bibinfo {year} {1996})\ pp.\ \bibinfo {pages} {215--237}\BibitemShut
  {NoStop}%
\bibitem [{\citenamefont {Svozil}(2011)}]{svozil-07-physical_unknowables}%
  \BibitemOpen
  \bibfield  {author} {\bibinfo {author} {\bibfnamefont {Karl}\ \bibnamefont
  {Svozil}},\ }\bibfield  {title} {\enquote {\bibinfo {title} {Physical
  unknowables},}\ }in\ \href {\doibase 10.1017/CBO9780511974236.013} {\emph
  {\bibinfo {booktitle} {{K}urt {G}{\"o}del and the Foundations of
  Mathematics}}},\ \bibinfo {editor} {edited by\ \bibinfo {editor}
  {\bibfnamefont {Matthias}\ \bibnamefont {Baaz}}, \bibinfo {editor}
  {\bibfnamefont {Christos~H.}\ \bibnamefont {Papadimitriou}}, \bibinfo
  {editor} {\bibfnamefont {Hilary~W.}\ \bibnamefont {Putnam}}, \bibinfo
  {editor} {\bibfnamefont {Dana~S.}\ \bibnamefont {Scott}}, \ and\ \bibinfo
  {editor} {\bibfnamefont {Charles~L.}\ \bibnamefont {{Harper, Jr}}}}\
  (\bibinfo  {publisher} {Cambridge University Press},\ \bibinfo {address}
  {Cambridge, UK},\ \bibinfo {year} {2011})\ pp.\ \bibinfo {pages} {213--251},\
  \Eprint {http://arxiv.org/abs/arXiv:physics/0701163} {arXiv:physics/0701163}
  \BibitemShut {NoStop}%
\bibitem [{\citenamefont {Norton}(2003)}]{Norton-2003-cafs}%
  \BibitemOpen
  \bibfield  {author} {\bibinfo {author} {\bibfnamefont {John~D.}\ \bibnamefont
  {Norton}},\ }\bibfield  {title} {\enquote {\bibinfo {title} {Causation as
  folk science},}\ }\href {http://hdl.handle.net/2027/spo.3521354.0003.004}
  {\bibfield  {journal} {\bibinfo  {journal} {Philosophers' Imprint}\ }\textbf
  {\bibinfo {volume} {3}},\ \bibinfo {pages} {1--22} (\bibinfo {year}
  {2003})}\BibitemShut {NoStop}%
\bibitem [{\citenamefont {Sloane}(2019)}]{Sloane_oeis.org/A033307}%
  \BibitemOpen
  \bibfield  {author} {\bibinfo {author} {\bibfnamefont {Neil James~Alexander}\
  \bibnamefont {Sloane}},\ }\href {https://oeis.org/A033307} {\enquote
  {\bibinfo {title} {{A033307} {D}ecimal expansion of {C}hampernowne constant
  (or {M}ahler's number), formed by concatenating the positive integers.}}\ }
  (\bibinfo {year} {2019}),\ \bibinfo {note} {accessed on Sep 8th,
  2019}\BibitemShut {NoStop}%
\bibitem [{\citenamefont {Dow}(1939)}]{dow_aristotlekleroteria_1939}%
  \BibitemOpen
  \bibfield  {author} {\bibinfo {author} {\bibfnamefont {Sterling}\
  \bibnamefont {Dow}},\ }\bibfield  {title} {\enquote {\bibinfo {title}
  {{A}ristotle, the {K}leroteria, and the courts},}\ }\href {\doibase
  10.2307/310590} {\bibfield  {journal} {\bibinfo  {journal} {Harvard Studies
  in Classical Philology}\ }\textbf {\bibinfo {volume} {50}},\ \bibinfo {pages}
  {1--34} (\bibinfo {year} {1939})}\BibitemShut {NoStop}%
\bibitem [{\citenamefont {Toffoli}(1978)}]{toffoli:79}%
  \BibitemOpen
  \bibfield  {author} {\bibinfo {author} {\bibfnamefont {Tommaso}\ \bibnamefont
  {Toffoli}},\ }\bibfield  {title} {\enquote {\bibinfo {title} {The role of the
  observer in uniform systems},}\ }in\ \href {\doibase
  10.1007/978-1-4757-0555-3\_29} {\emph {\bibinfo {booktitle} {Applied General
  Systems Research: Recent Developments and Trends}}},\ \bibinfo {editor}
  {edited by\ \bibinfo {editor} {\bibfnamefont {George~J.}\ \bibnamefont
  {Klir}}}\ (\bibinfo  {publisher} {Plenum Press, Springer US},\ \bibinfo
  {address} {New York, London, and Boston, MA},\ \bibinfo {year} {1978})\ pp.\
  \bibinfo {pages} {395--400}\BibitemShut {NoStop}%
\bibitem [{\citenamefont {Fredkin}(1990)}]{fredkin}%
  \BibitemOpen
  \bibfield  {author} {\bibinfo {author} {\bibfnamefont {Edward}\ \bibnamefont
  {Fredkin}},\ }\bibfield  {title} {\enquote {\bibinfo {title} {Digital
  mechanics. an informational process based on reversible universal cellular
  automata},}\ }\href {\doibase 10.1016/0167-2789(90)90186-S} {\bibfield
  {journal} {\bibinfo  {journal} {Physica}\ }\textbf {\bibinfo {volume}
  {D45}},\ \bibinfo {pages} {254--270} (\bibinfo {year} {1990})}\BibitemShut
  {NoStop}%
\bibitem [{\citenamefont {Svozil}(1995)}]{svozil-nat-acad}%
  \BibitemOpen
  \bibfield  {author} {\bibinfo {author} {\bibfnamefont {Karl}\ \bibnamefont
  {Svozil}},\ }\bibfield  {title} {\enquote {\bibinfo {title} {A constructivist
  manifesto for the physical sciences -- {C}onstructive re-interpretation of
  physical undecidability},}\ }in\ \href {\doibase
  10.1007/978-94-017-3327-4\_6} {\emph {\bibinfo {booktitle} {The Foundational
  Debate: Complexity and Constructivity in Mathematics and Physics}}},\
  \bibinfo {series and number} {Vienna Circle Institute Yearbook [1995] 3},\
  \bibinfo {editor} {edited by\ \bibinfo {editor} {\bibfnamefont
  {Werner~DePauli}\ \bibnamefont {Schimanovich}}, \bibinfo {editor}
  {\bibfnamefont {Eckehart}\ \bibnamefont {K{\"{o}}hler}}, \ and\ \bibinfo
  {editor} {\bibfnamefont {Friedrich}\ \bibnamefont {Stadler}}}\ (\bibinfo
  {publisher} {Kluwer Academic Publishers, Springer Netherlands},\ \bibinfo
  {address} {Dordrecht, Boston, London},\ \bibinfo {year} {1995})\ pp.\
  \bibinfo {pages} {65--88}\BibitemShut {NoStop}%
\bibitem [{\citenamefont {Bostrom}(2003)}]{Bostrom-sim}%
  \BibitemOpen
  \bibfield  {author} {\bibinfo {author} {\bibfnamefont {Nick}\ \bibnamefont
  {Bostrom}},\ }\bibfield  {title} {\enquote {\bibinfo {title} {Are we living
  in a computer simulation?}}\ }\href {\doibase 10.1111/1467-9213.00309}
  {\bibfield  {journal} {\bibinfo  {journal} {The Philosophical Quarterly}\
  }\textbf {\bibinfo {volume} {53}},\ \bibinfo {pages} {243--255} (\bibinfo
  {year} {2003})},\ \Eprint
  {http://arxiv.org/abs/https://www.simulation-argument.com/simulation.pdf}
  {https://www.simulation-argument.com/simulation.pdf} \BibitemShut {NoStop}%
\bibitem [{\citenamefont {Maxwell}(1882, 1999)}]{Campbell-1882}%
  \BibitemOpen
  \bibfield  {author} {\bibinfo {author} {\bibfnamefont {James~Clerk}\
  \bibnamefont {Maxwell}},\ }\bibfield  {title} {\enquote {\bibinfo {title}
  {Does the progress of physical science tend to give any advantage to the
  opinion of necessity (or determinism) over that of the contingency of events
  and the freedom of the will?}}\ }in\ \href
  {https://archive.org/details/lifeofjamesclerk00camprich} {\emph {\bibinfo
  {booktitle} {The life of {J}ames {C}lerk {M}axwell. {W}ith a selection from
  his correspondence and occasional writings and a sketch of his contributions
  to science}}},\ \bibinfo {editor} {edited by\ \bibinfo {editor}
  {\bibfnamefont {Lewis}\ \bibnamefont {Campbell}}\ and\ \bibinfo {editor}
  {\bibfnamefont {William}\ \bibnamefont {Garnett}}}\ (\bibinfo  {publisher}
  {MacMillan},\ \bibinfo {address} {London},\ \bibinfo {year} {1882, 1999})\
  \bibinfo {edition} {2nd}\ ed.\BibitemShut {Stop}%
\bibitem [{\citenamefont {Norton}(2008)}]{Norton-dome-2008}%
  \BibitemOpen
  \bibfield  {author} {\bibinfo {author} {\bibfnamefont {John~D.}\ \bibnamefont
  {Norton}},\ }\bibfield  {title} {\enquote {\bibinfo {title} {The dome: An
  unexpectedly simple failure of determinism},}\ }\href {\doibase
  10.1086/594524} {\bibfield  {journal} {\bibinfo  {journal} {Philosophy of
  Science}\ }\textbf {\bibinfo {volume} {75}},\ \bibinfo {pages} {786--798}
  (\bibinfo {year} {2008})},\ \Eprint
  {http://arxiv.org/abs/http://philsci-archive.pitt.edu/2943/}
  {http://philsci-archive.pitt.edu/2943/} \BibitemShut {NoStop}%
\bibitem [{\citenamefont {van Strien}(2014)}]{vanStrien2014}%
  \BibitemOpen
  \bibfield  {author} {\bibinfo {author} {\bibfnamefont {Marij}\ \bibnamefont
  {van Strien}},\ }\bibfield  {title} {\enquote {\bibinfo {title} {The {N}orton
  dome and the nineteenth century foundations of determinism},}\ }\href
  {\doibase 10.1007/s10838-014-9241-0} {\bibfield  {journal} {\bibinfo
  {journal} {Journal for General Philosophy of Science}\ }\textbf {\bibinfo
  {volume} {45}},\ \bibinfo {pages} {167--185} (\bibinfo {year}
  {2014})}\BibitemShut {NoStop}%
\bibitem [{\citenamefont {Smoluchowski}(1912)}]{Smoluchovski-1912}%
  \BibitemOpen
  \bibfield  {author} {\bibinfo {author} {\bibfnamefont {Marian}\ \bibnamefont
  {Smoluchowski}},\ }\bibfield  {title} {\enquote {\bibinfo {title}
  {{E}xperimentell nachweisbare, der {\"u}blichen {T}hermodynamik
  widersprechende {M}olekularph{\"a}nomene},}\ }\href
  {http://matwbn.icm.edu.pl/ksiazki/pms/pms2/pms2122.pdf} {\bibfield  {journal}
  {\bibinfo  {journal} {Physikalische Zeitschrift}\ }\textbf {\bibinfo {volume}
  {13}},\ \bibinfo {pages} {1069--1080} (\bibinfo {year} {1912})}\BibitemShut
  {NoStop}%
\bibitem [{\citenamefont {Uffink}(2011)}]{Uffink2011-UFFSPS}%
  \BibitemOpen
  \bibfield  {author} {\bibinfo {author} {\bibfnamefont {Jos}\ \bibnamefont
  {Uffink}},\ }\bibfield  {title} {\enquote {\bibinfo {title} {Subjective
  probability and statistical physics},}\ }in\ \href {\doibase
  10.1093/acprof:oso/9780199577439.003.0002} {\emph {\bibinfo {booktitle}
  {Probabilities in Physics}}},\ \bibinfo {editor} {edited by\ \bibinfo
  {editor} {\bibfnamefont {Claus}\ \bibnamefont {Beisbart}}\ and\ \bibinfo
  {editor} {\bibfnamefont {Stephan}\ \bibnamefont {Hartmann}}}\ (\bibinfo
  {publisher} {Oxford University Press},\ \bibinfo {address} {Oxford, UK},\
  \bibinfo {year} {2011})\ pp.\ \bibinfo {pages} {25--49}\BibitemShut {NoStop}%
\bibitem [{\citenamefont {Watzlawick}\ \emph {et~al.}(1967)\citenamefont
  {Watzlawick}, \citenamefont {Beavin},\ and\ \citenamefont
  {Jackson}}]{Watzlawick-1967}%
  \BibitemOpen
  \bibfield  {author} {\bibinfo {author} {\bibfnamefont {Paul}\ \bibnamefont
  {Watzlawick}}, \bibinfo {author} {\bibfnamefont {Janet~Helmick}\ \bibnamefont
  {Beavin}}, \ and\ \bibinfo {author} {\bibfnamefont {Don~D.}\ \bibnamefont
  {Jackson}},\ }\href@noop {} {\emph {\bibinfo {title} {Pragmatics of Human
  Communication: A Study of Interactional Patterns, Pathologies, and
  Paradoxes}}}\ (\bibinfo  {publisher} {W. W. Norton \& Company},\ \bibinfo
  {address} {New York},\ \bibinfo {year} {1967})\BibitemShut {NoStop}%
\bibitem [{\citenamefont {Hamilton}(1963)}]{Hamilton-1963}%
  \BibitemOpen
  \bibfield  {author} {\bibinfo {author} {\bibfnamefont {William~Donald}\
  \bibnamefont {Hamilton}},\ }\bibfield  {title} {\enquote {\bibinfo {title}
  {The evolution of altruistic behavior},}\ }\href {\doibase 10.1086/497114}
  {\bibfield  {journal} {\bibinfo  {journal} {The American Naturalist}\
  }\textbf {\bibinfo {volume} {97}},\ \bibinfo {pages} {354--356} (\bibinfo
  {year} {1963})}\BibitemShut {NoStop}%
\bibitem [{\citenamefont {Born}(1926)}]{born-26-1}%
  \BibitemOpen
  \bibfield  {author} {\bibinfo {author} {\bibfnamefont {Max}\ \bibnamefont
  {Born}},\ }\bibfield  {title} {\enquote {\bibinfo {title} {Zur
  {Q}uantenmechanik der {S}to{\ss}vorg{\"{a}}nge},}\ }\href {\doibase
  10.1007/BF01397477} {\bibfield  {journal} {\bibinfo  {journal} {Zeitschrift
  f\"{u}r Physik}\ }\textbf {\bibinfo {volume} {37}},\ \bibinfo {pages}
  {863--867} (\bibinfo {year} {1926})}\BibitemShut {NoStop}%
\bibitem [{\citenamefont {Hiebert}(2000)}]{Hiebert2000}%
  \BibitemOpen
  \bibfield  {author} {\bibinfo {author} {\bibfnamefont {Erwin~N.}\
  \bibnamefont {Hiebert}},\ }\bibfield  {title} {\enquote {\bibinfo {title}
  {Common frontiers of the exact sciences and the humanities},}\ }\href
  {\doibase 10.1007/s000160050034} {\bibfield  {journal} {\bibinfo  {journal}
  {Physics in Perspective}\ }\textbf {\bibinfo {volume} {2}},\ \bibinfo {pages}
  {6--29} (\bibinfo {year} {2000})}\BibitemShut {NoStop}%
\bibitem [{\citenamefont {St\"oltzner}(1999)}]{Stoeltzner-1999}%
  \BibitemOpen
  \bibfield  {author} {\bibinfo {author} {\bibfnamefont {Michael}\ \bibnamefont
  {St\"oltzner}},\ }\bibfield  {title} {\enquote {\bibinfo {title} {{V}ienna
  indeterminism: {M}ach, {B}oltzmann, {E}xner},}\ }\href {\doibase
  10.1023/a:1005243320885} {\bibfield  {journal} {\bibinfo  {journal}
  {Synthese}\ }\textbf {\bibinfo {volume} {119}},\ \bibinfo {pages} {85--111}
  (\bibinfo {year} {1999})}\BibitemShut {NoStop}%
\bibitem [{\citenamefont {Schweidler}(1906)}]{schweidler-1905}%
  \BibitemOpen
  \bibfield  {author} {\bibinfo {author} {\bibfnamefont {Egon~von}\
  \bibnamefont {Schweidler}},\ }\enquote {\bibinfo {title} {{\"U}ber
  {S}chwankungen der radioaktiven {U}mwandlung},}\ in\ \href
  {https://archive.org/details/premiercongrsin03unkngoog} {\emph {\bibinfo
  {booktitle} {Premier Congres international pour L'etude de la radiologie et
  de {I}'ionisation tenu a {L}iege du 12 au 14 {S}eptembre 1905}}}\ (\bibinfo
  {publisher} {H. Dunod \& E. Pinat},\ \bibinfo {address} {Paris},\ \bibinfo
  {year} {1906})\ pp.\ \bibinfo {pages} {German part, 1--3}\BibitemShut
  {NoStop}%
\bibitem [{\citenamefont {Exner}(1909, 2016)}]{Exner-1908}%
  \BibitemOpen
  \bibfield  {author} {\bibinfo {author} {\bibfnamefont {Franz~Serafin}\
  \bibnamefont {Exner}},\ }\href {http://phaidra.univie.ac.at/o:451413} {\emph
  {\bibinfo {title} {{\"U}ber {G}esetze in {N}aturwissenschaft und
  {H}umanistik: {I}naugurationsrede gehalten am 15. {O}ktober 1908}}}\
  (\bibinfo  {publisher} {H\"older, Ebooks on Demand Universit\"atsbibliothek
  Wien},\ \bibinfo {address} {Vienna},\ \bibinfo {year} {1909, 2016})\ \bibinfo
  {note} {handle https://hdl.handle.net/11353/10.451413, o:451413, Uploaded:
  30.08.2016}\BibitemShut {NoStop}%
\bibitem [{\citenamefont {Armstrong}(1983)}]{armstrong_1983}%
  \BibitemOpen
  \bibfield  {author} {\bibinfo {author} {\bibfnamefont {D.~M.}\ \bibnamefont
  {Armstrong}},\ }\href {\doibase 10.1017/CBO9781139171700} {\emph {\bibinfo
  {title} {What is a Law of Nature?}}},\ Cambridge Studies in Philosophy\
  (\bibinfo  {publisher} {Cambridge University Press},\ \bibinfo {address}
  {Cambridge},\ \bibinfo {year} {1983})\BibitemShut {NoStop}%
\bibitem [{\citenamefont {van Fraassen}(1989, 2003)}]{vanFraassen1989-VANLAS}%
  \BibitemOpen
  \bibfield  {author} {\bibinfo {author} {\bibfnamefont {Bas~C.}\ \bibnamefont
  {van Fraassen}},\ }\href {\doibase 10.1093/0198248601.001.0001} {\emph
  {\bibinfo {title} {Laws and Symmetry}}}\ (\bibinfo  {publisher} {Oxford
  University Press},\ \bibinfo {year} {1989, 2003})\BibitemShut {NoStop}%
\bibitem [{\citenamefont {Calude}\ and\ \citenamefont
  {Meyerstein}(1999)}]{calude-meyerstein}%
  \BibitemOpen
  \bibfield  {author} {\bibinfo {author} {\bibfnamefont {Cristian}\
  \bibnamefont {Calude}}\ and\ \bibinfo {author} {\bibfnamefont {F.~Walter}\
  \bibnamefont {Meyerstein}},\ }\bibfield  {title} {\enquote {\bibinfo {title}
  {Is the universe lawful?}}\ }\href {\doibase 10.1016/S0960-0779(98)00145-3}
  {\bibfield  {journal} {\bibinfo  {journal} {Chaos, Solitons \& Fractals}\
  }\textbf {\bibinfo {volume} {10}},\ \bibinfo {pages} {1075--1084} (\bibinfo
  {year} {1999})}\BibitemShut {NoStop}%
\bibitem [{\citenamefont {Rosen}(2010)}]{lawlses_rosen2010}%
  \BibitemOpen
  \bibfield  {author} {\bibinfo {author} {\bibfnamefont {Joe}\ \bibnamefont
  {Rosen}},\ }\href@noop {} {\emph {\bibinfo {title} {Lawless Universe: Science
  and the Hunt for Reality}}}\ (\bibinfo  {publisher} {The John Hopkins
  University Press},\ \bibinfo {address} {Balrtimreo, Maryland},\ \bibinfo
  {year} {2010})\BibitemShut {NoStop}%
\bibitem [{\citenamefont {Calude}\ \emph {et~al.}(2013)\citenamefont {Calude},
  \citenamefont {Meyerstein},\ and\ \citenamefont
  {Salomaa}}]{calude2013theeinai}%
  \BibitemOpen
  \bibfield  {author} {\bibinfo {author} {\bibfnamefont {Cristian~S.}\
  \bibnamefont {Calude}}, \bibinfo {author} {\bibfnamefont {Walter~F.}\
  \bibnamefont {Meyerstein}}, \ and\ \bibinfo {author} {\bibfnamefont {Arto}\
  \bibnamefont {Salomaa}},\ }\bibfield  {title} {\enquote {\bibinfo {title}
  {The universe is lawless or ``panton chrematon metron anthropon einai''},}\
  }in\ \href {\doibase 10.1142/9789814374309\_0026} {\emph {\bibinfo
  {booktitle} {Computable Universe: {U}nderstanding and Exploring Nature as
  Computation}}},\ \bibinfo {editor} {edited by\ \bibinfo {editor}
  {\bibfnamefont {Hector}\ \bibnamefont {Zenil}}}\ (\bibinfo  {publisher}
  {World Scientific},\ \bibinfo {address} {Singapore},\ \bibinfo {year}
  {2013})\ pp.\ \bibinfo {pages} {525--537}\BibitemShut {NoStop}%
\bibitem [{\citenamefont {Yanofsky}(2017)}]{chaos_multiverse2017}%
  \BibitemOpen
  \bibfield  {author} {\bibinfo {author} {\bibfnamefont {Noson}\ \bibnamefont
  {Yanofsky}},\ }\bibfield  {title} {\enquote {\bibinfo {title} {Chaos makes
  the multiverse unnecessary},}\ }\href
  {http://nautil.us/issue/49/the-absurd/chaos-makes-the-multiverse-unnecessary}
  {\bibfield  {journal} {\bibinfo  {journal} {Nautilus}\ ,\ \bibinfo {pages}
  {1--16}} (\bibinfo {year} {2017})}\BibitemShut {NoStop}%
\bibitem [{\citenamefont {Mueller}(2017)}]{Mueller-2017}%
  \BibitemOpen
  \bibfield  {author} {\bibinfo {author} {\bibfnamefont {Markus~P.}\
  \bibnamefont {Mueller}},\ }\href {https://arxiv.org/abs/1712.01816} {\enquote
  {\bibinfo {title} {Could the physical world be emergent instead of
  fundamental, and why should we ask? (short version)},}\ } (\bibinfo {year}
  {2017}),\ \Eprint {http://arxiv.org/abs/arXiv:1712.01816} {arXiv:1712.01816}
  \BibitemShut {NoStop}%
\bibitem [{\citenamefont {Cabello}(2019)}]{Cabello-2018-BornRule}%
  \BibitemOpen
  \bibfield  {author} {\bibinfo {author} {\bibfnamefont {Ad\'an}\ \bibnamefont
  {Cabello}},\ }\href {https://arxiv.org/abs/1801.06347} {\enquote {\bibinfo
  {title} {Physical origin of quantum nonlocality and contextuality},}\ }
  (\bibinfo {year} {2019}),\ \Eprint {http://arxiv.org/abs/arXiv:1801.06347}
  {arXiv:1801.06347} \BibitemShut {NoStop}%
\bibitem [{\citenamefont {Calude}\ and\ \citenamefont
  {Svozil}(2019)}]{svozil-2018-was}%
  \BibitemOpen
  \bibfield  {author} {\bibinfo {author} {\bibfnamefont {Cristian~S.}\
  \bibnamefont {Calude}}\ and\ \bibinfo {author} {\bibfnamefont {Karl}\
  \bibnamefont {Svozil}},\ }\bibfield  {title} {\enquote {\bibinfo {title}
  {Spurious, emergent laws in number worlds},}\ }\href {\doibase
  10.3390/philosophies4020017} {\bibfield  {journal} {\bibinfo  {journal}
  {Philosophies}\ }\textbf {\bibinfo {volume} {4}},\ \bibinfo {pages} {17}
  (\bibinfo {year} {2019})},\ \Eprint {http://arxiv.org/abs/arXiv:1812.04416}
  {arXiv:1812.04416} \BibitemShut {NoStop}%
\bibitem [{\citenamefont {Clark}(2017)}]{Clark-2017-GodAsCurler}%
  \BibitemOpen
  \bibfield  {author} {\bibinfo {author} {\bibfnamefont {Kelly~James}\
  \bibnamefont {Clark}},\ }\href@noop {} {\enquote {\bibinfo {title} {Is {G}od
  a bowler or a curler?}}\ } (\bibinfo {year} {2017}),\ \bibinfo {note}
  {presentation at the Randomness and Providence Workshop, May 9,
  2017}\BibitemShut {NoStop}%
\bibitem [{\citenamefont {Voltaire}(1764)}]{voltaire-dict}%
  \BibitemOpen
  \bibfield  {author} {\bibinfo {author} {\bibnamefont {Voltaire}},\ }\href
  {https://ebooks.adelaide.edu.au/v/voltaire/dictionary/} {\emph {\bibinfo
  {title} {A Philosophical Dictionary}}}\ (\bibinfo {year} {1764})\ \bibinfo
  {note} {derived from The Works of Voltaire, A Contemporary Version, (New
  York: E.R. DuMont, 1901)}\BibitemShut {NoStop}%
\bibitem [{\citenamefont {Galouye}(1964)}]{simula}%
  \BibitemOpen
  \bibfield  {author} {\bibinfo {author} {\bibfnamefont {Daniel~F.}\
  \bibnamefont {Galouye}},\ }\href@noop {} {\emph {\bibinfo {title} {Simulacron
  3}}}\ (\bibinfo  {publisher} {Bantam Books},\ \bibinfo {address} {New York},\
  \bibinfo {year} {1964})\BibitemShut {NoStop}%
\bibitem [{\citenamefont {Egan}(1994)}]{permutationcity}%
  \BibitemOpen
  \bibfield  {author} {\bibinfo {author} {\bibfnamefont {Greg}\ \bibnamefont
  {Egan}},\ }\href {http://www.gregegan.net/PERMUTATION/Permutation.html}
  {\emph {\bibinfo {title} {Permutation City}}}\ (\bibinfo {year} {1994})\
  \bibinfo {note} {accessed on January 4, 2017}\BibitemShut {NoStop}%
\bibitem [{\citenamefont {Svozil}(1994)}]{svozil-94}%
  \BibitemOpen
  \bibfield  {author} {\bibinfo {author} {\bibfnamefont {Karl}\ \bibnamefont
  {Svozil}},\ }\bibfield  {title} {\enquote {\bibinfo {title}
  {Extrinsic-intrinsic concept and complementarity},}\ }in\ \href {\doibase
  10.1007/978-3-642-48647-0\_15} {\emph {\bibinfo {booktitle} {Inside versus
  Outside}}},\ \bibinfo {series} {Springer Series in Synergetics},
  Vol.~\bibinfo {volume} {63},\ \bibinfo {editor} {edited by\ \bibinfo {editor}
  {\bibfnamefont {Harald}\ \bibnamefont {Atmanspacher}}\ and\ \bibinfo {editor}
  {\bibfnamefont {Gerhard~J.}\ \bibnamefont {Dalenoort}}}\ (\bibinfo
  {publisher} {Springer},\ \bibinfo {address} {Berlin Heidelberg},\ \bibinfo
  {year} {1994})\ pp.\ \bibinfo {pages} {273--288}\BibitemShut {NoStop}%
\bibitem [{\citenamefont {Eccles}(1990)}]{eccles:papal}%
  \BibitemOpen
  \bibfield  {author} {\bibinfo {author} {\bibfnamefont {John~C.}\ \bibnamefont
  {Eccles}},\ }\bibfield  {title} {\enquote {\bibinfo {title} {The mind-brain
  problem revisited: The microsite hypothesis},}\ }in\ \href {\doibase
  10.1017/S0960129510000344} {\emph {\bibinfo {booktitle} {The Principles of
  Design and Operation of the Brain}}},\ \bibinfo {editor} {edited by\ \bibinfo
  {editor} {\bibfnamefont {John~C.}\ \bibnamefont {Eccles}}\ and\ \bibinfo
  {editor} {\bibfnamefont {O.}~\bibnamefont {Creutzfeldt}}}\ (\bibinfo
  {publisher} {Springer},\ \bibinfo {address} {Berlin},\ \bibinfo {year}
  {1990})\ pp.\ \bibinfo {pages} {549--572}\BibitemShut {NoStop}%
\bibitem [{\citenamefont {Kochen}\ and\ \citenamefont
  {Specker}(1967)}]{kochen1}%
  \BibitemOpen
  \bibfield  {author} {\bibinfo {author} {\bibfnamefont {Simon}\ \bibnamefont
  {Kochen}}\ and\ \bibinfo {author} {\bibfnamefont {Ernst~P.}\ \bibnamefont
  {Specker}},\ }\bibfield  {title} {\enquote {\bibinfo {title} {The problem of
  hidden variables in quantum mechanics},}\ }\href {\doibase
  10.1512/iumj.1968.17.17004} {\bibfield  {journal} {\bibinfo  {journal}
  {Journal of Mathematics and Mechanics (now Indiana University Mathematics
  Journal)}\ }\textbf {\bibinfo {volume} {17}},\ \bibinfo {pages} {59--87}
  (\bibinfo {year} {1967})}\BibitemShut {NoStop}%
\bibitem [{\citenamefont {{Everett III}}(2012)}]{everett-collw}%
  \BibitemOpen
  \bibfield  {author} {\bibinfo {author} {\bibfnamefont {Hugh}\ \bibnamefont
  {{Everett III}}},\ }in\ \href {http://press.princeton.edu/titles/9770.html}
  {\emph {\bibinfo {booktitle} {The {E}verett Interpretation of Quantum
  Mechanics: Collected Works 1955-1980 with Commentary}}},\ \bibinfo {editor}
  {edited by\ \bibinfo {editor} {\bibfnamefont {Jeffrey~A.}\ \bibnamefont
  {Barrett}}\ and\ \bibinfo {editor} {\bibfnamefont {Peter}\ \bibnamefont
  {Byrne}}}\ (\bibinfo  {publisher} {Princeton University Press},\ \bibinfo
  {address} {Princeton, NJ},\ \bibinfo {year} {2012})\BibitemShut {NoStop}%
\bibitem [{\citenamefont {Jaynes}(1990)}]{Jaynes1990}%
  \BibitemOpen
  \bibfield  {author} {\bibinfo {author} {\bibfnamefont {Edwin~Thompson}\
  \bibnamefont {Jaynes}},\ }\bibfield  {title} {\enquote {\bibinfo {title}
  {Probability theory as logic},}\ }in\ \href {\doibase
  10.1007/978-94-009-0683-9\_1} {\emph {\bibinfo {booktitle} {Maximum Entropy
  and {B}ayesian Methods}}},\ \bibinfo {editor} {edited by\ \bibinfo {editor}
  {\bibfnamefont {Paul~F.}\ \bibnamefont {Foug{\`e}re}}}\ (\bibinfo
  {publisher} {Springer Netherlands},\ \bibinfo {address} {Dordrecht},\
  \bibinfo {year} {1990})\ pp.\ \bibinfo {pages} {1--16}\BibitemShut {NoStop}%
\bibitem [{\citenamefont {Freud}(1912, 1999)}]{Freud-1912}%
  \BibitemOpen
  \bibfield  {author} {\bibinfo {author} {\bibfnamefont {Sigmund}\ \bibnamefont
  {Freud}},\ }\bibfield  {title} {\enquote {\bibinfo {title}
  {{R}atschl{\"{a}}ge f{\"{u}}r den {A}rzt bei der psychoanalytischen
  {B}ehandlung},}\ }in\ \href
  {http://gutenberg.spiegel.de/buch/kleine-schriften-ii-7122/15} {\emph
  {\bibinfo {booktitle} {{G}esammelte {W}erke. {C}hronologisch geordnet.
  {A}chter {B}and. {W}erke aus den {J}ahren 1909--1913}}},\ \bibinfo {editor}
  {edited by\ \bibinfo {editor} {\bibfnamefont {Anna}\ \bibnamefont {Freud}},
  \bibinfo {editor} {\bibfnamefont {E.}~\bibnamefont {Bibring}}, \bibinfo
  {editor} {\bibfnamefont {W.}~\bibnamefont {Hoffer}}, \bibinfo {editor}
  {\bibfnamefont {E.}~\bibnamefont {Kris}}, \ and\ \bibinfo {editor}
  {\bibfnamefont {O.}~\bibnamefont {Isakower}}}\ (\bibinfo  {publisher}
  {Fischer},\ \bibinfo {address} {Frankfurt am Main},\ \bibinfo {year} {1912,
  1999})\ pp.\ \bibinfo {pages} {376--387}\BibitemShut {NoStop}%
\bibitem [{\citenamefont {Freud}(1912, 1958)}]{Freud-1912-e}%
  \BibitemOpen
  \bibfield  {author} {\bibinfo {author} {\bibfnamefont {Sigmund}\ \bibnamefont
  {Freud}},\ }\bibfield  {title} {\enquote {\bibinfo {title} {Recommendations
  to physicians practising psycho-analysis},}\ }in\ \href
  {https://www.pep-web.org/document.php?id=se.012.0109a\#p0109} {\emph
  {\bibinfo {booktitle} {The Standard Edition of the Complete Psychological
  Works of Sigmund Freud, Volume XII (1911-1913): The Case of {S}chreber,
  Papers on Technique and Other Works}}},\ \bibinfo {editor} {edited by\
  \bibinfo {editor} {\bibfnamefont {Anna Freud~Anna}\ \bibnamefont {Freud}},
  \bibinfo {editor} {\bibfnamefont {Alix}\ \bibnamefont {Strachey}}, \ and\
  \bibinfo {editor} {\bibfnamefont {Alan}\ \bibnamefont {Tyson}}}\ (\bibinfo
  {publisher} {The Hogarth Press and the Institute of Psycho-Analysis},\
  \bibinfo {address} {London},\ \bibinfo {year} {1912, 1958})\ pp.\ \bibinfo
  {pages} {109--120}\BibitemShut {NoStop}%
\bibitem [{\citenamefont {Hahn}(1930)}]{Hahn1930}%
  \BibitemOpen
  \bibfield  {author} {\bibinfo {author} {\bibfnamefont {Hans}\ \bibnamefont
  {Hahn}},\ }\bibfield  {title} {\enquote {\bibinfo {title} {{D}ie {B}edeutung
  der wissenschaftlichen {W}eltauffassung, insbesondere f{\"u}r {M}athematik
  und {P}hysik},}\ }\href {\doibase 10.1007/BF00208612} {\bibfield  {journal}
  {\bibinfo  {journal} {Erkenntnis}\ }\textbf {\bibinfo {volume} {1}},\
  \bibinfo {pages} {96--105} (\bibinfo {year} {1930})}\BibitemShut {NoStop}%
\bibitem [{\citenamefont {Carnap}(1931)}]{Carnap1931}%
  \BibitemOpen
  \bibfield  {author} {\bibinfo {author} {\bibfnamefont {Rudolf}\ \bibnamefont
  {Carnap}},\ }\bibfield  {title} {\enquote {\bibinfo {title} {{\"U}berwindung
  der {M}etaphysik durch logische {A}nalyse der {S}prache},}\ }\href {\doibase
  10.1007/BF02028153} {\bibfield  {journal} {\bibinfo  {journal} {Erkenntnis}\
  }\textbf {\bibinfo {volume} {2}},\ \bibinfo {pages} {219--241} (\bibinfo
  {year} {1931})}\BibitemShut {NoStop}%
\bibitem [{\citenamefont {Carnap}(1959)}]{Carnap-1931-engl}%
  \BibitemOpen
  \bibfield  {author} {\bibinfo {author} {\bibfnamefont {Rudolf}\ \bibnamefont
  {Carnap}},\ }\bibfield  {title} {\enquote {\bibinfo {title} {The elimination
  of metaphysics through logical analysis of language},}\ }in\ \href@noop {}
  {\emph {\bibinfo {booktitle} {Logical Positivism}}},\ \bibinfo {editor}
  {edited by\ \bibinfo {editor} {\bibfnamefont {Alfred~Jules}\ \bibnamefont
  {Ayer}}}\ (\bibinfo  {publisher} {Free Press},\ \bibinfo {address} {New
  York},\ \bibinfo {year} {1959})\ pp.\ \bibinfo {pages} {60--81},\ \bibinfo
  {note} {translated by Arthur Arp}\BibitemShut {NoStop}%
\bibitem [{\citenamefont {G{\"{o}}del}(1931)}]{godel1}%
  \BibitemOpen
  \bibfield  {author} {\bibinfo {author} {\bibfnamefont {Kurt}\ \bibnamefont
  {G{\"{o}}del}},\ }\bibfield  {title} {\enquote {\bibinfo {title} {{\"{U}}ber
  formal unentscheidbare {S}\"{a}tze der {P}rincipia {M}athematica und
  verwandter {S}ysteme},}\ }\href {\doibase 10.1007/s00605-006-0423-7}
  {\bibfield  {journal} {\bibinfo  {journal} {Monatshefte f{\"{u}}r Mathematik
  und Physik}\ }\textbf {\bibinfo {volume} {38}},\ \bibinfo {pages} {173--198}
  (\bibinfo {year} {1931})}\BibitemShut {NoStop}%
\bibitem [{\citenamefont {{T}uring}(1936-7 and 1937)}]{turing-36}%
  \BibitemOpen
  \bibfield  {author} {\bibinfo {author} {\bibfnamefont {Alan~M.}\ \bibnamefont
  {{T}uring}},\ }\bibfield  {title} {\enquote {\bibinfo {title} {On computable
  numbers, with an application to the {E}ntscheidungsproblem},}\ }\href
  {\doibase 10.1112/plms/s2-42.1.230, 10.1112/plms/s2-43.6.544} {\bibfield
  {journal} {\bibinfo  {journal} {Proceedings of the London Mathematical
  Society, Series 2}\ }\textbf {\bibinfo {volume} {42, 43}},\ \bibinfo {pages}
  {230--265, 544--546} (\bibinfo {year} {1936-7 and 1937})}\BibitemShut
  {NoStop}%
\bibitem [{\citenamefont {Smullyan}(1992,2020)}]{smullyan-92}%
  \BibitemOpen
  \bibfield  {author} {\bibinfo {author} {\bibfnamefont {Raymond~M.}\
  \bibnamefont {Smullyan}},\ }\href {\doibase
  10.1093/oso/9780195046724.001.0001} {\emph {\bibinfo {title} {{G}{\"{o}}del's
  Incompleteness Theorems}}},\ Oxford Logic Guides 19\ (\bibinfo  {publisher}
  {Oxford University Press},\ \bibinfo {address} {New York, Oxford},\ \bibinfo
  {year} {1992,2020})\BibitemShut {NoStop}%
\bibitem [{\citenamefont {Smullyan}(1993,2020)}]{Smullyan1993-SMURTF}%
  \BibitemOpen
  \bibfield  {author} {\bibinfo {author} {\bibfnamefont {Raymond~M.}\
  \bibnamefont {Smullyan}},\ }\href {\doibase
  10.1093/oso/9780195082326.001.0001} {\emph {\bibinfo {title} {Recursion
  Theory for Metamathematics}}},\ Oxford Logic Guides 22\ (\bibinfo
  {publisher} {Oxford University Press},\ \bibinfo {address} {New York,
  Oxford},\ \bibinfo {year} {1993,2020})\BibitemShut {NoStop}%
\bibitem [{\citenamefont {Smullyan}(1994)}]{book:486992}%
  \BibitemOpen
  \bibfield  {author} {\bibinfo {author} {\bibfnamefont {Raymond~M.}\
  \bibnamefont {Smullyan}},\ }\href@noop {} {\emph {\bibinfo {title}
  {Diagonalization and Self-Reference}}},\ \bibinfo {series} {Oxford Logic
  Guides}, Vol.~\bibinfo {volume} {27}\ (\bibinfo  {publisher} {Clarendon
  Press},\ \bibinfo {address} {New York, Oxford},\ \bibinfo {year}
  {1994})\BibitemShut {NoStop}%
\bibitem [{\citenamefont {Chaitin}(1987,2003)}]{chaitin3}%
  \BibitemOpen
  \bibfield  {author} {\bibinfo {author} {\bibfnamefont {Gregory~J.}\
  \bibnamefont {Chaitin}},\ }\href {\doibase 10.1017/CBO9780511608858} {\emph
  {\bibinfo {title} {Algorithmic Information Theory}}},\ \bibinfo {edition}
  {revised edition}\ ed.,\ Cambridge Tracts in Theoretical Computer Science,
  Volume 1\ (\bibinfo  {publisher} {Cambridge University Press},\ \bibinfo
  {address} {Cambridge},\ \bibinfo {year} {1987,2003})\BibitemShut {NoStop}%
\bibitem [{\citenamefont {{Augustine of
  Hippo}}(2019)}]{Augustinus-Confessiones}%
  \BibitemOpen
  \bibfield  {author} {\bibinfo {author} {\bibnamefont {{Augustine of
  Hippo}}},\ }\href
  {https://www.hackettpublishing.com/confessions-williams-translation-4222}
  {\emph {\bibinfo {title} {Confessions}}}\ (\bibinfo  {publisher} {Hackett
  Publishing Company, Inc.},\ \bibinfo {address} {Indianapolis, Indiana, USA},\
  \bibinfo {year} {2019})\ \bibinfo {note} {williams edition, translated by
  Thomas Williams}\BibitemShut {NoStop}%
\end{thebibliography}

\else


%

\fi

\end{document}